\documentclass[astrosymb,twocolumn,trackchanges]{aastex701}
\usepackage{amsmath}

\newcommand{\kms}{km\,s$^{-1}$}
\newcommand{\msun}{${\rm M}_\sun$}
\newcommand{\lsun}{${\rm L}_\sun$}
\newcommand{\methanol}{CH$_3$OH}
\newcommand{\metcyn}{CH$_3$CN}
\newcommand{\mettrans}{$J_{K_a,K_c}=18_{3,15}-17_{4,14}\,A,\,v_t=0$}
\newcommand{\metcyntrans}{$J_K = 12_K - 11_K$}
\newcommand{\metcyntransk}{$J_K = 12_3 - 11_3$}

\defcitealias{Olguin2021}{Paper I}
\defcitealias{Olguin2022}{Paper II}
\defcitealias{Taniguchi2023}{Paper III}
\defcitealias{Ishihara2024}{Paper IV}
\defcitealias{Sakai2025}{Paper V}
\defcitealias{Luo2026}{Paper VI}

\begin{document}

\title{Digging into the Interior of Hot Cores with ALMA (DIHCA). VII. Disk candidates around high-mass stars and evidence of anisotropic infall}

\correspondingauthor{Fernando A. Olguin}

\author[0000-0002-8250-6827]{Fernando A. Olguin} 
\email[show]{f.olguin@yukawa.kyoto-u.ac.jp}
\affiliation{Center for Gravitational Physics, Yukawa Institute for Theoretical Physics, Kyoto University, Kitashirakawa Oiwakecho, Sakyo-ku, Kyoto 606-8502, Japan}
\affiliation{National Astronomical Observatory of Japan, National Institutes of Natural Sciences, 2-21-1 Osawa, Mitaka, Tokyo 181-8588, Japan}
\affil{Institute of Astronomy and Department of Physics, National Tsing Hua University, Hsinchu 30013, Taiwan} 

\author[0000-0002-7125-7685]{Patricio Sanhueza} 
\email{psanhueza@astron.s.u-tokyo.ac.jp}
\affiliation{Department of Astronomy, School of Science, The University of Tokyo, 7-3-1 Hongo, Bunkyo, Tokyo 113-0033, Japan}

\author[0000-0002-0197-8751]{Yoko Oya}
\email{yoko.oya@yukawa.kyoto-u.ac.jp}
\affiliation{Center for Gravitational Physics, Yukawa Institute for Theoretical Physics, Kyoto University, Kitashirakawa Oiwakecho, Sakyo-ku, Kyoto 606-8502, Japan}

\author[0000-0001-6431-9633]{Adam Ginsburg}
\email{adam.g.ginsburg@gmail.com}
\affiliation{Department of Astronomy, University of Florida, P.O. Box 112055, Gainesville, FL, USA}

\author[0000-0003-3315-5626]{Maria T. Beltr\'an}
\email{maria.beltran@inaf.it}
\affiliation{INAF-Osservatorio Astrofisico di Arcetri, Largo E. Fermi 5, 50125 Firenze, Italy}

\author[0000-0002-6752-6061]{Kaho Morii}
\email{kaho.morii@cfa.harvard.edu}
\affiliation{Center for Astrophysics $|$ Harvard \& Smithsonian, 60 Garden Street, Cambridge, MA 02138, USA}

\author[0000-0003-1480-4643]{Roberto Galv\'an-Madrid}
\email{robertogalvanmadrid@gmail.com}
\affiliation{Instituto de Radioastronom\'ia y Astrof\'isica, Universidad Nacional Aut\'onoma de M\'exico, Antigua Carretera a P\'atzcuaro 8701, Ex-Hda. San Jos\'e de la Huerta, Morelia, Michoac\'an, M\'exico C.P. 58089}

\author[0000-0002-9774-1846]{Huei-Ru Vivien Chen}
\email{hchen@phys.nthu.edu.tw}
\affil{Institute of Astronomy and Department of Physics, National Tsing Hua University, Hsinchu 30013, Taiwan}

\author[0000-0003-4506-3171]{Qiuyi Luo} 
\email{luoqiuyi233@gmail.com}
\affiliation{Institute of Astronomy, Graduate School of Science, The University of Tokyo, 2-21-1 Osawa, Mitaka, Tokyo 181-0015, Japan}
\affiliation{Department of Astronomy, School of Science, The University of Tokyo, 7-3-1 Hongo, Bunkyo, Tokyo 113-0033, Japan}

\author[0000-0002-6907-0926]{Kei E. I. Tanaka}
\email{kt503i@gmail.com}
\affiliation{Department of Earth and Planetary Sciences, Institute of Science Tokyo, Meguro, Tokyo, 152-8551, Japan}

\author[0000-0002-8389-6695]{Suinan Zhang}
\email{suinan.zhang@gmail.com}
\affiliation{Shanghai Astronomical Observatory, Chinese Academy of Sciences, 80 Nandan Road, Shanghai 200030, People’s Republic of China}
\affiliation{Department of Earth and Planetary Sciences, Institute of Science Tokyo, Meguro, Tokyo, 152-8551, Japan}
\affiliation{National Astronomical Observatory of Japan, National Institutes of Natural Sciences, 2-21-1 Osawa, Mitaka, Tokyo 181-8588, Japan}

\author[0000-0002-8691-4588]{Yu Cheng}
\email{ycheng.astro@gmail.com}
\affiliation{National Astronomical Observatory of Japan, National Institutes of Natural Sciences, 2-21-1 Osawa, Mitaka, Tokyo 181-8588, Japan}

\author[0000-0001-5431-2294]{Fumitaka Nakamura}
\email{fumitaka.nakamura@nao.ac.jp}
\affiliation{National Astronomical Observatory of Japan, National Institutes of Natural Sciences, 2-21-1 Osawa, Mitaka, Tokyo 181-8588, Japan}
\affiliation{Astronomical Science Program, The Graduate University for Advanced Studies, SOKENDAI, 2-21-1 Osawa, Mitaka, Tokyo 181-8588, Japan}

\author[0000-0003-1275-5251]{Shanghuo Li}
\email{shli@nju.edu.cn}
\affiliation{School of Astronomy and Space Science, Nanjing University, Nanjing, China}
\affiliation{Key Laboratory of Modern Astronomy and Astrophysics, Nanjing University, Ministry of Education, Nanjing, China}

\author[0000-0003-4402-6475]{Kotomi Taniguchi}
\email{kotomi.taniguchi@nao.ac.jp}
\affiliation{National Astronomical Observatory of Japan, National Institutes of Natural Sciences, 2-21-1 Osawa, Mitaka, Tokyo 181-8588, Japan}

\author[0000-0003-1649-7958]{Guido Garay}
\email{guido@das.uchile.cl}
\affiliation{Departamento de Astronomía, Universidad de Chile, Casilla 36-D, Santiago, Chile}
\affiliation{Chinese Academy of Sciences South America Center for Astronomy, National Astronomical Observatories, Chinese Academy of Sciences, Beijing, 100101, People’s Republic of China}

\author[0000-0003-2384-6589]{Qizhou Zhang}
\email{qzhang@cfa.harvard.edu}
\affiliation{Center for Astrophysics $|$ Harvard \& Smithsonian, 60 Garden Street, Cambridge, MA 02138, USA}

\author[0000-0003-0769-8627]{Masao Saito}
\email{masao.saito@nao.ac.jp}
\affiliation{National Astronomical Observatory of Japan, National Institutes of Natural Sciences, 2-21-1 Osawa, Mitaka, Tokyo 181-8588, Japan}
\affiliation{Astronomical Science Program, The Graduate University for Advanced Studies, SOKENDAI, 2-21-1 Osawa, Mitaka, Tokyo 181-8588, Japan.}

\author[0000-0003-4521-7492]{Takeshi Sakai}
\email{takeshi.sakai@uec.ac.jp}
\affiliation{Graduate School of Informatics and Engineering, The University of Electro-Communications, Chofu, Tokyo 182-8585, Japan.}

\author[0000-0003-2619-9305]{Xing Lu}
\email{xinglu@shao.ac.cn}
\affiliation{Shanghai Astronomical Observatory, Chinese Academy of Sciences, 80 Nandan Road, Shanghai 200030, People’s Republic of China}

\author[]{Jixiang Weng}
\email{wengjixiang24@mails.ucas.ac.cn}
\affiliation{Shanghai Astronomical Observatory, Chinese Academy of Sciences, 80 Nandan Road, Shanghai 200030, People’s Republic of China}

\author[0000-0003-0990-8990]{Andr\'es E. Guzm\'an}
\email{andres.guzman.fernandez@gmail.com}
\affiliation{Joint Alma Observatory (JAO), Alonso de C\'ordova 3107, Vitacura, Santiago, Chile.}

\begin{abstract}
We study the kinematics of condensations in 30 fields forming high-mass stars with ALMA at a high-resolution of ${\sim}0\farcs08$ on average (${\sim}230$\,au). 
The presence of disks is important for feeding high-mass stars without feedback halting growth as their masses increase.
In the search for velocity gradients resembling rotation that can reveal the presence of disks, we analyze the emission of gas tracers in 49 objects using \methanol, \metcyn, and tentative detections of HNCO and cis-HCOOH. 
Most of the velocity distributions show velocity gradients indicative of rotation.
We reveal a total of 32 disk candidates, the largest sample to date that has been uniformly analyzed at a few hundred au scales in the high-mass regime.
Their position-velocity maps are generally asymmetric with one side brighter than the opposite.
We successfully fit a power law to the position-velocity maps of the disk candidates and find indices between $-0.5$ (Keplerian rotation) and $-1$ (rotation under specific angular momentum conservation) with a median of $-0.7$.
Under Keplerian rotation assumption, we estimate central masses, uncorrected for inclination, ranging between 7 to 45\,\msun.
Excluding outliers, the disk candidates are relatively more compact ($<$200\,au) and less massive ($<$5\,\msun) than previous results at coarser angular resolution.
We calculate an average Toomre-$Q$ parameter and find that most are gravitationally unstable (median of 0.5).
We conclude that these observations offer the first opportunity to separate the disk and envelope components of hot cores on a statistically significant sample, and confirm that anisotropic collapse plays an role in feeding high-mass (proto)stars.
\end{abstract}


\section{Introduction}\label{sec:intro}

The current paradigm of high-mass star formation requires non-spherical accretion of gas \citep[e.g.,][]{Yorke1999,Krumholz2009,Cunningham2011}.
This accretion flow can take the form of a disk: a flattened structure formed as a result of the conservation of angular momentum.
Early search for these structures at (sub)mm wavelengths resulted in relatively large disks and/or toroidal structures with radii of the order of 1000\,au \citep[see][and references therein]{Beltran2016}.
In addition to large sizes, these structures are massive, making them unstable and thus susceptible to fragmentation (allowing the formation of companions) or to develop substructures (e.g., spiral arms, \citealp{Johnston2020}).
While these observations have been supported by simulations \citep[e.g.,][]{Kuiper2018,Commercon2022,Meyer2017}, a newer picture has started to emerge with the raising number of high-resolution observations.

As the number of antennas of the Atacama Large Millimeter/submillimeter Array (ALMA) increased, the resulting higher sensitivity have allowed us to peek into scales smaller than 1000\,au in a timely manner.
These observations have allowed to find that disks in Galactic high-mass star-forming regions traced by dust emission can be more compact (radius $<500$\,au; e.g., \citealp{Ilee2018,Maud2019,Goddi2020,Tanaka2020,Guzman2020,Fernandez-Lopez2023}).
Confirming that such structures indeed correspond to rotationally supported or Keplerian disks has only been done in a few cases compared to those detected in low-mass sources.
An early search for disks around O-type star candidates by  \citet[][]{Cesaroni2017} only confirmed one Keplerian disk from  kinematics, reveal rotation signatures around another two, and provided three additional tentative candidates.
Case studies, as the one of \citet[][]{Maud2019} in G17.64+0.16, have provided additional examples of Keplerian disks around O-type stars.
\citet{Goddi2020} summarize disks-like structures observed by ALMA toward 14 hot cores forming O- and B-type stars and whose kinematics can be described by Keplerian rotation (their Appendix E).
Most recent results have expanded in some of these regions.
For instance, \citet{Zhang2022} found individual disks surrounded by a larger disk for the binary source G35.20-0.74 N core B initially studied by \citet{Sanchez-Monge2013}.
Newer observations have also allowed detailed modeling of the kinematics with multiple molecular tracers to derive the disk properties \citep[e.g., G339.88--1.23;][]{Zhang2019}.
Inside disk scales, tracers like water and salts have revealed Keplerian motions close to the forming stars \citep[e.g.,][]{Tanaka2020,Ginsburg2023}.
In the northern hemisphere at latitudes only reachable by the IRAM/NOEMA interferometer, the CORE large program has provided 13 disk candidates from 800\,au resolution observations \citep{Ahmadi2023}. 
Higher spatial resolution observations with IRAM/NOEMA targeting the Cygnus-X cloud found two Keplerian disks out of seven fragments \citep[][300\,au resolution]{Pan2025}.
Nevertheless, a systematic search for rotating structures in high-mass star forming regions with ALMA at higher resolutions is lacking.

\subsection{The DIHCA project}

In order to study the kinematics of gas feeding high-mass (proto)stars we have observed 30 fields at high-angular resolution as part of the Digging into the Interior of Hot Cores with ALMA (DIHCA) project.
These observations combine compact and extended ALMA 12m array configurations (see \S\ref{sec:obs}), and provide the continuum and line information necessary for detailed study of the gas distribution and kinematics of the inner 1000\,au of hot cores.
Case studies presenting the kinematics include G335.579--0.272 \citep[][hereafter Paper I and II, respectively]{Olguin2021,Olguin2022} and G336.01--0.82 \citep{Olguin2023}.
Based on the compact configuration data, \citet[][hereafter Paper III]{Taniguchi2023} determined the systemic velocity and temperatures of the main cores, while \citet[][hereafter Paper IV]{Ishihara2024} cataloged the continuum cores.
Using the same configuration, \citet[][hereafter Paper V]{Sakai2025} studied the deuterium fractionation of \methanol\ and determined the gas temperatures at peak $^{13}$CH$_3$OH emission.
The DIHCA sample selection is detailed in \citetalias{Ishihara2024}, and it comprises bright nearby sources which have the potential to form high-mass stars as determined by previous studies. 
The sample list is presented in Table~\ref{tab:obsprops:ch3oh} and among its targets include the well studied fields G11.92--0.61 \citep{Cyganowski2017,Ilee2018,Cyganowski2022,Sanhueza2025,Zhang2024}, G29.96--0.02 \citep{Cesaroni2017}, G333.23--0.06 \citep{Li2024}, G333.46--0.16 \citep{Saha2024}, G335.579--0.272 (\citealp{Avison2021}, \citetalias{Olguin2021}, \citetalias{Olguin2022}), G343.12--0.06 \citep{Tanaka2020,Zapata2024}, G35.03+0.35 A \citep{Beltran2014}, G35.20--0.74 N \citep{Sanchez-Monge2014,Zhang2022}, G351.77--0.54 \citep{Beuther2019,Beuther2025,Ginsburg2023}, G5.89--0.37 \citep{Zapata2020,Fernandez-Lopez2021}, IRAS 16562--3959 \citep{Guzman2020}, IRAS 18089--1732 \citep{Beuther2004,Sanhueza2021}, IRAS 18162--2048 \citep[GGD 27--MM1,][]{Anez-Lopez2020,Fernandez-Lopez2023,Carrasco-Gonzalez2012}, NGC 6334I \citep{Cortes2024}, NGC 6334I(N) \citep[][]{Cortes2021,Li2025}, and W33A \citep{Maud2017}.
Complementary to these studies, here we study the main condensations in these fields as well as secondary sources in them (if any).
These condensations are over-dense structures where a single or a close multiple system is expected to form within cores. 
While cores can fragment into wider multiple systems, i.e., host one or more condensations, multiple systems in condensations are expected to form due to disk fragmentation.
As such, we expect condensations to consist of at least one (proto)star with a gas envelope potentially hosting a circumstellar disk, hence the gas kinematics are increasingly dominated by the gravitational potential of the central source and rotation induced by the initial angular momentum of the core.
In the DIHCA sample, low- to high-mass condensations in the cores identified in \citetalias{Ishihara2024} are cataloged in \citet[][in press; hereafter Paper VI]{Luo2026}.
This catalog uses \textsc{PyBDSF}\footnote{\url{https://pybdsf.readthedocs.io/en/latest/}} \citep{2015ascl.soft02007M} to identify and fit 2-D Gaussians to the condensations from the continuum maps resulting in positions and radii, and exploits previous DIHCA efforts \citepalias[][]{Taniguchi2023} to assign temperatures to each condensation.

\begin{deluxetable*}{lcLccc}
\tablecaption{\methanol\ \mettrans\ observations\label{tab:obsprops:ch3oh}}
\tablewidth{\textwidth}
\tablehead{
\colhead{Source} & \colhead{D} & \colhead{Beam} & \colhead{Res.} & \colhead{rms} & \colhead{Chan.}\\
\colhead{} & \colhead{(kpc)} & \colhead{maj$\times$min (P.A.)} & \colhead{(au)} & \colhead{(mJy\,beam$^{-1}$)} & \colhead{(kHz)}\\
\colhead{} & \colhead{(1)} & \colhead{(2)} & \colhead{(3)} & \colhead{(4)} & \colhead{(5)}
}
\startdata
G10.62--0.38     & 4.95 & 0\farcs075\times0\farcs055\,(76\degr)  & 320 & 1.2 & 976.452 \\
G11.1--0.12      & 3.00 & 0\farcs074\times0\farcs052\,(74\degr)  & 190 & 1.2 & 976.452 \\
G11.92--0.61     & 3.37 & 0\farcs074\times0\farcs060\,(88\degr)  & 220 & 1.2 & 976.452 \\
G14.22--0.50 S   & 1.90 & 0\farcs072\times0\farcs052\,(68\degr)  & 120 & 1.5 & 488.285 \\
G24.60+0.08      & 3.45 & 0\farcs112\times0\farcs085\,(-45\degr) & 340 & 2.4 & 488.284 \\
G29.96--0.02     & 5.26 & 0\farcs082\times0\farcs072\,(-80\degr) & 400 & 1.7 & 488.295 \\
G333.12--0.56    & 3.30 & 0\farcs111\times0\farcs068\,(-32\degr) & 290 & 2.2 & 488.284 \\
G333.23--0.06    & 5.20 & 0\farcs070\times0\farcs041\,(55\degr)  & 280 & 1.7 & 488.297 \\
G333.46--0.16    & 2.90 & 0\farcs109\times0\farcs068\,(-32\degr) & 250 & 2.0 & 488.284 \\
G335.579--0.272  & 3.25 & 0\farcs070\times0\farcs043\,(50\degr)  & 180 & 1.8 & 488.298 \\
G335.78+0.17     & 3.20 & 0\farcs109\times0\farcs067\,(-42\degr) & 270 & 2.0 & 488.284 \\
G336.01--0.82    & 3.10 & 0\farcs107\times0\farcs069\,(-43\degr) & 270 & 2.2 & 488.285 \\
G34.43+0.24 MM1  & 3.03 & 0\farcs078\times0\farcs069\,(77\degr)  & 220 & 1.6 & 488.294 \\
G34.43+0.24 MM2  & 3.03 & 0\farcs128\times0\farcs082\,(-56\degr) & 310 & 1.7 & 976.577 \\
G343.12--0.06    & 2.90 & 0\farcs102\times0\farcs072\,(-79\degr) & 250 & 1.4 & 488.292 \\
G35.03+0.35 A    & 2.32 & 0\farcs078\times0\farcs071\,(82\degr)  & 170 & 1.6 & 488.294 \\
G35.13--0.74     & 2.20 & 0\farcs126\times0\farcs083\,(-56\degr) & 230 & 1.8 & 976.578 \\
G35.20--0.74 N   & 2.19 & 0\farcs124\times0\farcs107\,(-73\degr) & 250 & 1.7 & 488.294 \\
G351.77--0.54    & 2.00 & 0\farcs139\times0\farcs079\,(-65\degr) & 210 & 2.1 & 488.238 \\
G5.89--0.37      & 2.99 & 0\farcs075\times0\farcs055\,(70\degr)  & 190 & 1.5 & 976.458 \\
IRAS 16562--3959 & 2.39 & 0\farcs098\times0\farcs069\,(86\degr)  & 200 & 1.7 & 488.291 \\
IRAS 18089--1732 & 2.34 & 0\farcs077\times0\farcs055\,(71\degr)  & 150 & 1.4 & 976.449 \\
IRAS 18151--1208 & 3.00 & 0\farcs094\times0\farcs054\,(87\degr)  & 210 & 1.2 & 976.456 \\
IRAS 18162--2048 & 1.30 & 0\farcs075\times0\farcs060\,(88\degr)  &  90 & 1.5 & 976.451 \\
IRAS 18182--1433 & 3.58 & 0\farcs101\times0\farcs054\,(-87\degr) & 260 & 1.4 & 976.457 \\
IRAS 18337--0743 & 3.80 & 0\farcs108\times0\farcs084\,(-42\degr) & 360 & 2.3 & 488.285 \\
IRDC 18223--1243 & 3.40 & 0\farcs096\times0\farcs055\,(-88\degr) & 250 & 1.3 & 976.455 \\
NGC 6334I        & 1.35 & 0\farcs087\times0\farcs065\,(80\degr)  & 100 & 1.7 & 488.286 \\
NGC 6334I(N)     & 1.35 & 0\farcs087\times0\farcs067\,(82\degr)  & 100 & 1.6 & 488.286 \\
W33A             & 2.53 & 0\farcs074\times0\farcs055\,(78\degr)  & 160 & 1.2 & 976.449 \\ 
\enddata
\tablecomments{(1) Distance to the source. 
(2) Beam sizes correspond to the common beam within the selected range used for the moment maps. In general, the beam varies less than 1\% within the selected ranges.
(3) Spatial resolution derived from the geometric mean of the beam axes at the source distance. 
(4) Noise level estimated as the median absolute deviation of the data cube.
(5) Channel width. A channel width of ${\sim}448.3$\,kHz corresponds to a velocity width of ${\sim}0.63-0.67$\,\kms. The spectral resolution is ${\sim}976.5$\,kHz.
}
\end{deluxetable*}

In this paper we present an overview of the kinematics at few hundred au resolution of the 30 observed fields in the DIHCA project with a focus on the rotation of gas structures.
Except for a few individual cases, previous results lack the resolution to separate disk (a few hundred au in radius) and envelope (few hundred to thousands of au in radius) kinematics. 
This DIHCA study is designed to conduct the first systematic and statistical study on disk rotation in high-mass hot cores with sufficient resolution.
A detailed modeling of the kinematics with 2-D axisymmetric models will be the focus of a subsequent paper.
The paper is structured as follows: \S\ref{sec:obs} presents a summary of the observations and molecular gas tracers used, \S\ref{sec:results} presents the kinematics of the sources, \S\ref{sec:discussion} discusses our results, and \S\ref{sec:conclusions} summarizes and presents our main conclusions.

\section{Observations} \label{sec:obs}

The observations of the 30 fields were carried out by ALMA in band 6 (1.3\,mm, 220\,GHz; Project IDs: 2016.1.01036.S, 2017.1.00237.S; PI: Patricio Sanhueza) utilizing two array configurations.
The compact configuration (C-5) observations were made during November 2016, May and July 2017, January and November 2018.
The extended configuration (C-8) observations were performed during September and November 2017, and July 2019.
A detailed description of the compact configuration data is presented in \citetalias{Ishihara2024}, while \citetalias{Luo2026} presents a summary of the extended configuration observations.
We calibrated the data following the standard procedure with the appropriate CASA pipeline release (see \citetalias{Ishihara2024}, \citetalias{Luo2026}).
Due to limitations on the data transfer on earlier observations, the channels of some spectral windows were averaged.
Hence, some observations have a channel width of ${\sim}488.3$\,kHz (${\sim}0.63-0.67$\,\kms), while others have channel widths equivalent to the spectral resolution (${\sim}976.5$\,kHz${\sim}1.26-1.34$\,\kms).

With the exception of four fields (IRAS 18162--2048, NGC~6334I, NGC~6334I(N) and W33A), the continuum-subtracted visibilities were obtained using the procedure described in \citetalias{Olguin2021} \citep[\textsc{GoContinuum} v2.0.0,][]{2020zndo...4302846O}.
The phase centers of the visibilities for the extended and compact configurations were aligned (if needed) and then the visibilities were concatenated.
For the remaining four fields, image artifacts were observed after imaging the continuum-subtracted visibilities in the form of negatives bowls in a specific direction common to all the bright continuum sources in some cases or curved spectra in others, probably introduced by the CASA task \textsc{uvcontsub}.
To produce continuum-subtracted cubes for these four fields, we utilized \textsc{statcont} \citep{Sanchez-Monge2018} on the cubes produced by imaging the combined non continuum-subtracted visibilities with the corrected sigma-clip algorithm and a noise level of 1.5\,mJy\,beam$^{-1}$.
Although variations of the sigma-clip algorithm is used by both \textsc{GoContinuum} (asymmetric sigma-clipping) and \textsc{statcont}, \citetalias{Olguin2021} found that similar results are obtained.
We use YCLEAN version v2.3.0 \citep{olguin_2025_17197133} to image \methanol\ \mettrans\ (233.795666\,GHz, $E_u/k_B=447$\,K) for the 30 fields (Table~\ref{tab:obsprops:ch3oh}).
The YCLEAN code automatically and iteratively produces incremental masks and cleaning thresholds based on the beam secondary lobe level \citep{Contreras18} which are then used by the CASA \textsc{tclean} task to produce clean cubes.
For the \textsc{tclean} task, we used the Hogbom deconvolver with Briggs weighting and a robust parameter of 0.5.
In addition, for a subset of 22 fields we image an entire spectral window where the \metcyn\ \metcyntrans\ $K$ ladder is located (hereafter 220\,GHz spectral window) using YCLEAN, which is the only spectral window with the two channels per spectral resolution element for all fields (Table~\ref{tab:obsprops:ch3cn}) .
The chosen \methanol\ transition has shown good results not only tracing rotating structures but also substructures like accretion flows \citep[e.g.,][]{Olguin2023}, even considering that its $E_u/k_B$ is relatively high.
\metcyn\ is a well known kinematics tracer, however it may trace more extended gas and could easily become optically thick toward the central sources \citep[e.g.,][]{Olguin2022} and has slightly lower angular resolution in our observation setup (typically a 20\,au difference in spatial resolution and up to 80\,au in the worst case, cf. Tables~\ref{tab:obsprops:ch3oh} and \ref{tab:obsprops:ch3cn}).
In particular cases, we use additional lines in an attempt to extract as much information about the kinematics as possible.
Hence, when the \methanol\ transition is marginally or not detected or more channels per line are needed to explore the kinematics we use \metcyn\ \metcyntransk\ (220.7090165\,GHz, $E_u/k_B=133$\,K). 
When these two fail, we explore other lines in the 220\,GHz spectral window (Table~\ref{tab:obsprops:add}).
See \S\ref{sec:discussion:tracers} for a discussion of the implications on the derived source properties when using different tracers.

\begin{deluxetable*}{lLccc}
\tablecaption{\metcyn\ \metcyntransk\ observations\label{tab:obsprops:ch3cn}}
\tablewidth{\textwidth}
\tablehead{
\colhead{Source} & \colhead{Beam} & \colhead{Res.} & \colhead{rms} & \colhead{Chan.}\\
\colhead{} & \colhead{maj$\times$min (P.A.)} & \colhead{(au)} & \colhead{(mJy\,beam$^{-1}$)} & \colhead{(kHz)}
}
\startdata
G10.62--0.38     & 0\farcs081\times0\farcs065\,(90\degr)  & 360 & 1.5 & 488.226\\ 
G11.1--0.12      & 0\farcs081\times0\farcs059\,(79\degr)  & 210 & 1.2 & 488.226\\ 
G11.92--0.61     & 0\farcs101\times0\farcs076\,(-68\degr) & 300 & 1.2 & 488.225\\ 
G14.22--0.50 S   & 0\farcs076\times0\farcs054\,(64\degr)  & 120 & 1.6 & 488.285\\ 
G24.60+0.08      & 0\farcs115\times0\farcs091\,(-43\degr) & 350 & 2.2 & 488.284\\ 
G333.12--0.56    & 0\farcs116\times0\farcs077\,(-34\degr) & 310 & 1.9 & 488.284\\ 
G333.23--0.06    & 0\farcs077\times0\farcs043\,(55\degr)  & 300 & 1.6 & 488.297\\ 
G333.46--0.16    & 0\farcs114\times0\farcs076\,(-33\degr) & 270 & 2.0 & 488.284\\ 
G34.43+0.24 MM2  & 0\farcs120\times0\farcs079\,(-57\degr) & 300 & 1.8 & 488.288\\ 
G35.03+0.35 A    & 0\farcs085\times0\farcs077\,(-88\degr) & 190 & 1.9 & 488.294\\ 
G35.13--0.74     & 0\farcs119\times0\farcs079\,(-57\degr) & 210 & 1.9 & 488.289\\ 
G5.89--0.37      & 0\farcs088\times0\farcs077\,(-34\degr) & 250 & 1.7 & 488.229\\ 
IRAS 16562--3959 & 0\farcs108\times0\farcs065\,(82\degr)  & 200 & 1.5 & 488.291\\ 
IRAS 18089--1732 & 0\farcs096\times0\farcs078\,(-36\degr) & 200 & 1.7 & 488.225\\ 
IRAS 18151--1208 & 0\farcs099\times0\farcs057\,(86\degr)  & 230 & 1.3 & 488.228\\ 
IRAS 18162--2048 & 0\farcs082\times0\farcs067\,(-86\degr) & 100 & 1.4 & 488.225\\ 
IRAS 18182--1433 & 0\farcs107\times0\farcs057\,(-87\degr) & 280 & 1.4 & 488.229\\ 
IRAS 18337--0743 & 0\farcs111\times0\farcs090\,(-40\degr) & 380 & 2.2 & 488.285\\ 
IRDC 18223--1243 & 0\farcs102\times0\farcs058\,(-89\degr) & 260 & 1.3 & 488.227\\ 
NGC 6334I        & 0\farcs093\times0\farcs065\,(75\degr)  & 110 & 1.9 & 488.286\\ 
NGC 6334I(N)     & 0\farcs095\times0\farcs066\,(78\degr)  & 110 & 1.5 & 488.286\\ 
W33A             & 0\farcs081\times0\farcs061\,(84\degr)  & 180 & 1.8 & 488.224\\ 
\enddata
\tablecomments{Columns are the same as in Table~\ref{tab:obsprops:ch3oh}, except for the distance column.
}
\end{deluxetable*}

\begin{deluxetable*}{llLccLc}
\tablecaption{Additional lines used from the 220\,GHz spectral window\label{tab:obsprops:add}}
\tablewidth{\textwidth}
\tablehead{
\colhead{Source} & \colhead{Molecule} & \colhead{Transition} & \colhead{Rest Freq.} & \colhead{$E_u/k_B$} & \colhead{Beam} & \colhead{Res.} \\
\colhead{} & \colhead{} & \colhead{} & \colhead{(GHz)} & \colhead{(K)} & \colhead{maj$\times$min (P.A.)} & \colhead{(au)} 
}
\startdata
G333.46--0.16    & HNCO    & J_{K_a,K_c}=10_{0,10}-9_{0,9} & 219.79832   &  58 & 0\farcs115\times0\farcs076\,(-33\degr) & 270 \\
IRAS 18162--2048 & c-HCOOH & J_{K_a,K_c}=10_{4,6}-9_{4,5}  & 219.9729675 & 116 & 0\farcs082\times0\farcs066\,(-89\degr) & 100 \\
NGC 6334I        & HNCO    & J_{K_a,K_c}=10_{3,8}-9_{3,7}  & 219.6567695 & 433 & 0\farcs096\times0\farcs065\,(76\degr)  & 110 \\
\enddata
\tablecomments{The molecular line transitions assigned to the line emission are tentative.
Noise levels and channel widths per source are the same as in Table~\ref{tab:obsprops:ch3cn}.
}
\end{deluxetable*}

\subsection{Ancillary data}

We use the DIHCA compact configuration data targeting the SiO $J=5-4$ transition (217.10498\,GHz) to search for outflows not reported in the literature.
Although these data have lower angular resolution ($>0\farcs5$), outflows are expected to be extended thus its emission can be better recovered by the compact configuration. 
These data are only used to determine the direction of potential outflows and support the rotation origin of any velocity gradient detected in the molecules above.
A detailed study is necessary to confirm these outflows.
Nevertheless, these data were reduced in the same way as those in \citetalias{Olguin2021} using YCLEAN with \textsc{tclean}'s multiscale deconvolver (scales 0, 5, 15) and Briggs weighting with a robust parameter of 0.5.

\section{Results} \label{sec:results}

We follow the condensation naming and positions from the source catalog presented in \citetalias{Luo2026}.
Here we analyze the line emission from 49 condensations in \methanol\ \mettrans\ (19 condensations), \metcyn\ \metcyntransk\ (26 condensations), HNCO $J_{K_a,K_c}=10_{0,10}-9_{0,9}$ (1 condensation) and $J_{K_a,K_c}=10_{3,8}-9_{3,7}$ (2 condensations), and c-HCOOH $J_{K_a,K_c}=10_{4,6}-9_{4,5}$ (1 condensation).
Hereafter, we will refer to sources and condensations indistinctly.

\subsection{Line emission}

Table~\ref{tab:props} lists all the observed fields and their respective condensations (if any) where molecular line emission was detected over the 5$\sigma$ level (with $\sigma$ typically in the 1.0--2.5\,mJy\,beam$^{-1}$ range).
Of the 30 imaged fields, molecular line emission over the 5$\sigma$ level from the aforementioned transitions was detected toward 25 fields.
The total number of condensations with line emission in these 25 fields is 49.
The fields not listed in Table~\ref{tab:props} are: G24.60+0.08, G34.43+0.24 MM2, IRAS 18151--1208, IRAS 18337--0743, and IRDC 18223--1243.
Of these, \metcyn\ emission is marginally detected over the 5$\sigma$ level in a few cores in IRAS 18151--1208, G24.60+0.08 and IRDC 18223--1243, but not extended enough to resolve their kinematics.
Although line emission was detected toward some of these fields in the compact configuration data at core scales \citepalias[e.g., IRDC 18223--1243;][]{Taniguchi2023}, the combined dataset does not provide enough valid pixels for a study of their kinematics.  
In G34.43+0.24 MM2 and IRAS 18337--0743 neither molecule is detected over the 5$\sigma$ level toward the cataloged condensations, likely due to the low sensitivity of the observation setup combined with fainter line emission in these sources.
Other sources in the fields listed in Table~\ref{tab:props} that may be forming high-mass stars also do not show significant line emission to properly study their kinematics (e.g., G14.22--0.50 S ALMA2).
Additional observations with higher sensitivity may help to study the kinematics in these sources in the future.

\begin{figure*}
\plotone{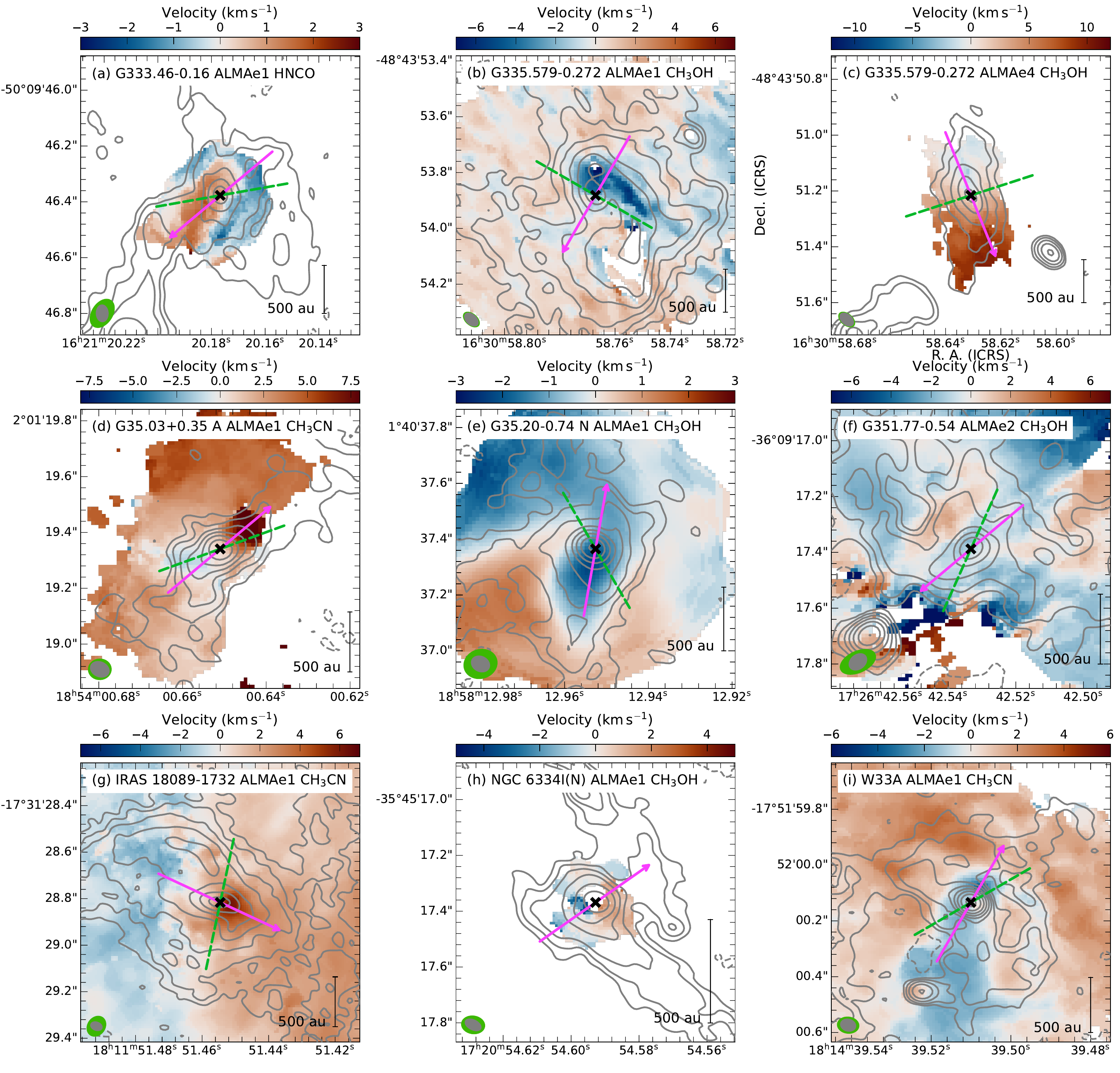}
\caption{First order moment maps from the labeled molecule emission (color scale) and continuum emission (contours) of the detected condensations.
The contour levels are 5, 10, 20, ... $\times \sigma$ where $\sigma$ is the noise level estimated as the standard deviation from the median absolute deviation of the continuum maps.
The black crosses mark the position of the condensations in Table~\ref{tab:props}, while the pink arrows show the direction of the velocity gradient used to calculate the position-velocity maps in Figure~\ref{fig:pvmaps}.
The green dashed lines show the direction of the (tentative) outflows.
The synthesized beam of the continuum (gray ellipse) and the moment maps (green ellipse; see Tables~\ref{tab:obsprops:ch3oh}-\ref{tab:obsprops:add}) are shown in the bottom left corner.
The complete figure set (48 images) is available in the online journal.
}\label{fig:mom1}
\end{figure*}

Figure~\ref{fig:mom1} shows the first order moment maps (intensity weighted velocity) toward individual condensations.
As expected from their upper-level energies $E_u/k_B$, the selected transition of \methanol\ is relatively more compact than the \metcyn\ one, with the former generally tracing closer to the continuum emission.
There are sources, however, where \methanol\ is more extended than the continuum at the 5$\sigma$ level (e.g., G335.579--0.272 ALMAe1, G35.20--0.74 N ALMAe1).
On the other hand, there are sources whose velocity gradient is not clear (e.g., NGC 6334I(N) ALMAe1) or first moment map emission is compact (e.g., IRAS 16562--3959 ALMAe5).
These may be the result of line emission becoming optically thick toward the peak continuum intensity or absorbed by the continuum, or the sensitivity is insufficient.
Nevertheless, we keep these sources to further analyze their emission and kinematics through position-velocity (PV) maps.
A few fields have extended emission covering several cores cataloged in \citetalias{Ishihara2024} where it is not possible to assess if the emission belongs to a specific source, and thus we exclude these sources in the analysis.
These cases include G5.89--0.37, where part of the emission is associated with an \ion{H}{2} region, and NGC 6334I where extended emission is found to be connecting different sources.
Similarly, for W33A, we found that the kinematics associated with the object labeled as ALMAe4 could only be partially separated from the cloud emission.

\startlongtable
\begin{deluxetable*}{lccRRCCCc}
\tablecaption{Source detections\label{tab:props}}
\tabletypesize{\footnotesize }
\tablehead{
\colhead{Source} & \colhead{ALMAe} & \colhead{Mol.} & \colhead{R. A.}     & \colhead{Decl.}                     & \colhead{$v_{LSR}$} & \colhead{P.A.} & \colhead{Out. P.A.} & \colhead{P.A.} \\
\colhead{}       & \colhead{Source} &               & \colhead{$[h:m:s]$} & \colhead{$[\degr:\arcmin:\arcsec]$} &  \colhead{(\kms)}      & \colhead{(deg)} & \colhead{(deg)} & \colhead{Ref.} \\
\colhead{}       & \colhead{(1)}  & \colhead{(2)}  & \colhead{(3)}       & \colhead{(4)}                        & \colhead{(5)}      & \colhead{(6)} & \colhead{(7)} & \colhead{(8)}
}
\startdata
G10.62--0.38     & 1      & \metcyn   & 18:10:28.60755 & -19:55:49.4614 &  0.85   & 0   & \nodata & \nodata \\  
                 & 2      & \metcyn   & 18:10:28.69080 & -19:55:49.8281 &  -1.51  & -30 & \nodata & \nodata \\ 
                 & 3      & \metcyn   & 18:10:28.73003 & -19:55:49.4837 &  -1.35  & 180 & \nodata & \nodata \\ 
G11.1--0.12      & 1      & \metcyn   & 18:10:28.24854 & -19:22:30.3222 &  31.85  & -65 & \nodata & \nodata \\ 
G11.92--0.61     & 1      & \metcyn   & 18:13:58.11097 & -18:54:20.2010 &  35.1   & 129 & 53      & [1]     \\  
                 & 4      & \metcyn   & 18:13:58.13491 & -18:54:16.2731 &  35.7   & 225 & \nodata\tablenotemark{a} & \nodata \\ 
                 & 6      & \metcyn   & 18:13:58.12719 & -18:54:20.7199 &  37.0   & -15 & \nodata & \nodata \\ 
G14.22--0.50 S   & 3      & \methanol & 18:18:12.86247 & -16:57:20.3666 &  22.0   & 170 & 75      & \nodata \\   
G29.96--0.02     & 1      & \methanol & 18:46:03.77860 & -02:39:22.3746 &  97.25  & 155\tablenotemark{b} & 142 & [2]   \\   
G333.12--0.56    & 1      & \methanol & 16:21:35.37589 & -50:40:56.6044 &  -58.47 & 135 & 40      & \nodata \\ 
                 & 8      & \metcyn   & 16:21:36.25475 & -50:40:47.2340 &  -54.10 & 90  & -35     & \nodata \\ 
G333.23--0.06    & 6      & \metcyn   & 16:19:50.88363 & -50:15:10.5272 &  -87.04 & 220 & 133\tablenotemark{c}& [3]    \\ 
                 & 17     & \metcyn   & 16:19:50.87824 & -50:15:10.6063 &  -87.04 & 125 & 221\tablenotemark{c}& [3]    \\ 
G333.46--0.16    & 1      & HNCO      & 16:21:20.17674 & -50:09:46.3768 &  -43.55 & 130 & 100     & \nodata \\ 
                 & 2      & \metcyn   & 16:21:20.17290 & -50:09:48.8762 &  -43.2  & 130 & 209     & [4]     \\ 
G335.579--0.272  & 1      & \methanol & 16:30:58.76705 & -48:43:53.8816 &  -46.9  & 150 & 240     & [5]     \\ 
                 & 4      & \methanol & 16:30:58.63085 & -48:43:51.2159 &  -53.1  & 202 & -72     & [6, 7]  \\ 
G335.78+0.17     & 1      & \methanol & 16:29:47.33461 & -48:15:52.2666 &  -48.94 & 215 & -30     & \nodata \\ 
                 & 2      & \methanol & 16:29:46.12974 & -48:15:49.9512 &  -50.51 & 65  & \nodata & \nodata \\ 
G336.01--0.82    & 3      & \methanol & 16:35:09.25854 & -48:46:47.6623 &  -47.2  & 125 & 205     & [8]     \\ 
G34.43+0.24 MM1  & 2      & \methanol & 18:53:18.00684 & +01:25:25.4228 &  58.44  & 280 & 200     & \nodata \\ 
G343.12--0.06    & 1      & \methanol & 16:58:17.20735 & -42:52:07.4161 &  -33.71 & 230 & 135     & [9, 10] \\ 
G35.03+0.35 A    & 1      & \metcyn   & 18:54:00.65099 & +02:01:19.3410 &  45.94  & 310 & 110     & \nodata \\ 
G35.13--0.74     & 1      & \metcyn   & 18:58:06.13626 & +01:37:07.4306 &  35.93  & 45  & 145     & \nodata \\ 
                 & 2      & \metcyn   & 18:58:06.16874 & +01:37:08.1602 &  34.39  & 75  & \nodata & \nodata \\ 
                 & 7      & \metcyn   & 18:58:06.27972 & +01:37:07.2029 &  35.53  & 95  & 0       & \nodata \\ 
G35.20--0.74 N   & 1      & \methanol & 18:58:12.95261 & +01:40:37.3652 &  32.11  & -10\tablenotemark{d} & 30 & [11]   \\ 
                 & 2      & \methanol & 18:58:13.03690 & +01:40:35.9257 &  30.0   & -23 & 3       & [11, 12]\\ 
G351.77--0.54    & 2      & \methanol & 17:26:42.53329 & -36:09:17.3878 &  -3.83  & 130 & -24     & [13, 14]\\ 
G5.89--0.37      & 1      & \metcyn   & 18:00:30.63209 & -24:04:03.0057 &  7.92   & -30 & \nodata & \nodata \\ 
IRAS 16562--3959 & 1      & \metcyn   & 16:59:41.62564 & -40:03:43.6385 &  -16.86 & 10  & 100     & [15]    \\ 
                 & 5      & \metcyn   & 16:59:41.08769 & -40:03:39.0848 &  -11.65 & 225 & 285     & \nodata \\ 
IRAS 18089--1732 & 1      & \metcyn   & 18:11:51.45454 & -17:31:28.8163 &  32.88  & 245 & -12\tablenotemark{e} &\nodata\\ 
                 & 2      & \metcyn   & 18:11:51.39962 & -17:31:29.9457 &  33.02  & 315 & 35      & \nodata \\ 
IRAS 18162--2048 & 1      & c-HCOOH   & 18:19:12.09528 & -20:47:30.9566 &  45.69  & -70 & 20      & [16, 17]\\ 
IRAS 18182--1433 & 1      & \metcyn   & 18:21:08.97822 & -14:31:47.5970 &  60.64  & 0   & 90      & \nodata \\ 
                 & 2      & \metcyn   & 18:21:09.04928 & -14:31:47.7925 &  62.78  & 40  & 155     & \nodata \\ 
                 & 5      & \metcyn   & 18:21:09.12462 & -14:31:48.5998 &  61.12  & 10  & 130     & \nodata \\ 
                 & 11     & \metcyn   & 18:21:09.01613 & -14:31:47.9401 &  62.7   & 40  & \nodata & \nodata \\ 
NGC 6334 I       & 1      & HNCO      & 17:20:53.41869 & -35:46:57.8969 &  -8.0   & 60  & 0       & \nodata \\ 
                 & 3      & \methanol & 17:20:53.64398 & -35:46:54.8823 &  -4.7   & 50  & \nodata & \nodata \\ 
                 & 4      & HNCO      & 17:20:53.16652 & -35:46:59.1704 &  -6.92  & 20  & 250     & \nodata \\ 
NGC 6334I(N)     & 1      & \methanol & 17:20:54.59280 & -35:45:17.3686 &  -3.9   & -55 & \nodata & \nodata \\ 
                 & 2      & \methanol & 17:20:55.19138 & -35:45:03.9525 &  -1.77  & 310 & 35      & \nodata \\ 
                 & 8      & \metcyn   & 17:20:54.61962 & -35:45:08.6715 &  -7.2   & 130 & \nodata & \nodata \\ 
                 & 9      & \metcyn   & 17:20:54.87122 & -35:45:06.4309 &  -6.7   & 130 & 40      & \nodata \\ 
                 & 14     & \metcyn   & 17:20:54.83871 & -35:45:06.0527 &  -5.0   & 160 & \nodata & \nodata \\ 
W33A             & 1      & \metcyn   & 18:14:39.51004 & -17:52:00.1338 &  37.9   & -30 & 120     & [18]    \\ 
                 & 4      & \metcyn   & 18:14:39.52227 & -17:52:00.4500 &  37.9   & 90  & \nodata & \nodata \\ 
\enddata
\tablecomments{(1) Extended configuration condensation index.
(2) Molecule analyzed.
(3) \& (4) Continuum position in ICRS system.
(5) Systemic velocity.
(6) Velocity gradient P.A. (east from north) as shown in the fist order moment map in Figure~\ref{fig:mom1}.
(7) Outflow direction P.A.
(8) Reference for the P.A. of the velocity gradient, the outflow or both as specified in the Table References. 
Note that some outflow angles are not quoted in the references, but they are measured from the observations presented there.}
\tablenotetext{a}{\citet{Cyganowski2011} lists an outflow that may be associated to this source, however we could not determine its P.A.}
\tablenotetext{b}{\citet{Cesaroni2017} used a P.A. orthogonal to the outflow direction. However \citet{Sanchez-Monge2019} showed that there is an accretion flow close to the outflow axis.}
\tablenotetext{c}{\citet{Li2024} found several outflow candidates in this region using DIHCA compact configuration data (angular resolution ${\sim}0\farcs5$), here we assign them based on the direction with respect to the \metcyn\ velocity gradient.}
\tablenotetext{d}{\citet{Sanchez-Monge2014} derived a P.A. of 10\degr\ while \citet{Zhang2022} used a value of 0\degr, the value in the table is more appropriate from the first moment map in Figure~\ref{fig:mom1}.}
\tablenotetext{e}{\citet{Beuther2004} found a P.A. of roughly 20\degr\ from SiO emission at ${\sim}2\arcsec$ resolution, while SiO emission at ${\sim}0\farcs5$ resolution from DIHCA favors a slightly eastern direction.}
\tablerefs{
[1] \citet[][]{Ilee2018}; 
[2] outflow: \citet{Cesaroni2017};
[3] outflow: \citet{Li2024};
[4] outflow: \citet{Saha2024};
[5] \citetalias{Olguin2022}; 
[6] rotation: \citetalias{Olguin2021}; 
[7] outflow: \citet{Avison2021};
[8] \citet{Olguin2023};
[9] rotation: \citet{Tanaka2020};
[10] outflow: \citet{Zapata2019};
[11] outflow: \citet{Zhang2022};
[12] rotation: \citet{Sanchez-Monge2014};
[13] rotation: \citet{Ginsburg2023};
[14] outflow: \citet{Beuther2019};
[15] \citet{Guzman2020};
[16] rotation: \citet{Anez-Lopez2020};
[17] outflow: \citet{Carrasco-Gonzalez2012};
[18] outflow: \citet{Maud2017}.
}
\end{deluxetable*}

\figsetstart
\figsetnum{1}
\figsettitle{First order moment maps}

\figsetend

\subsection{Source kinematics}\label{subsec:kinematics}

Of the 49 sources with line emission, most exhibit velocity gradients as shown in the first moment maps in Figure~\ref{fig:mom1}.
These moment maps were produced from emission over the 5$\sigma$ level.
The direction of the velocity gradients was determined visually from the first moment maps and channel-by-channel, but informed by the discrete first derivative of the first moment maps around the sources and the continuum morphology, for the sources whose rotation direction is not in the literature (Table~\ref{tab:props}).
To confirm that the gradient directions are likely those of a disk-outflow system, Figure~\ref{fig:mom1} shows the direction of the outflows in the literature (see Table~\ref{tab:props} for the references) or the tentative directions from DIHCA SiO $J=5-4$ data.
A 60\% of the sources with (tentative) outflow detections (33 sources) have gradient-outflow angles within 20\degr\ of the expected for disk-outflow systems (i.e., $90\degr\pm20\degr$).
This fraction increases to 80\% for the $90\degr\pm45\degr$ range.
Note that some of the directions of velocity gradients in the literature are chosen to be orthogonal to the outflow direction.
For G335.579--0.272 ALMAe1 we use the gradient direction from \citetalias{Olguin2022}, which better represents the rotation close to the main source as traced by vinyl cyanide (CH$_2$CHCN) emission. 
\citet{Sanchez-Monge2014} derived a P.A. of 10\degr\ for G35.20--0.74 N ALMAe1 while \citet{Zhang2022} favors a P.A. of 0\degr.
Here we use a P.A. of --10\degr, which follows closer the orientation of the continuum elongation.
The kinematics in G351.77--0.54 are rather complex and the position angle uncertain due to possible contamination from molecular line emission in the outflow \citep{Beuther2017}.
Thus, we list the position angle of ALMAe2 derived from the disk traced by salts \citep[e.g.,][]{Ginsburg2023}, while the large scale velocity gradient is closer to the east-west direction \citep[e.g.,][]{Beuther2025}.
Finally, the position angle from \citet[][]{Guzman2020} for IRAS 16562--3959 ALMAe1 was derived from rotation traced by hydrogen recombination lines.
Its orientation is orthogonal to that of the gradient detected in \metcyn, however, this may be the result of \metcyn\ contamination in the outflow cavities or significant misalignment of the rotation direction from the large envelope to the small scale ionized disk \citep[e.g.,][]{Bate2018}.

To confirm the rotation nature of the first moment maps, we produce PV maps along (pink arrows in Figure~\ref{fig:mom1}) and across the velocity gradients, with an slit width of 0\farcs05 (this is the expected resolution of the continuum, and corresponds to 5 pixels).
The PV maps are presented in Figures~\ref{fig:pvmaps} and \ref{fig:pvmaps_noedge}.
To facilitate the modeling below, we orient the slits from blue- to red-shifted emission, i.e., the arrows in Figure~\ref{fig:mom1} are oriented toward the P.A. in Table~\ref{tab:props} from the source position.
Figure~\ref{fig:pvmaps} shows the PV maps whose velocity gradient can be modeled as a power law (32 sources, see below).
These velocity gradients show distributions resembling Keplerian-like rotation (e.g., IRAS 18089--1732 ALMAe1) and rotation with infall (e.g., G335.579--0.272 ALMAe1) rather than outflows.
Most strikingly, many sources (around a third) show asymmetric PV maps with a lobe approximately Keplerian and another that is not.
These asymmetries are also expressed in the form of different peak intensity between lobes. 
We count 14 sources where the blue-shifted lobe dominates, 7 sources where the red-shifted lobe dominates and 11 sources where neither lobe dominates.
The possible origins for these asymmetries is discussed in \S~\ref{sec:discussion:anysotropies}.
The sources that do not show clear or resolved velocity gradients, or with velocity gradients resembling solid body rotation ($v\propto r$)  are presented in Figure~\ref{fig:pvmaps_noedge}.
On the other hand, the PV maps in the orthogonal direction allow the identification of possible infall and outflow motions.

\figsetstart
\figsetnum{2}
\figsettitle{PV maps with fitted edge}

\figsetgrpstart
\figsetgrpnum{2.14}
\figsetgrptitle{G335.78+0.17 ALMAe1}
\figsetplot{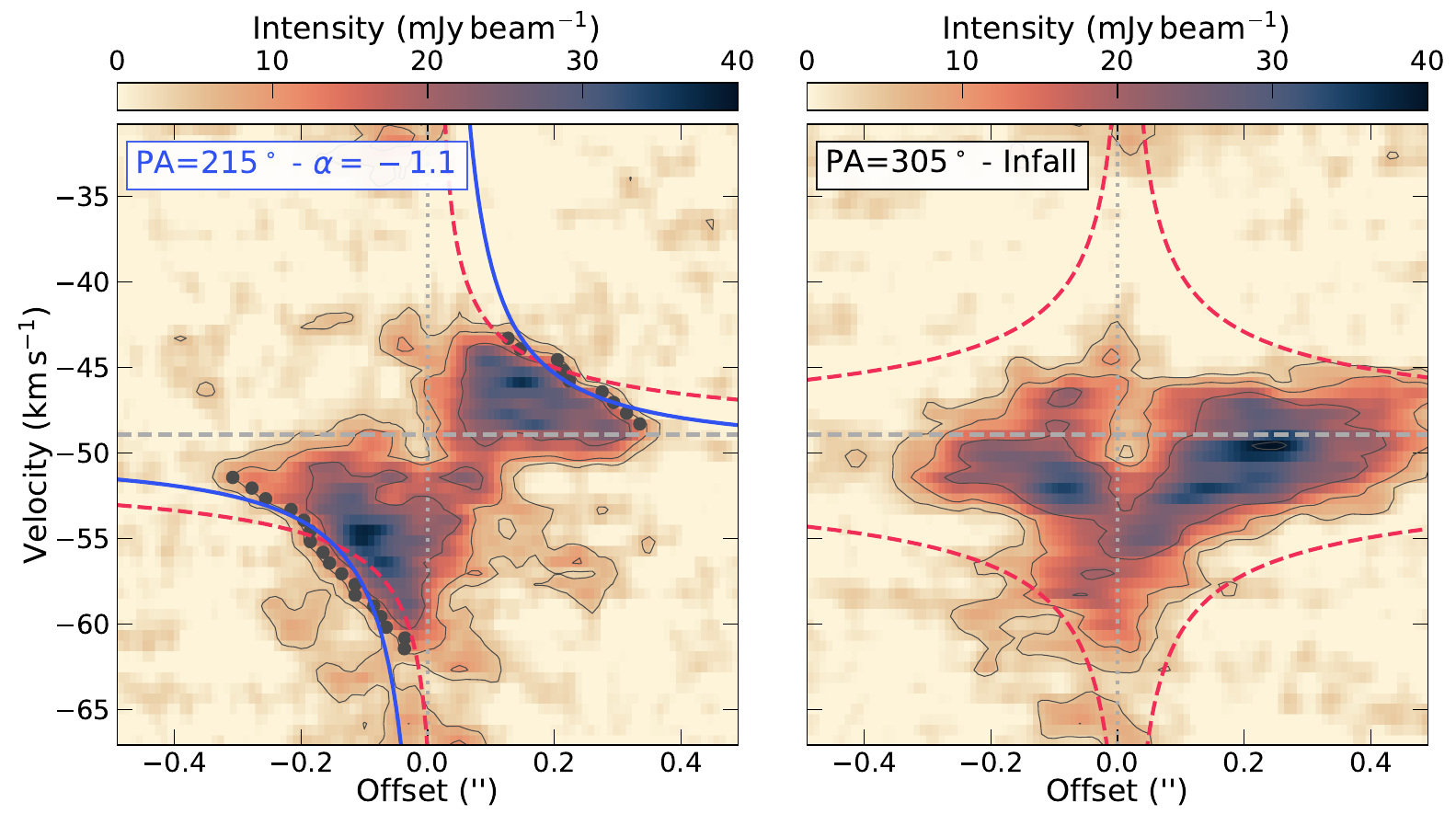}
\figsetgrpnote{Position velocity maps along (left) and orthogonal (right) to the velocity gradient direction for each condensation with edge calculation.
The molecular line emission used for the maps is listed in Table~\ref{tab:props} (see also Figure~\ref{fig:mom1}).
Left: The black dots mark the edge points, and the continuous colored line and dashed red line show the boundless and Keplerian power law fits to the edge points, respectively (see \S\ref{subsec:kinematics}).
The colored lines and labels group sources in those with $\alpha>-0.7$ (purple), $-0.7\ge\alpha\ge-0.8$ (green), $\alpha\le-0.8$ (blue) and outliers ($\alpha<-2$, black).
Right: The red dashed lines correspond to a free-fall velocity profile, i.e., $\sqrt{2}$ times the Keplerian power law in the left panel.
Note that these are plotted in the four quadrants because the location of the closest half of the disk is unknown along the PV slit.
Profiles resembling infall (blue skewed or inverse p-Cygni) or outflows (e.g., Hubble-like expansion) are labeled.
The dotted vertical and dashed horizontal gray lines indicate the zero position offset and the systemic velocity (Table~\ref{tab:props}), respectively.}
\figsetgrpend

\figsetgrpstart
\figsetgrpnum{2.23}
\figsetgrptitle{IRAS 18089-1732 ALMAe1}
\figsetplot{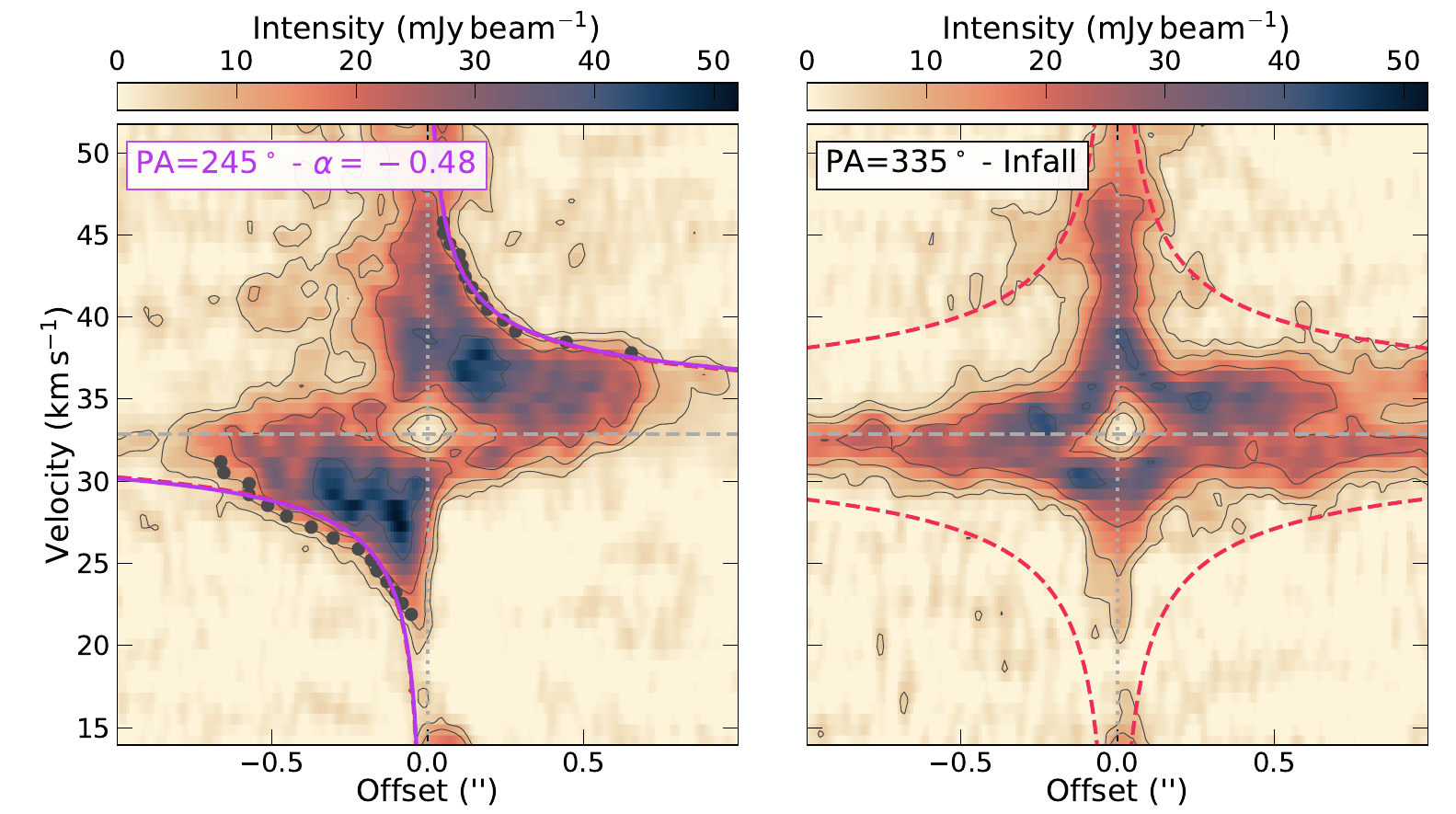}
\figsetgrpnote{Position velocity maps along (left) and orthogonal (right) to the velocity gradient direction for each condensation with edge calculation.
The molecular line emission used for the maps is listed in Table~\ref{tab:props} (see also Figure~\ref{fig:mom1}).
Left: The black dots mark the edge points, and the continuous colored line and dashed red line show the boundless and Keplerian power law fits to the edge points, respectively (see \S\ref{subsec:kinematics}).
The colored lines and labels group sources in those with $\alpha>-0.7$ (purple), $-0.7\ge\alpha\ge-0.8$ (green), $\alpha\le-0.8$ (blue) and outliers ($\alpha<-2$, black).
Right: The red dashed lines correspond to a free-fall velocity profile, i.e., $\sqrt{2}$ times the Keplerian power law in the left panel.
Note that these are plotted in the four quadrants because the location of the closest half of the disk is unknown along the PV slit.
Profiles resembling infall (blue skewed or inverse p-Cygni) or outflows (e.g., Hubble-like expansion) are labeled.
The dotted vertical and dashed horizontal gray lines indicate the zero position offset and the systemic velocity (Table~\ref{tab:props}), respectively.}
\figsetgrpend

\figsetgrpstart
\figsetgrpnum{2.26}
\figsetgrptitle{IRAS 18182-1433 ALMAe5}
\figsetplot{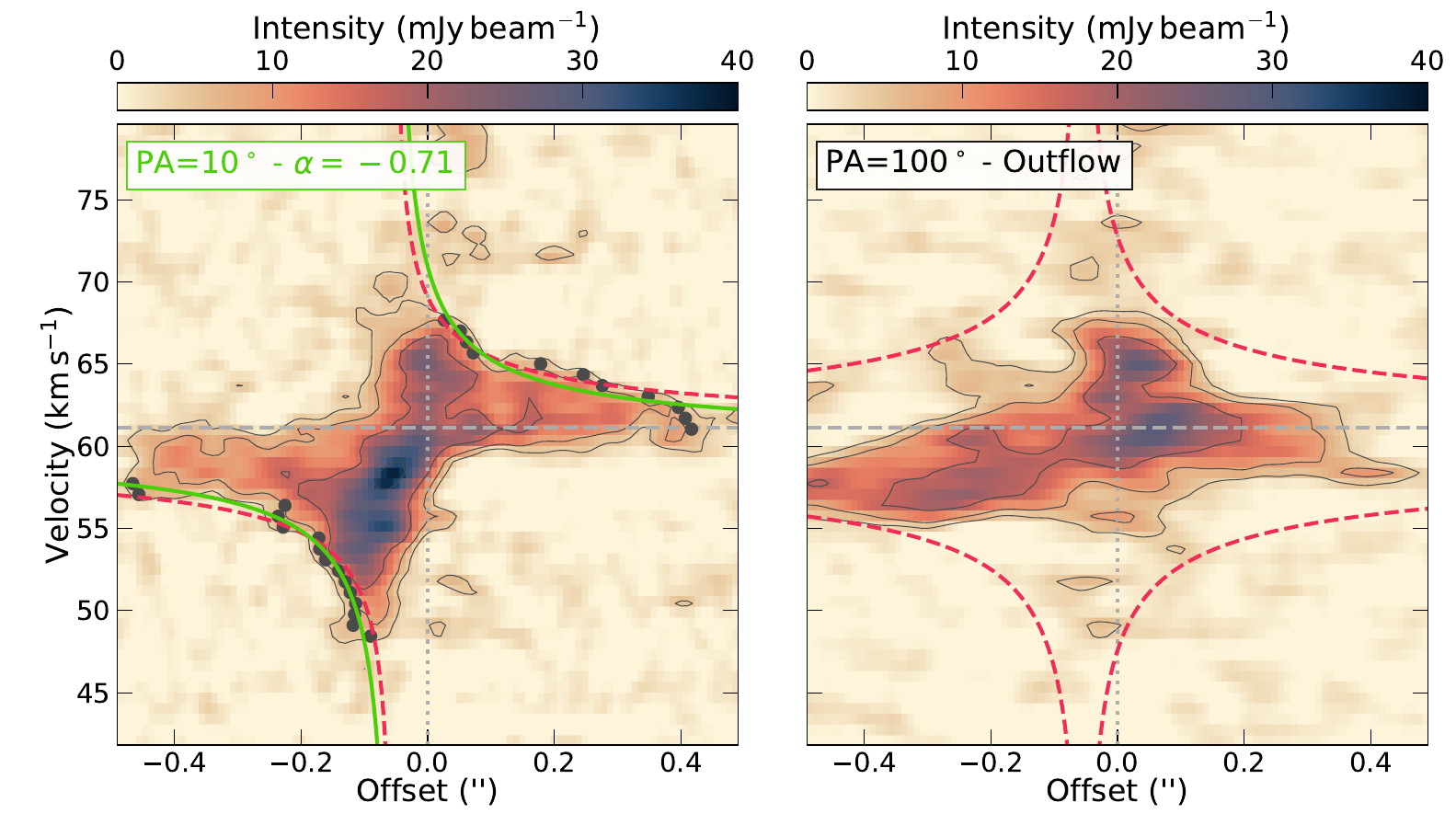}
\figsetgrpnote{Position velocity maps along (left) and orthogonal (right) to the velocity gradient direction for each condensation with edge calculation.
The molecular line emission used for the maps is listed in Table~\ref{tab:props} (see also Figure~\ref{fig:mom1}).
Left: The black dots mark the edge points, and the continuous colored line and dashed red line show the boundless and Keplerian power law fits to the edge points, respectively (see \S\ref{subsec:kinematics}).
The colored lines and labels group sources in those with $\alpha>-0.7$ (purple), $-0.7\ge\alpha\ge-0.8$ (green), $\alpha\le-0.8$ (blue) and outliers ($\alpha<-2$, black).
Right: The red dashed lines correspond to a free-fall velocity profile, i.e., $\sqrt{2}$ times the Keplerian power law in the left panel.
Note that these are plotted in the four quadrants because the location of the closest half of the disk is unknown along the PV slit.
Profiles resembling infall (blue skewed or inverse p-Cygni) or outflows (e.g., Hubble-like expansion) are labeled.
The dotted vertical and dashed horizontal gray lines indicate the zero position offset and the systemic velocity (Table~\ref{tab:props}), respectively.}
\figsetgrpend

\figsetend

\begin{figure*}
\gridline{\fig{f2_14.pdf}{0.55\linewidth}{G335.78+0.17 ALMAe1}}
\gridline{\fig{f2_26.pdf}{0.55\linewidth}{IRAS 18182--1433 ALMAe5}}
\gridline{\fig{f2_23.pdf}{0.55\linewidth}{IRAS 18089--1732 ALMAe1}}
\caption{Position velocity maps along (left) and orthogonal (right) to the velocity gradient direction for each condensation with edge calculation.
The molecular line emission used for the maps is listed in Table~\ref{tab:props} (see also Figure~\ref{fig:mom1}).
Left: The black dots mark the edge points, and the continuous colored line and dashed red line show the boundless and Keplerian power law fits to the edge points, respectively (see \S\ref{subsec:kinematics}).
The colored lines and labels group sources in those with $\alpha>-0.7$ (purple), $-0.7\ge\alpha\ge-0.8$ (green), $\alpha\le-0.8$ (blue) and outliers ($\alpha<-2$, black).
Right: The red dashed lines correspond to a free-fall velocity profile, i.e., $\sqrt{2}$ times the Keplerian power law in the left panel.
Note that these are plotted in the four quadrants because the location of the closest half of the disk is unknown along the PV slit.
Profiles resembling infall (blue skewed or inverse p-Cygni) or outflows (e.g., Hubble-like expansion) are labeled.
The dotted vertical and dashed horizontal gray lines indicate the zero position offset and the systemic velocity (Table~\ref{tab:props}), respectively.
The complete figure set (32 images) is available in the online journal.
}\label{fig:pvmaps}
\end{figure*}

\figsetstart
\figsetnum{3}
\figsettitle{PV maps without edge}

\figsetgrpstart
\figsetgrpnum{3.5}
\figsetgrptitle{G343.12-0.06 ALMAe1}
\figsetplot{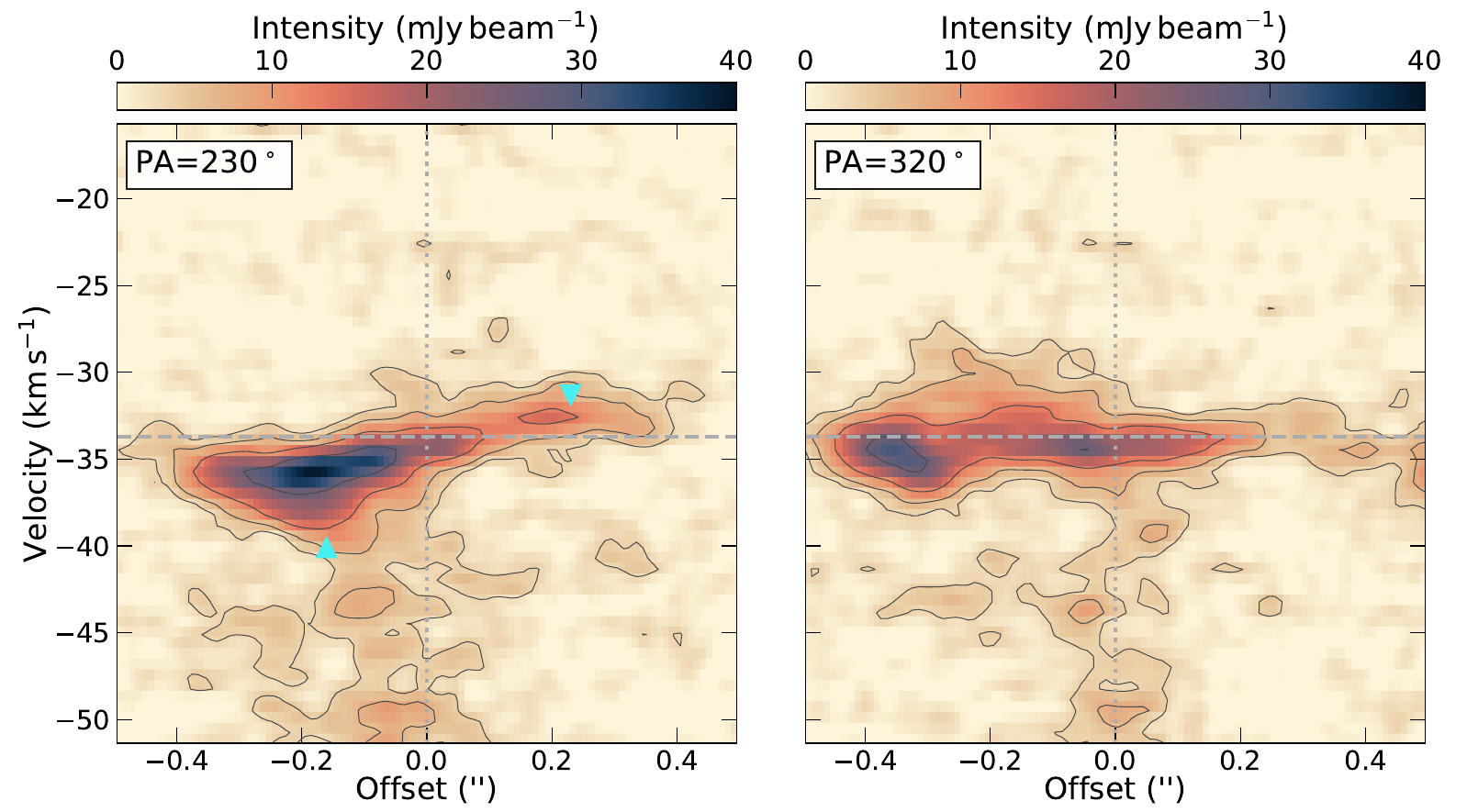}
\figsetgrpnote{Position velocity maps along and orthogonal to the velocity gradient direction for each condensation where the edge could not be determined.
The cyan triangles indicate the position of the velocity extrema determined from the edge points.
The dotted vertical and dashed horizontal gray lines indicate the zero position offset and the systemic velocity (Table~\ref{tab:props}), respectively.}
\figsetgrpend

\figsetend

\begin{figure}
\fig{f3_5.pdf}{\linewidth}{G343.12-0.06 ALMAe1}
\caption{Position velocity maps along and orthogonal to the velocity gradient direction for each condensation where the edge could not be determined.
The cyan triangles indicate the position of the velocity extrema determined from the edge points.
The dotted vertical and dashed horizontal gray lines indicate the zero position offset and the systemic velocity (Table~\ref{tab:props}), respectively.
The complete figure set (17 images) is available in the online journal.
}\label{fig:pvmaps_noedge}
\end{figure}

In order to obtain the physical properties of the condensations (e.g., central source masses) we derive the edge of the PV on each lobe \citep[e.g.,][]{Aso2015,Seifried2016,Sai2020}.
Here, the edge is defined as the points at a certain contour level (see Appendix~\ref{sec:appendix:edge} for details on contour selection).
To derive the edge offset corresponding to each velocity value, we first find the offset with the closest intensity to the selected contour level (initial offset).
Then we perform an intensity weighted average of pixels around the initial offset. 
For this average we use a total of 5 offset values centered in the initial offset (i.e., 2 values on each side). 
Similar to \citet{Aso2015}, we fit a power law to the velocity along the line of sight, $v_{los}$, as a function of the offset, $\theta$, of the form:
\begin{equation}\label{eq:powerlaw}
    v_{los} = v_{sys} +
    \begin{cases}
-v_{100} \left(\frac{|\theta - \theta_{off}|}{100\,{\rm au}}\right)^{\alpha} & \text{if } \theta < \theta_{off} \\
v_{100} \left(\frac{\theta - \theta_{off}}{100\,{\rm au}}\right)^{\alpha}  & \text{if } \theta \ge \theta_{off}
    \end{cases}
    \;,
\end{equation}
where $v_{100}$ is the velocity at 100\,au, and $v_{sys}$ and $\theta_{off}$ determine the zero velocity and offset values, respectively.
In this definition, a Keplerian disk would have a power law index $\alpha=-0.5$.
Note that free-fall also has a power law index of --0.5 but the velocity gradient in the orthogonal direction of the PV map is expected to follow the same distribution, while pure rotation should be flat as the velocity along the line of sight should be zero in the orthogonal direction.
For a rotating and infalling envelope under specific angular momentum conservation an $\alpha$ value of --1 is expected \citep[e.g.,][]{Oya2022}.
The fitting is performed in three steps, first we fit an initial Keplerian distribution with initial parameters from the data and a restricted $\theta_{off}$ value between the closest points where the distribution changes to the opposite quadrant.
Without this first step solutions tend to diverge for some sources in the following steps.
The results are then used as initial parameters for a fit with all parameters free and without boundary conditions, henceforth referred to as the boundless fit.
Finally, we perform a final Keplerian fit ($\alpha$ set to --0.5) with only the velocity $v_{100}$ as a free parameter and the remaining parameters ($v_{sys}$ and $\theta_{off}$) set to the boundless fit results.
We successfully fitted the edges of 32 sources (${\sim}60$\% of the sources with line detection) and the results of the fit are listed in Table~\ref{tab:plfit}.
The first Keplerian fit is performed using the Sequential Least Squares Programming fitter from \texttt{astropy}, which allows for boundary conditions for the input parameters.
The latter two fits are performed using the Orthogonal Distance Regression fitter from \texttt{scipy}, which allow for data with errors in both axes.
In this case, we estimate the error in offset from the propagation of uncertainty for the weighted average with an error per pixel equal to the beam size (geometric mean in Gaussian dispersion form).
The error in velocity is one channel, i.e., half the velocity resolution.
The error values in Table~\ref{tab:plfit} are the standard deviation of the fitted parameters calculated by the regression algorithm.
From the Keplerian power law, we estimate an enclosed mass within 100\,au and uncorrected for source inclination from the fit results:
\begin{equation}
    M_c\sin^2 i = 100\,{\rm au} \frac{v_{100}^2}{G} \;,
\end{equation}
with $G$ the gravitational constant. 
We use $v_{100}$ from the Keplerian fit, which is not the best for all sources.
Particularly for sources with $\alpha=-1$, infall and rotation cannot be separated from just the direction of the velocity gradient alone and the equivalent $M_c\sin^2 i$ would need to fit both PV slit directions \citep[e.g.,][]{Momose1998}.
The errors in mass are derived from the propagation of uncertainty.
Note that by fitting the edge of the PV maps the masses will be overestimated compared to fitting the peak intensity \citep[the so-called ridge; e.g.,][]{Sai2020}.
The anisotropic nature of the PV morphologies and/or non-Gaussian line shape precludes a good ridge fitting.
Since the ridge includes emission from the bulk of the envelope, the measured power law from the PV maps can have contamination leading to a shallower index \citep{Mori2024}.
For instance, the central source mass can be overestimated by a factor ${\sim}2$ if Keplerian rotation is assumed but the underlying velocity distribution follows that of a infalling and rotating envelope under specific angular momentum conservation.

\begin{figure*}
\begin{center}
\includegraphics[angle=0,width=\textwidth]{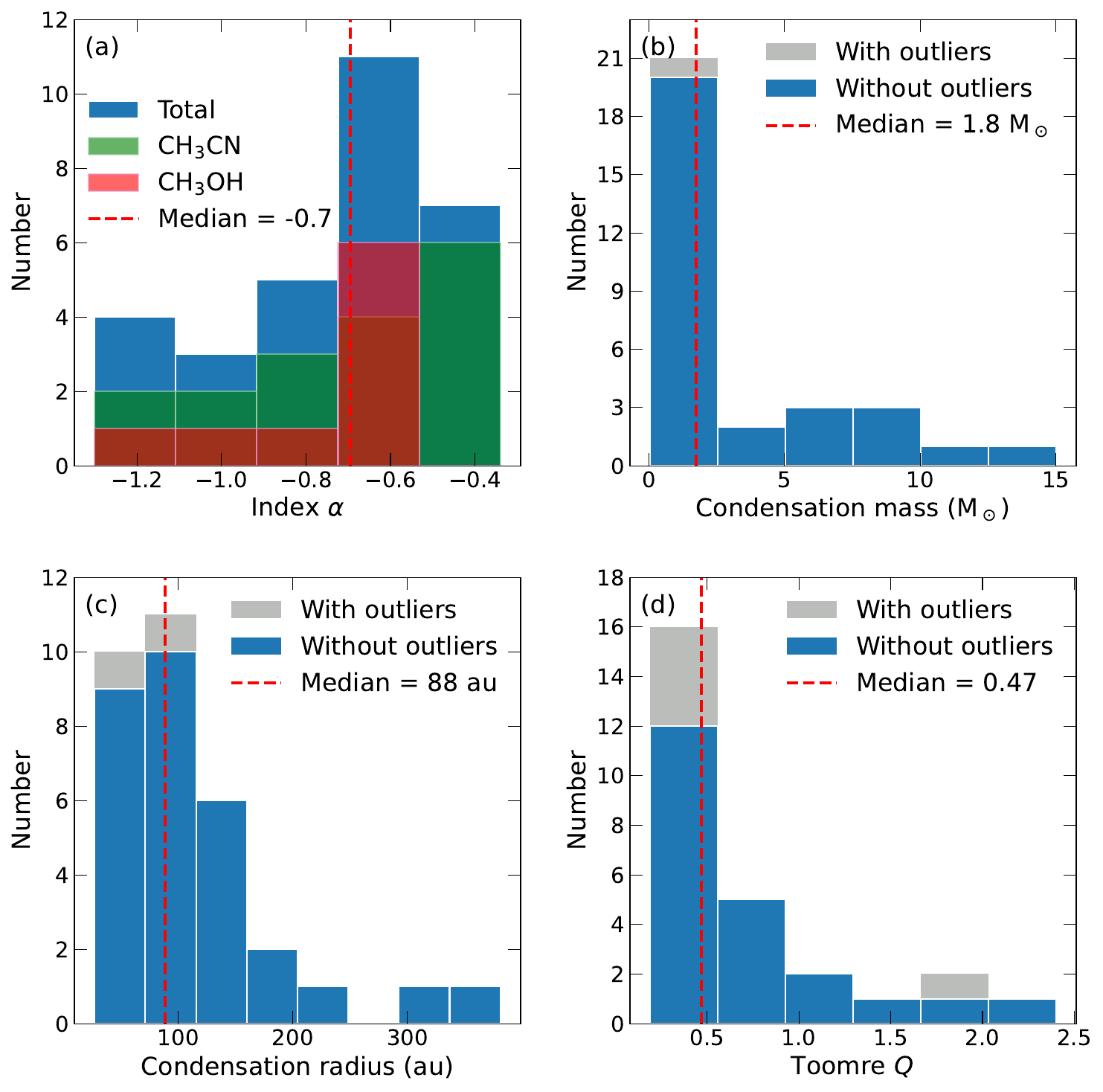}
\end{center}
\caption{Histograms of (a) power law indices $\alpha$ values, (b) condensation masses, (c) condensation radii and (d) Toomre-$Q$ parameter. 
Outliers with $\alpha<-2$ are not considered in (a), while the outliers for the other parameters are defined in \S\ref{subsec:relations}.
A further outlier source in (b) is located at 40\,\msun, while two additional sources in (d) are located at 3.1 and 12 (see Table~\ref{tab:condprops}).
Note that in (a) the total include three sources studied in other lines than \methanol\ and \metcyn.
}\label{fig:alpha_hist}
\end{figure*}

In general relatively good fits are obtained by the boundless power law fit.
Figure~\ref{fig:alpha_hist}(a) shows a histogram of the alpha values without outliers ($\alpha<-2$) binned using the Freedman Diaconis estimator.
The histogram peaks around an $\alpha$ value of --0.6.
Some of the $\alpha$ values are close to Keplerian (e.g., G333.46--0.16 ALMAe2), while others seem to follow rotation under specific angular momentum conservation power law (e.g., G335.78+0.17 ALMAe1).
We group the sources by using $\alpha$ ranges (color coded in Figure~\ref{fig:pvmaps}): $\alpha>-0.7$ for Keplerian, $-0.8\le\alpha\le-0.7$ for mixed and $\alpha<-0.8$ as infall and rotation under angular momentum conservation.
Values between --0.5 and --1 can be the result of changes in $\alpha$ with radius as expected from the transition from envelope to disk (e.g., G11.92--0.61 ALMAe1), unresolved disks embedded in an envelope, or asymmetric distributions of the edges at opposite lobes (e.g., G35.03+0.35 A ALMAe1).
Sources G10.62--0.38 ALMAe1, G11.92--0.61 ALMAe1, G29.96--0.02 ALMAe1, G333.12--0.56 ALMAe1, G335.579--0.272 ALMAe1, G5.89--0.37 ALMAe1, IRAS 18089--1732 ALMAe1, NGC 6334I(N) ALMAe8 and ALMAe9, and W33A ALMAe1 have PV maps that are similar in both slit directions and could also be fitted by a power law, hence in these cases a contribution from infall (e.g., $\alpha=-1$ for a rotating and infalling envelope \citealp{Oya2022}) is expected in the measured $\alpha$ values.
As a guide, the right panels of Figure~\ref{fig:pvmaps} show the expected curve from free-fall for a source with the same mass derived from the Keplerian assumption, i.e., the free-fall velocity at 100\,au is $v_{100,ff}=\sqrt{2} v_{100}$ with $v_{100}$ from the Keplerian fit.
We classify the PV maps in the orthogonal direction on whether they resemble outflows (Hubble-like expansion $v_{los}\propto r$) or infall profiles (blue skewed profiles or inverse p-Cygni).
Figure~\ref{fig:peak_spec} in Appendix~\ref{app:figs} shows the peak spectra toward the 32 sources to confirm the infalling nature of some of the sources in the sample.
Of the sources with large $\alpha$ values, G11.1--0.12 ALMAe1 seems to follow a power law in one lobe while the other lobe has a complex intensity distribution that precludes a good fit.
G336.01--0.82 ALMAe3 is also not well fitted by a power law distribution, however, the detailed modeling presented in \citet{Olguin2023} shows that the emission can be well described by a rotating and infalling envelope.

\begin{deluxetable*}{lcCCCCcCC}
\tablecaption{PV map edge fitting and central source masses\label{tab:plfit}}
\tablewidth{\textwidth}
\tablehead{
\colhead{Source} & \colhead{ALMAe} & \multicolumn{4}{c}{Boundless} & & \multicolumn{2}{c}{Keplerian} \\
\cline{3-6}\cline{8-9}
\colhead{} & \colhead{Source} & \colhead{$\alpha$} & \colhead{$v_{100}$} & \colhead{$v_{sys}$} & \colhead{$\theta_{off}$} &
&\colhead{$v_{100}$} &  \colhead{$M_c \sin^2 i$}\\
\colhead{} & \colhead{} & \colhead{} & \colhead{(\kms)} & \colhead{(\kms)} & \colhead{(mas)} & & \colhead{(\kms)} &  \colhead{(\msun)}
}
\startdata
G10.62--0.38     & 1  &  -1.2\pm0.5  &  36\pm25  &   -2.3\pm0.9 &  82\pm9  & &   12\pm1   &   16\pm3   \\ 
                 & 3  &  -1.0\pm0.2  &  14\pm4   &   -2.2\pm0.3 &  41\pm9  & &  6.6\pm0.5 &  5.0\pm0.8 \\ 
G11.1--0.12      & 1  &    -3\pm1    & 120\pm170 &   27.5\pm0.8 &  66\pm13 & &  7.8\pm0.8 &    7\pm2   \\ 
G11.92--0.61     & 1  & -0.91\pm0.05 &  43\pm4   &   35.1\pm0.2 &  10\pm6  & & 19.9\pm0.7 &   45\pm3   \\ 
                 & 4  & -0.58\pm0.07 &  11\pm1   &   36.0\pm0.2 &  -7\pm8  & &  9.9\pm0.3 & 11.0\pm0.6 \\ 
G29.96--0.02     & 1  & -0.77\pm0.08 &  37\pm7   &   97.8\pm0.3 &  16\pm11 & & 19.9\pm0.9 &   45\pm4   \\ 
G333.12--0.56    & 1  & -0.56\pm0.06 &  13\pm1   &  -58.1\pm0.2 & -36\pm7  & & 11.6\pm0.2 & 15.2\pm0.6 \\ 
G333.23--0.06    & 6  & -0.86\pm0.07 &  18\pm2   &  -87.5\pm0.2 &  28\pm4  & & 10.1\pm0.5 &   11\pm1   \\ 
                 & 17 &  -1.2\pm0.4  &  20\pm10  &  -87.7\pm0.5 &  24\pm8  & &  7.6\pm0.8 &    7\pm1   \\ 
G333.46--0.16    & 1  &  -0.8\pm0.1  &  10\pm1   &  -43.6\pm0.3 &  32\pm9  & &  6.9\pm0.4 &  5.4\pm0.6 \\ 
                 & 2  & -0.45\pm0.05 & 9.5\pm0.6 &  -41.8\pm0.2 & -18\pm8  & & 10.1\pm0.3 & 11.6\pm0.6 \\ 
G335.579--0.272  & 1  & -0.62\pm0.06 & 9.4\pm0.7 &  -46.5\pm0.2 & -23\pm5  & &  8.2\pm0.3 &  7.6\pm0.5 \\ 
                 & 4  & -0.64\pm0.05 &  21\pm2   &  -48.8\pm0.2 &  45\pm6  & & 16.6\pm0.3 &   31\pm1   \\ 
G335.78+0.17     & 1  &  -1.1\pm0.1  &  33\pm8   &  -50.0\pm0.3 &  15\pm11 & & 12.2\pm0.7 &   17\pm2   \\ 
                 & 2  &  -0.7\pm0.2  &  12\pm3   &  -51.6\pm0.5 &  11\pm15 & &  9.3\pm0.6 &   10\pm1   \\ 
G336.01--0.82    & 3  &  -2.1\pm0.5  & 180\pm150 &  -45.4\pm0.6 &   4\pm15 & &   11\pm1   &   14\pm3   \\ 
G34.43+0.24 MM1  & 2  & -0.58\pm0.08 &  16\pm3   &   58.7\pm0.2 & -21\pm16 & & 13.9\pm0.4 &   22\pm1   \\ 
G35.03+0.35 A    & 1  &  -0.7\pm0.4  &  18\pm10  &     51\pm2   & -20\pm50 & & 13.3\pm0.7 &   20\pm2   \\ 
G35.13--0.74     & 1  & -0.46\pm0.06 & 7.5\pm0.5 &   35.8\pm0.2 & -38\pm8  & &  7.7\pm0.2 &  6.8\pm0.4 \\ 
                 & 2  &  -1.0\pm0.4  &   6\pm1   &   34.0\pm0.7 & -14\pm15 & &  4.9\pm0.3 &  2.7\pm0.4 \\ 
G35.20--0.74 N   & 1  &  -1.2\pm0.2  &  15\pm5   &   31.8\pm0.3 & -34\pm20 & &  5.9\pm0.6 &  3.9\pm0.8 \\ 
G5.89--0.37      & 1  &  -0.4\pm0.1  &   8\pm2   &    8.2\pm0.4 &  33\pm24 & &  8.6\pm0.8 &    8\pm2   \\ 
IRAS 18089--1732 & 1  & -0.48\pm0.04 &15.3\pm0.8 &   33.5\pm0.2 & -10\pm8  & & 15.6\pm0.3 &   28\pm1   \\ 
                 & 2  & -0.80\pm0.09 &  11\pm1   &   33.2\pm0.3 &   1\pm7  & &  8.1\pm0.4 &  7.5\pm0.7 \\ 
IRAS 18162--2048 & 1  & -0.34\pm0.09 & 8.8\pm0.8 &   44.3\pm0.4 &  20\pm40 & & 10.3\pm0.3 & 11.9\pm0.6 \\ 
IRAS 18182--1433 & 5  & -0.71\pm0.06 &  17\pm2   &   60.2\pm0.2 & -54\pm7  & & 12.4\pm0.4 &   17\pm1   \\ 
NGC 6334I        & 1  &  -0.6\pm0.1  &  10\pm2   &   -6.5\pm0.3 &  -4\pm30 & &  8.8\pm0.5 &  8.7\pm0.9 \\ 
                 & 4  &  -1.3\pm0.2  &  24\pm6   &   -8.1\pm0.2 &  13\pm17 & &  8.1\pm0.7 &    7\pm1   \\ 
NGC 6334I(N)     & 2  &  -0.6\pm0.2  &  13\pm3   &   -3.6\pm0.5 & 120\pm40 & & 10.9\pm0.5 &   13\pm1   \\ 
                 & 8  & -0.44\pm0.05 & 9.1\pm0.4 &   -3.7\pm0.2 & -33\pm9  & &  9.4\pm0.2 & 10.0\pm0.5 \\ 
                 & 9  & -0.69\pm0.08 & 7.5\pm0.5 &   -5.1\pm0.3 &   3\pm10 & &  6.7\pm0.3 &  5.0\pm0.4 \\ 
W33A             & 1  & -0.53\pm0.06 &  13\pm1   &   37.3\pm0.3 &  16\pm9  & & 12.6\pm0.5 &   18\pm2   \\ 
\enddata
\end{deluxetable*}

As shown in the first moment maps (Figure~\ref{fig:mom1}), many sources are embedded in extended line emission from the surrounding clouds (e.g., G10.62--0.38 ALMAe2), making it difficult to disentangle the kinematics of the individual condensations.
In addition to the sources whose kinematics cannot be resolved, we obtain an upper limit for the source mass from their PV map distribution.
For these sources we apply the same algorithm to derive the edge, but select one (e.g., when the individual gas kinematics cannot be disentangled from the cloud) or two (e.g., when the kinematics cannot be fitted by a power law) points which represent the maximum line of sight velocity $v_{los}$.
These sources are listed in Table~\ref{tab:upperlims} and the points are represented by triangles in Figure~\ref{fig:pvmaps_noedge}.
Assuming that the infall and rotation contribution is the same, the upper limit of the mass of the source within a radius $r$ can be calculated as \citep[][]{Oya2014}:
\begin{equation}
    M_c \sin^2 i = \frac{r}{G} \Delta v^2 \,,
\end{equation}
where $\Delta v$ is either $|v_{los}-v_{LSR}|$ for sources with only one measurement and $|v_{los,1} - v_{los,2}|/2$ otherwise.
Similarly, the radius for sources with one measurement is the distance to zero offset, while for those with two measurements is half the distance between both measurements.
The latter choice is justified by the fact that many of the sources are asymmetric and that the $v_{LSR}$ may not be well estimated. 

\begin{deluxetable}{lcCCC}
\tablecaption{PV map edge extrema\label{tab:upperlims}}
\tablewidth{0pt}
\tablehead{
\colhead{Source} & \colhead{ALMAe} & \colhead{$\Delta v$} & \colhead{$r$} & \colhead{$M_c \sin^2 i$} \\
\colhead{} & \colhead{Source} & \colhead{(\kms)} & \colhead{(au)} & \colhead{(\msun)}
}
\startdata
G11.92--0.61     & 6  & 4.0 & 233 &  4 \\
G14.22--0.50 S   & 3  & 4.1 &  88 &  2 \\
G333.12--0.56    & 8  & 4.7 & 310 &  8 \\
G343.12--0.06    & 1  & 4.4 & 570 & 12 \\
G35.13--0.74     & 7  & 4.7 & 170 &  4 \\
G35.20--0.74 N   & 2  & 3.5 & 380 &  5 \\
IRAS 16562--3959 & 5  & 2.7 & 110 &  0.9 \\
IRAS 18182--1433 & 1  & 3.3 & 200 &  3 \\ 
                 & 2  & 2.7 & 170 &  1 \\ 
                 & 11 & 3.0 &  64 &  0.6 \\ 
NGC 6334I        & 3  & 2.5 &  68 &  0.5 \\
NGC 6334I(N)     & 1  & 4.1 & 180 &  3 \\ 
                 & 14 & 5.1 & 120 &  3 \\ 
W33A             & 4  & 6.0 & 430 & 18 \\  
\enddata
\end{deluxetable}

\section{Discussion} \label{sec:discussion}

\subsection{Anisotropic collapse}\label{sec:discussion:anysotropies}

The PV maps presented in Figure~\ref{fig:pvmaps} show a variety of rotation curves.
They show intensity distributions that are not as smooth as some of those detected toward low-mass disks \citep[e.g.,][]{Aso2015}. 
These asymmetries result in PV maps seemingly having two different distributions on opposite sides (expressed in power laws fitting one side better than the other), one side brighter than the other, and/or shifts in offset from the continuum source position.
The observed intensity is an interplay of density distribution and temperature gradient in the system.  
In the case of a hot disk embedded in a colder collapsing envelope, a weaker red-shifted lobe in the PV map is expected due to absorption caused by receding colder gas in the near side of the envelope  \citep[e.g.,][]{Chen2016}.
However, such a simple model cannot account for cases with a stronger red-shifted lobe. 
We speculate that anisotropic infall occurs in more than half of our sample.
Anisotropic collapse can be the result of the initial turbulence and magnetic field in these regions, which would allow the formation of structures like infall streamers \citep[e.g.,][]{Seifried2015} as those confirmed in some of the sources in the sample (see next paragraph).
A clumpy environment can also result in anisotropic collapse, as tidal events disrupts these gas clumps allowing the formation of asymmetric configurations \citep[e.g.,][]{Dullemond2019}.
As these flows feed the midplane/disk defined by the rotation axis at different radius projected in the plane of the sky, the enhancement of density and/or temperature (e.g., due to accretion shocks) would result in asymmetric distributions of line emission and potentially dust continuum emission.
In addition, since massive stars are frequently born in clustered environments \citep[][and references therein]{Beuther2025rev}, it is expected that core-core interactions are more frequent. 
These interactions or collisions result in shock-compressed gas allowing the formation of asymmetric streamers or gas stripping depending on the speed of the interaction \citep[e.g.,][]{Yano2024}.
Therefore the observed asymmetries can be in part the result of such interactions.
Finally, the development of gravitational instabilities as a result of rapid infall can produce anisotropies within ${\sim}1000$\,au scales \citep{2007ApJ...656..959K,2007ApJ...665..478K}. 

For instance, the edges of sources like G335.78+0.17 ALMAe1 or NGC 6334I(N) ALMAe8 clearly show different slopes and probably different power law on either PV lobe.
Most sources exhibit a brighter lobe (e.g., IRAS 18089--1732 ALMAe1 and ALMAe2), with some extreme cases where one lobe is much weaker (e.g., G35.03+0.35 A ALMAe1, also reported in \citealp{Beltran2014}).
This has also been observed in other sources with asymmetrical gas distributions of rotating structures (e.g., IRAS 20126+4104 \citealp{Cesaroni2014}, AFGL 4176 \citealp{Johnston2015}), and in the sample of \citet{Ahmadi2023} with the majority of its sources gravitationally unstable.
These differences in intensity and potentially gas distributions are also detected in the continuum emission, which in most cases is asymmetric and, in some cases,  with a distribution resembling spiral arms (the aforementioned streamers).
Well-known examples of this in our sample are G336.01-0.82 ALMAe3, whose kinematics have been studied in detail as part of the DIHCA project \citep{Olguin2023}, IRAS 18089--1732 ALMAe1 \citep{Sanhueza2021}, IRAS 18162-2048 ALMAe1 \citep{Fernandez-Lopez2023} and the streamer feeding the multiple system in W33A which includes ALMAe1 and ALMAe4 \citep{Izquierdo2018}.
Those studies found that the (putative) disks are fed by streamers which can have different infall rates than the bulk of the envelope, thus explaining the differences in intensity/gas distribution.

In addition, a few sources show offsets between the position of the continuum source and the center of symmetry of the PV maps.
One clear case is G335.579--0.272 ALMAe4, with an offset of roughly 50\,mas (Table~\ref{tab:plfit}).
This case seems to indicate that the high-mass source is not at the center of the disk/envelope either because it is part of a wide binary system or a previous core-core interaction with another source (second continuum peak seen in Figure~\ref{fig:mom1}).
This is similar to the low-mass case IRAS 16293--2422 Source A \citep{Oya2020}, where a forming multiple system produces a shift of the rotation centroid of the common envelope/disk.
If the offset represents the position of the center mass of a potential binary system with respect to the primary and the mass in Table~\ref{tab:plfit} is the total mass of the system, the primary and secondary stars would have a mass of roughly 27 and 4\,\msun, respectively, for a plane of the sky distance of 355\,mas (${\sim}1200$\,au) between them (measured from the continuum peak positions).
An additional interpretation would be that the star is moving with respect to its natal cloud as a result of cluster dynamics, resulting in a shift of the source position with respect to the gas in its immediate surroundings and fed through Bondi-Hoyle-like accretion.
On the other hand, G35.03+0.35 A ALMAe1 also has a slight offset of 20\,mas, but in this case the reason seems to be filamentary accretion as proposed by \citet[][]{Beltran2014}  and supported by the extended continuum emission toward the north-east (see Figure~\ref{fig:mom1}).

\subsection{Scaling relationships}\label{subsec:relations}

\begin{deluxetable*}{lcCCCCCCCCCc}
\tablecaption{Disk candidates properties\label{tab:condprops}}
\tablewidth{\textwidth}
\tabletypesize{\scriptsize}
\tablehead{
\colhead{Source} & \colhead{ALMAe} & \colhead{$T$} & \colhead{$L_{\rm core}$} & \colhead{$S_{1.3{\rm mm}}$} & \colhead{$I_{1.3{\rm mm}}$} & \colhead{$\tau_{1.3{\rm mm}}$} & \colhead{$R$} & \colhead{$M_g$} & \colhead{$\Sigma$} & \colhead{Toomre $Q$} & \colhead{Disk} \\
\colhead{} & \colhead{} & \colhead{(K)} & \colhead{(\lsun)} & \colhead{(mJy)} & \colhead{(mJy)} & \colhead{} & \colhead{(au)} & \colhead{(\msun)} & \colhead{($\times10^2$\,g\,cm$^{-2}$)} & \colhead{} & \colhead{motion}
}
\startdata
G10.62--0.38     &  1 & 202                 & 4.1\times10^5 & 155.7 & 13.9 &   1.1 & 149                  &   40 &   51 &  0.07 & Infall  \\ 
                 &  3 & 202                 & 4.1\times10^5 &  11.3 &  5.8 &  0.32 &  73                  &  2.1 &   11 &  0.33 & Infall  \\ 
G11.1--0.12      &  1 &  62                 & 3.4\times10^2 &   4.4 &  1.4 &  0.26 & 100                  &  1.0 &  2.9 &  0.47 & Outlier \\ 
G11.92--0.61     &  1 & 212                 & 7.9\times10^3 &  55.8 & 11.3 &  0.84 & 149                  &  5.8 &  7.4 &  0.47 & Infall  \\ 
                 &  4 &  54\tablenotemark{a}& 2.9\times10^2 &   2.9 &  2.0 &  0.56 &  54                  &  1.1 &   11 &  0.36 & Keplerian \\ 
G29.96--0.02     &  1 & 270                 & 1.4\times10^5 &  90.6 & 16.0 &  0.47 & 217                  &   15 &  9.1 &  0.27 & Mixed \\ 
G333.12--0.56    &  1 & 193                 & 1.4\times10^3 &  56.5 & 12.0 &  0.88 & 133                  &  6.3 &   10 &  0.25 & Keplerian \\ 
G333.23--0.06    &  6 & 200                 & 1.4\times10^3 &   5.0 &  3.1 &  0.18 &  94                  & 0.97 &  3.1 &   1.1 & Infall \\ 
                 & 17 & 200                 & 1.4\times10^3 &   2.4 &  1.4 & 0.077 &<106\tablenotemark{b} & 0.44 &  1.1 &   1.9 & Infall \\ 
G333.46--0.16    &  1 & 150                 & 1.9\times10^4 &  79.5 &  5.2 &  0.35 & 381                  &  7.0 &  1.4 &  0.26 & Mixed  \\ 
                 &  2 & 113                 & 7.7\times10^2 &   6.4 &  3.6 &  0.32 &  59                  & 0.75 &  6.1 &  0.83 & Keplerian \\ 
G335.579--0.272  &  1 & 250                 & 1.5\times10^4 &  29.4 & 17.6 &   1.2 &  65                  &  2.8 &   19 &  0.31 & Keplerian \\ 
                 &  4 & 290\tablenotemark{c}& 2.5\times10^3 &  29.2 &  5.5 &  0.21 & 109                  &  1.5 &  3.7 &   1.4 & Keplerian \\ 
G335.78+0.17     &  1 & 250                 & 8.3\times10^3 & 116.4 & 20.8 &   1.5 & 156                  &   12 &   14 &  0.19 & Infall \\ 
                 &  2 & 246                 & 2.7\times10^3 &  29.1 & 13.5 &  0.72 &  75                  &  2.2 &   11 &  0.46 & Mixed \\ 
G336.01--0.82    &  3 & 218                 & 2.1\times10^3 &   6.1 &  4.1 &  0.18 &  57                  & 0.39 &  3.4 &   2.4 & Outlier \\ 
G34.43+0.24 MM1  &  2 & 151                 & 2.4\times10^3 &  17.3 & 10.5 &  0.58 &  76                  &  1.8 &  9.0 &  0.62 & Keplerian \\ 
G35.03+0.35 A    &  1 & 218                 & 3.4\times10^3 &   9.8 &  9.2 &  0.29 & <63\tablenotemark{b} & 0.37 &  2.6 &   3.1 & Mixed \\ 
G35.13--0.74     &  1 & 100                 & 7.8\times10^2 &  44.8 & 19.5 &\nodata&  70                  &  2.9 &   17 &  0.19 & Keplerian \\ 
                 &  2 & 175                 & 4.9\times10^2 &  25.0 & 12.7 &  0.66 &  66                  &  1.2 &  8.1 &  0.37 & Infall \\ 
G35.20--0.74 N   &  1 & 202                 & 9.0\times10^3 &  91.3 & 31.2 &   2.6 &  81                  &  8.0 &   35 &  0.12 & Infall \\ 
G5.89--0.37      &  1 & 100                 & 1.0\times10^3 &   8.5 &  3.6 &  0.49 &  83                  &  1.3 &  5.3 &  0.46 & Keplerian \\ 
IRAS 18089--1732 &  1 & 247                 & 1.5\times10^4 & 109.5 & 20.1 &   2.2 & 114                  &  7.9 &   17 &  0.27 & Keplerian \\ 
                 &  2 &  79                 & 4.9\times10^2 &  16.8 &  4.2 &  0.95 &  73                  &  2.5 &   13 &  0.21 & Mixed \\ 
IRAS 18162--2048 &  1 &  94                 & 1.2\times10^4 & 365.4 & 69.6 &\nodata& 162                  &  8.9 &  9.6 &  0.14 & Keplerian \\ 
IRAS 18182--1433 &  5 & 278                 & 3.0\times10^3 &  38.6 &  3.2 &   0.1 & 312                  &  2.4 &  0.7 &   1.1 & Mixed \\ 
NGC 6334I        &  1 & 278\tablenotemark{d}& 2.8\times10^4 & 233.5 & 33.5 &\nodata&  71                  &  2.0 &   11 &   0.5 & Keplerian \\ 
                 &  4 & 400\tablenotemark{d}& 1.1\times10^4 &   8.4 &  7.0 &  0.18 &  32                  &0.054 &  1.5 &    12 & Infall \\ 
NGC 6334I(N)     &  2 & 176                 & 7.6\times10^3 & 122.0 & 20.0 &\nodata& 165                  &  1.7 &  1.7 &  0.84 & Keplerian \\ 
                 &  8 & 201                 & 4.0\times10^2 &  38.7 &  9.7 &  0.64 & 130                  & 0.62 &  1.0 &   1.8 & Keplerian \\ 
                 &  9 & 182\tablenotemark{d}& 8.0\times10^2 &  72.4 &  9.2 &  0.68 & 137                  &  1.3 &  2.0 &  0.65 & Keplerian \\ 
W33A             &  1 & 199                 & 1.1\times10^4 &  29.1 & 24.9 &\nodata&  27                  &  1.2 &   48 &  0.56 & Keplerian \\ 
\enddata
\tablecomments{Disk motion types are: ``Infall'' for infall and rotation under angular momentum conservation, ``Keplerian'' for those with power law $\alpha>-0.7$, ``Mixed'' for those with $-0.7\ge\alpha\ge-0.8$, and ``Outliers''. Fluxes are measured in the primary beam corrected continuum maps.}
\tablenotetext{a}{Brightness temperature (see Figure~\ref{fig:peak_spec}).}
\tablenotetext{b}{Beam size ($\sqrt{\theta_{\rm min} \theta_{\rm max}}/(2\sqrt{2\ln 2})$ where $\theta$ values are the minor and major beam FWHM) at the source distance.}
\tablenotetext{c}{From \citetalias{Olguin2021}.}
\tablenotetext{d}{From \citetalias{Sakai2025}.}
\end{deluxetable*}

\begin{figure}
\begin{center}
\includegraphics[angle=0,width=\linewidth]{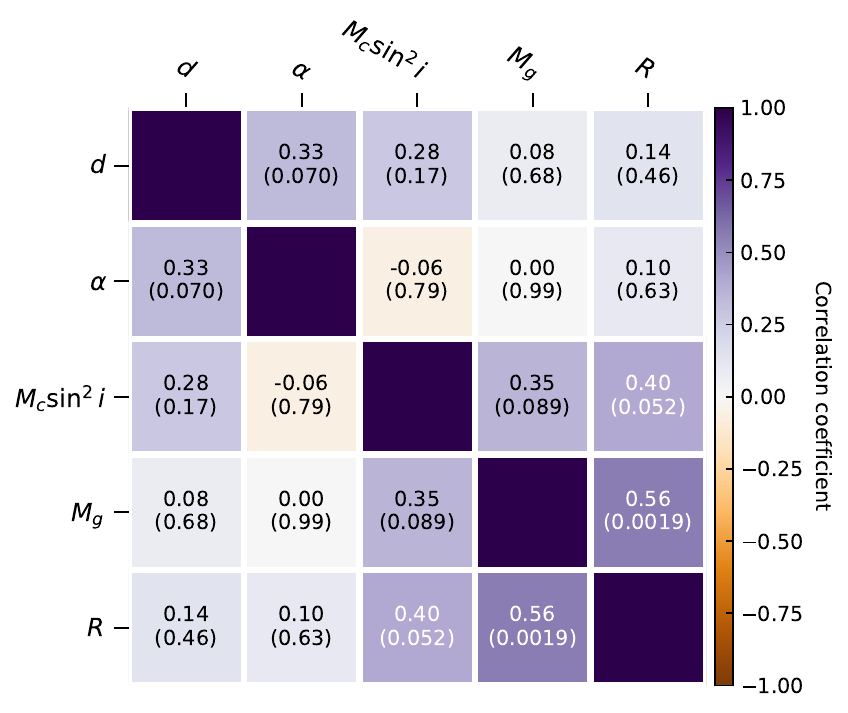}
\end{center}
\caption{Correlations between selected source properties: distance $d$, power law index $\alpha$, central source mass $M_c\sin^2i$, disk candidate mass $M_g$, and disk candidate radius $R$.
Spearman correlation coefficients and p-values (in parentheses) are calculated excluding outliers (see \S\ref{subsec:relations} for details).
}\label{fig:correl_map}
\end{figure}

In this section we search for relations between different source parameters and other related quantities.
To calculate the Spearman correlation coefficients, we exclude outlier values depending on the parameter or their combination.
In addition to the aforementioned outliers in the power law index $\alpha$ values (see \S~\ref{subsec:kinematics}, $\alpha<-2$), relations involving the central source mass $M_c \sin^2i$ exclude close to face-on sources G335.579--0.272 ALMAe1 whose central mass is probably underestimated due to its inclination angle \citepalias{Olguin2022}, and G35.20-0.74 N ALMAe1, that also seem to be closer to face-on (see shape of continuum and \methanol\ emission in Figure~\ref{fig:mom1}, and the masses determined from molecules other than \methanol\ and \metcyn.
The disk candidate radius, $R$, is derived from the deconvolved geometric mean of the semi-major and minor axes (FWHMs) of a 2-D Gaussian fitted during the source identification procedure utilizing \textsc{PyBDSF} \citepalias{Luo2026}, and individual values are listed in Table~\ref{tab:condprops} and a histogram is shown in Figure~\ref{fig:alpha_hist}(c).
In relations involving the radius, we exclude G333.23--0.06 ALMAe17 and G35.03+0.35 A ALMAe1 whose deconvolved radius is undefined and instead we provide an upper limit defined by the beam size in Table~\ref{tab:condprops}.
Finally, for relations involving the circumstellar gas mass and source luminosity (both described below), we exclude G10.62--0.38 ALMAe1 and ALMAe3 because they are likely contaminated by free-free emission. 

Figure~\ref{fig:correl_map} summarizes the correlation coefficients between the parameters described above, except for the luminosity, without outliers.
A very strong correlation has an absolute value of the Spearman coefficient between 0.8 and 1.0, a strong correlation between 0.6 and 0.79, a moderate correlation between 0.4 and 0.59, and weak and very weak correlations below 0.4.
We only find (very) weak correlations between the distance and the source properties, but with relatively large p-values particularly for the gas mass and source radius.
There is a moderate correlation between the source radius and the circumstellar gas mass (discussed below), while there is a borderline moderate correlation between the central source mass and the radius.

\begin{figure*}
%\figurenum{6}
\begin{center}
\includegraphics[angle=0,width=\textwidth]{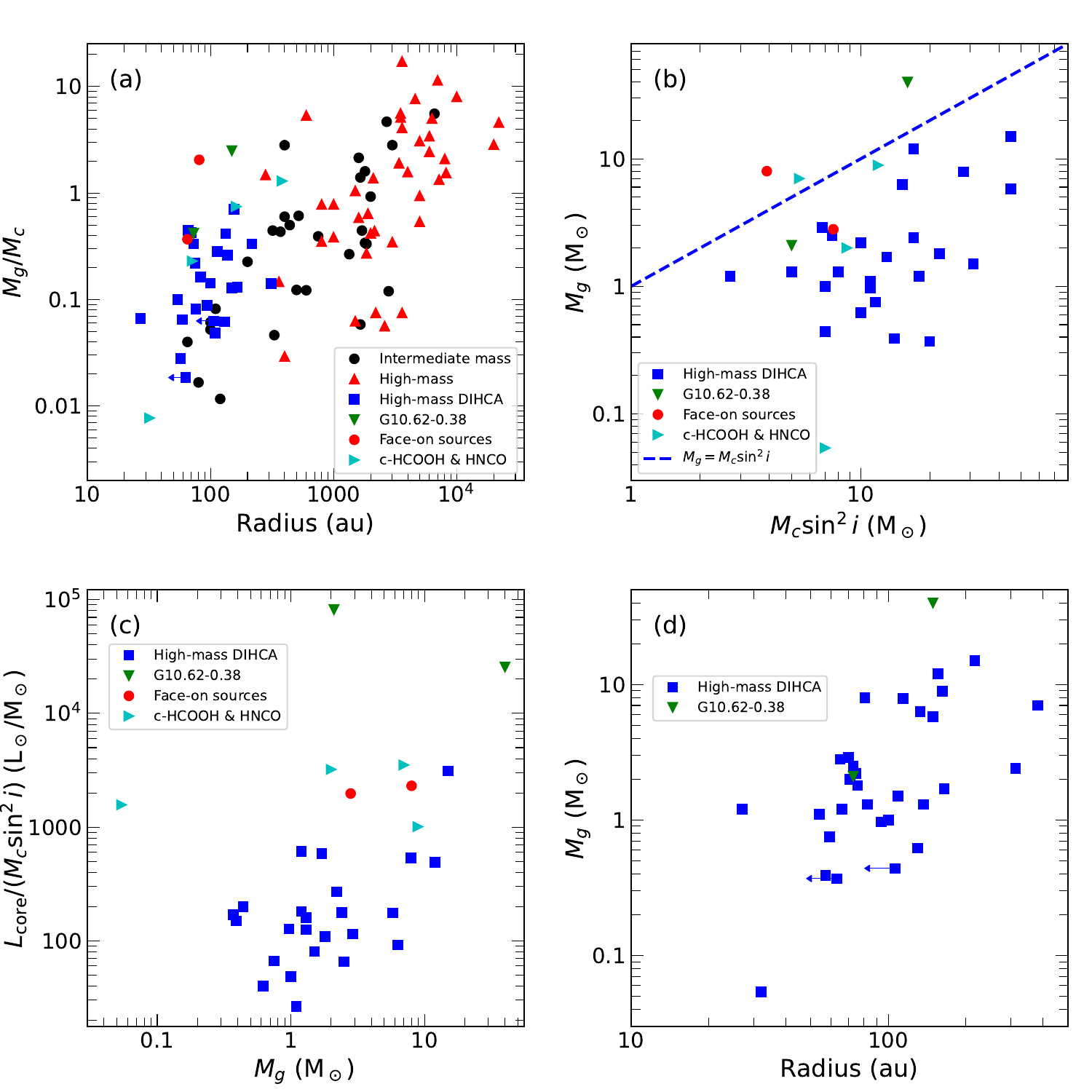}
\end{center}
\caption{Relations between measured quantities. 
The blue squares show the results from the DIHCA sample without sources considered as outliers.
(a) The $M_g/M_\star$ relation, with $M_\star=M_c\sin^2 i$, as a function of condensation radius.
The black circles and red triangles show the results summarized by \citet{Beltran2016} for intermediate-mass and high-mass star forming regions, respectively.
(b) The gas mass as a function of central source mass. 
The dashed blue line shows the 1:1 ratio.
(c) The evolutionary tracer $L_\star/M_\star$ versus gas mass.
Outliers are indicated in red circles, and cyan and green triangles (see details in \S\ref{subsec:relations}).
(d) The disk candidate mass as a function of its radius.
The green triangles show the outliers in G10.62--0.38, whose 1.3\,mm flux density is likely contaminated by free-free emission.
}\label{fig:rel}
\end{figure*}

\citet{Beltran2016} showed a correlation between the gas-to-stellar mass ratio, $M_g/M_\star$, and source radius derived from observations of intermediate and high-mass star-forming regions.
They attributed this relation to the lack of enough angular resolution to separate disk and envelope components, and hence determine the nature of the rotation and the masses of the putative disks.
From the flux densities of the condensations, $S_\nu$, and peak continuum intensity, $I_\nu$, measured by \citetalias{Luo2026} and listed in Table~\ref{tab:condprops}, we estimate the circumstellar gas mass corrected for the optical depth \citep[e.g.,][]{Pouteau2022}: 
\begin{equation}
    M_g = \tau_\nu \frac{\Omega_{\rm beam} S_{\nu} d^2}{R_{dg} \kappa_{\nu} I_\nu} = \tau_\nu M_g^{\rm thin} \frac{\Omega_{\rm beam} B_\nu(T)}{I_\nu}\,,
\end{equation}
with $\Omega_{\rm beam}$ the beam area, $d$ the distance in Table~\ref{tab:props}, $R_{dg}=0.01$ the dust-to-gas ratio, $B_\nu$ the Planck black-body function, and $M_g^{\rm thin}$ the gas mass assuming optically thin emission.
The optical depth is defined as
\begin{equation}
    \tau_\nu = -\ln\left( 1 - \frac{I_\nu}{\Omega_{\rm beam} B_\nu(T)}\right)\,.
\end{equation}
We assume that the dust temperature is the same as the gas temperature estimated in \citetalias{Taniguchi2023} by fitting the $^{13}$CH$_3$CN $J_K=13_K-11_K$ $K$-ladder from the compact configuration DIHCA data.
If the core associated to the condensation is not in \citetalias{Taniguchi2023} we use the $^{13}$CH$_3$OH rotational temperatures from \citetalias{Sakai2025}, with the exception of G335.579--0.272 ALMAe4 where we use the temperature from \citetalias{Olguin2021}.
Note that the $^{13}$CH$_3$OH temperatures are measured at the peak molecular line emission, hence we use the nearest value to the peak continuum emission.
In the absence of any temperature measurement, we use the brightness temperature at the peak of the line (Figure~\ref{fig:peak_spec}).
As the emission toward the central source becomes optically thick, the brightness temperature tends to the excitation temperature.
Since complex organic molecules are detected in the sample, we expect that water ice is sublimated and the dust grains are bare silicates.
Thus, we adopt a dust opacity value of $\kappa_\nu = 0.24$\,cm$^2$\,g$^{-1}$ at 1.3\,mm, following \citet{Yamamuro2025},
who evaluated the millimeter opacity of astronomical-silicate grains in massive proto-stellar disks using \textsc{OpTool} \citep{Dominik2021}.
Sources G35.13--0.74 ALMAe1, IRAS 18162--2048 ALMAe1, NGC 6334I ALMAe1, NGC 6334I(N) ALMAe2 and W33A ALMAe1 have undefined $\tau_\nu$ values.
In these cases, we use the $M_g^{\rm thin}$ values.

The individual properties of each condensation with central mass estimates are listed in Table~\ref{tab:condprops}, while a histogram of condensation masses is presented in Figure~\ref{fig:alpha_hist}(b).
To calculate the gas-to-stellar mass ratio, we assume that the central mass within 100\,au is dominated by the stellar mass, thus  $M_\star\approx M_c \sin^2 i$ (i.e., ignoring the inclination angle correction, see below).
Figure~\ref{fig:rel}(a) presents the gas-to-stellar mass ratio vs. radius relation from the DIHCA sample and the intermediate and high-mass points in \citet{Beltran2016}.
Note that there is an overlap between the DIHCA sample and the data compiled by \citet{Beltran2016}, but their stellar masses are estimated from simulated clusters and based on lower angular resolution observations than DIHCA.
The dependence of the central mass on inclination  would make our estimation of $M_\star$ a lower limit.
Similarly, the dust opacity is relatively low compared with other opacity laws (e.g., 1\,cm$^2$\,g$^{-1}$ from \citealp{Ossenkopf1994}, 1.9\,cm$^2$\,g$^{-1}$ for proto-planetary disks in \citealp{Birnstiel2018}) and the dust temperature can also be higher since the used value may represent the colder envelope component of the core, hence the estimation of $M_g$ would be an upper limit for the sources corrected for the optical depth.
The masses without optical depth correction are slightly underestimated (a few solar masses) compared to those with the correction.
However, \citet{Anez-Lopez2020} estimated a disk mass of 5\,\msun\ for IRAS 18162--2048 ALMAe1 from radiative transfer modeling, which is slightly lower than $M_g^{\rm thin}$.
Thus, the estimated $M_g/M_\star$ would be an upper limit in most cases.

Our results at ${\sim}230$\,au spatial resolution are located in the area of the most compact intermediate sources and separated from the previous high-mass results, thus confirming the lack of angular resolution hypothesis.
Omitting the \ion{H}{2} region source G10.62--0.38, 22 out of 30 condensations (73\%) have gas masses that are lower than 5\,\msun.
This is a significant fraction of the sources compared to the 7 out 13 sources below 5\,\msun\ in \citet[][54\%]{Ahmadi2023}, but in line with the mass of the fragments in Cygnus-X from \citet[][$<3$\,\msun]{Pan2025}.
On the other hand, the condensations are slightly more compact than the fragments in \citet[][effective radii $>200$\,au]{Pan2025}, with 27 out of 30 with radius smaller than 200\,au (90\%).
Given that many of the condensations have Keplerian-like rotation and that their radii is smaller than our angular resolution, they constitute a good sample of disk candidates.
In the more nearby sources with Keplerian power-laws, our observations may be mapping the outer regions of the disk, while power-laws between $-0.5$ and $-1.0$ may indicate that the observations are mapping a transition zone.
In this transition zone the superposition of the motions of an unresolved Keplerian disk and the rotating and infalling envelope would explain the observed power-law indices.

Figure~\ref{fig:rel}(b) indicates that the distribution of condensation masses is relatively flat compared to that of the central source masses, with only a weak dependency (Figure~\ref{fig:correl_map}).
Hence, massive disks may not necessarily be needed to form the most massive stars.
This could be the result of high-accretion rates which can be attained in part from accretion burst due to fragmentation or large scale streamers, or feedback may have already started to destroy the dust thus resulting in smaller gas masses on the sources with \ion{H}{2} regions.
In addition, we explore whether the condensation masses vary with evolution, by using the evolutionary indicator $L_{\star}/M_\star$, where $L_\star$ is the stellar luminosity.
For proto-stellar masses higher than about 5\,\msun, $L_{\star}/M_\star$ increases as the stellar mass increases, with the rate of increment a function of the accretion rate \citep{Zinnecker2007}.
Since most sources in Table~\ref{tab:plfit} are the only condensation with line emission associated to its respective core, we assume that the luminosity from the condensation with line emission dominates that of the core, $L_{\rm core}$, and that $L_\star \approx L_{\rm core}$.
We use the luminosity calculated as part of the ATLASGAL survey \citep{Urquhart2018}.
This luminosity is derived from single-dish observations (from mid-IR to sub-mm), and thus represents the clump luminosity, $L_{\rm clump}$ , containing multiple cores resolved in \citetalias{Ishihara2024}.
As an approximation, we scale the luminosity by the 1.3\,mm flux density contribution of each core, $S_{\rm 1.3\,mm}^{\rm core}$, in a given clump/field:
\begin{equation}
    L_{\rm core} =  \frac{S_{\rm 1.3\,mm}^{\rm core} L_{\rm clump}}{\Sigma S_{\rm 1.3\,mm}^{\rm core}}\,,
\end{equation}
where $\Sigma S_{\rm 1.3\,mm}^{\rm core}$ is the sum of the flux density values of all the cores in a clump/field.
Figure~\ref{fig:rel}(c) shows the $L_{\rm core}/M_\star$ ratio as a function of the gas mass.
Excluding outliers, we found that there is a weak positive correlation between the evolutionary indicator and the condensation mass (Spearman correlation coefficient of 0.33 and a p-value of 0.11).
Given the presence of gas close to the central sources we could assume that they are still in their accretion phase, as confirmed by the evidence of outflows.
The gas masses are thus high enough to avoid feedback effects (e.g., radiation, ionization, winds), particularly for those with masses larger than 20\,\msun\ \citep{Rosen2022}.
If the gas mass available is independent of evolution, as the weak correlation in Figure~\ref{fig:rel}(c) seems to show, the disk-candidates may be constantly being fed by the larger envelope. 
In turn, disks can become more massive at any point of the formation process depending on the infall rate and the available gas in the larger scales.
Similarly, we do not find a correlation between $L_\star/M_\star$ and the radius of the disk candidates (Spearman correlation coefficient of 0.33, p-value of 0.14), indicating that their size do not depend on the source evolutionary stage.
On the other hand, the moderate correlation between the sizes and masses of the disk candidates (Figures~\ref{fig:correl_map} and \ref{fig:rel}(d)) seems to imply that inflows bring gas that can build up massive and large disks.  
These large disks can be thus transitory as the low frequency of these type of sources in the DIHCA sample shows (cf. histograms in Figure~\ref{fig:alpha_hist}).
These evidences indicate that anisotropic accretion through, e.g., streamers can play an important role in feeding the disks at a rate sufficient to maintain the high densities needed to overcome the feedback and feed the stars as demonstrated by recent case studies \citep{Olguin2025,Mai2025}.

\begin{figure}
%\figurenum{7}
\begin{center}
\includegraphics[angle=0,width=\linewidth]{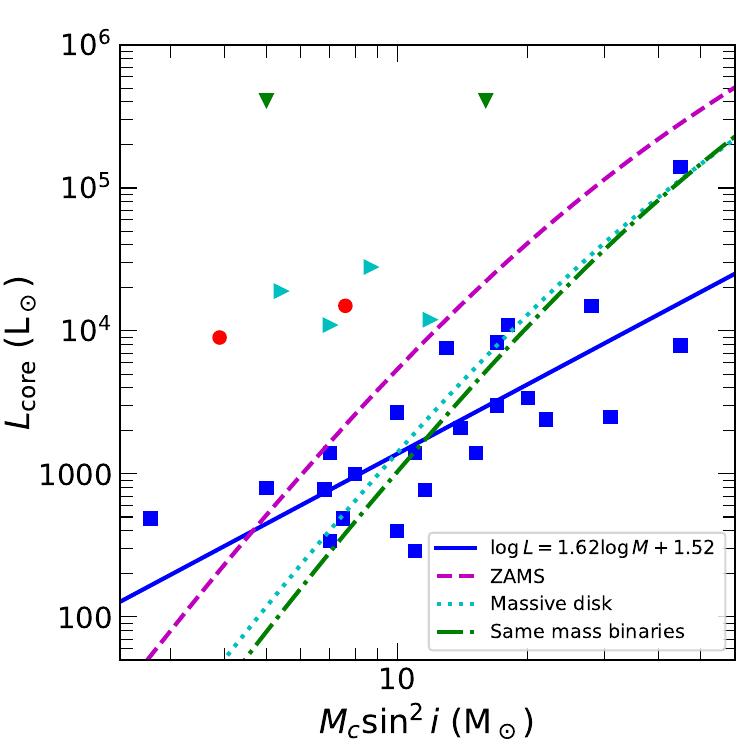}
\end{center}
\caption{Mass-luminosity relation for the sources in Table~\ref{tab:plfit}.
The point shapes and colors are the same as in Figure~\ref{fig:rel}(c).
The dashed magenta line shows the mass-luminosity relation for main sequence stars from the stellar models of \citet{Ekstrom2012}, the blue line show the power law fit to the blue squares, the dotted cyan line shows the relation expected for single stars with massive disks and the dash-dotted green line shows the relation expected for equal-mass binary systems.
}\label{fig:rel:lm}
\end{figure}

We explore whether a relation between the central/stellar masses and the luminosity, similar to that of main-sequence stars, holds at this stage or not.
Figure~\ref{fig:rel:lm} shows that most of the sources have a luminosity lower than the ZAMS luminosity \citep[from][]{Ekstrom2012} at the estimated dynamical masses.
The luminosity at the formation stage is expected to be higher than ZAMS at a given central source mass due to the contribution of the accretion luminosity \citep[e.g.,][]{Tanaka2017}.
Hence this is strikingly problematic, particularly at the higher central mass end where this trend is present even when the clump luminosity is considered (see Figure~\ref{fig:rel:aux} in Appendix~\ref{app:figs}).  
This may suggest that the total dynamical mass is not confined to single stars, but rather distributed in multiple systems or stars surrounded by massive, optically thick disks at smaller scales.
As illustrative cases, we plot the expected curves for an equal-mass binary and a single star with a massive disk, which better match the observed mass–luminosity relation.
To derive these relations, we first interpolate the ZAMS curve from \citet[][]{Ekstrom2012} to derive the $L_{\rm ZAMS}(M_\star)$ function, then an equal-mass binary system will have a luminosity $L_\star=2L_{\rm zams}(M_\star/2)$ and the luminosity of a single star with a massive disk is $L_\star=L_{\rm ZAMS}(M_\star/1.5)$.
Given the high fraction of close companions among massive stars \citep{Sana2012}, the presence of unresolved proto-stellar companions or compact disks at $<100–300$\,au scales is to be expected.
Alternatively, the luminosity may have been underestimated due to source morphology whereby radiation escapes through a different line of sight, hence affecting the emission a the peak of the spectral energy distribution.
Excluding outliers, we obtain a Spearman correlation coefficient of 0.77 (p-value of $10^{-5}$), indicative of a strong correlation, and we fit a power law of the form $\log L_\star = 1.62 \log M_c + 1.52$ (dash-dotted green line in Figure~\ref{fig:rel:lm}).

\subsection{Gravitational stability}\label{sec:discussion:stability}

In order to assess whether the sources can fragment, allowing the formation of binary/multiple systems, or develop substructures like spiral arms, we calculate the Toomre-$Q$ stability parameter.
Following \citet[, e.g.,][]{Ahmadi2023}:
\begin{equation}
    Q = \frac{c_s \Omega}{\pi G \Sigma}
\end{equation}
with the speed of sound
\begin{equation}
    c_s = \sqrt{\frac{\gamma k_B T}{\mu m_H}}
\end{equation}
where $k_B$ is the Boltzmann constant and $m_H$ the hydrogen mass.
Similar to \citet[][]{Ahmadi2023} and \citet{Pan2025}, we use an adiabatic index $\gamma=7/5$ and a mean molecular weight $\mu=2.8$.
The angular velocity is
\begin{equation}
    \Omega = \sqrt{\frac{G (M_g + M_c \sin^2 i)}{R^3}}
\end{equation}
and the surface density $\Sigma=M_g/(\pi R^2)$ with $R$ the condensation radius.
Figure~\ref{fig:alpha_hist}(d) shows a histogram of the Toomre-$Q$ values listed in Table~\ref{tab:condprops}.
Although the condensations in DIHCA at ${\sim}230$\,au scales differ from the those in the CORE program at 1000\,au \citep{Beuther2018,Ahmadi2023}, at the smaller scales most sources remain gravitationally unstable (ignoring outliers, 17 out of 22).
Sources in the DIHCA sample have Toomre-$Q$ values in the 0.07-12 range with a median of ${\sim}0.5$.
The extreme values seem to be outliers with the lowest one corresponding to G10.62--0.38 ALMAe1, whose gas mass is uncertain, and the highest one corresponding to NGC 6334I ALMAe4, which has the lowest gas mass in the sample like due to it relatively high gas temperature compared with the other sources.
Among the stable sources are G335.579--0.272 ALMAe4, which may have a companion, and G336.01--0.82 ALMAe3, which indicates that it may still hold a small stable disk fed by streamers.
While G35.03+0.35 A ALMAe1 seems also to be stable, its radius is not well constrained.

\subsection{Kinematics tracers}\label{sec:discussion:tracers}

\begin{figure}
\begin{center}
\includegraphics[angle=0,width=0.7\linewidth]{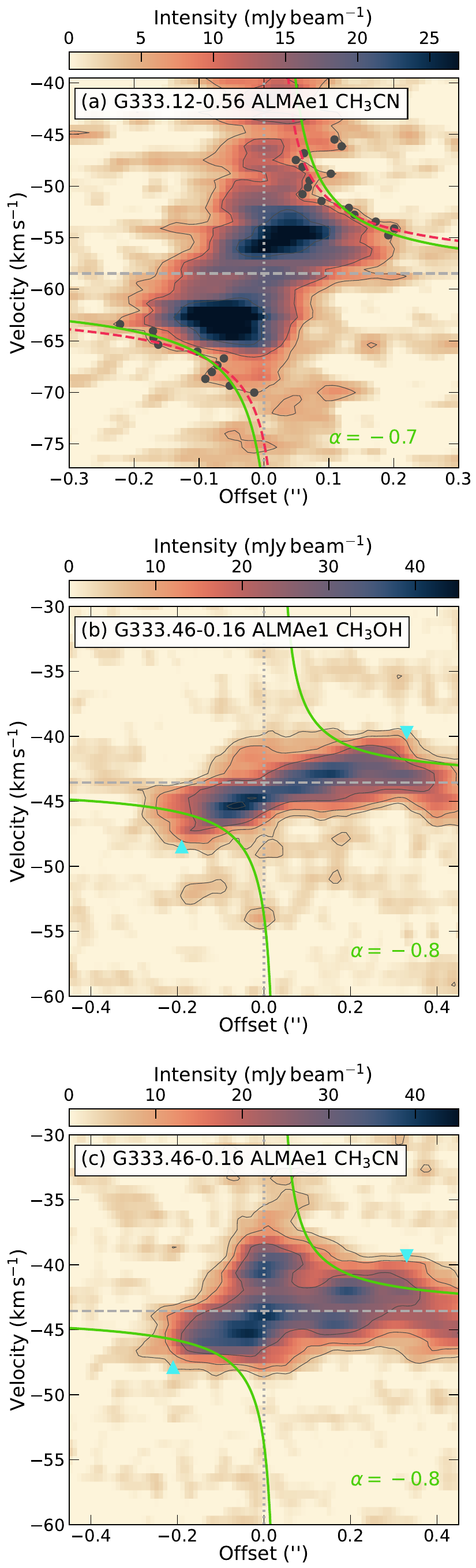}
\end{center}
\caption{Position-velocity maps from \methanol\ and \metcyn\ emission for selected sources along the rotation direction.
(a) The cyan circles indicate the edge points, and the continuous gray and dashed red lines show the boundless and Keplerian power law fits to the edge points, respectively.
(b) The cyan triangles indicate the position of the velocity extrema determined from the edge points.
The gray lines show the boundless model fitted to the HNCO emission in Figure~\ref{fig:pvmaps}.
The dotted vertical and dashed horizontal green lines indicate the zero position offset and the systemic velocity in Table~\ref{tab:props}, respectively.
}\label{fig:pvcompare}
\end{figure}

In this work we have used different molecules to trace the kinematics of the hot cores close to the forming stars.
As the explored transitions have different upper level energies, they may be tracing different layers of gas.
Here we explore what is the effect of the molecular line selection on the resulting source properties of G333.12--0.56 ALMAe1 and G333.46--0.16 ALMAe1.
We select these sources because they have kinematics resolved in \methanol\ and \metcyn, while G333.46--0.16 ALMAe1 also has HNCO (one of the additional molecules used), with the same channel width.
The PV maps in Figure~\ref{fig:pvmaps} show that they have a relatively well fitted Keplerian profile.

Figure~\ref{fig:pvcompare}(a) shows the PV map from \metcyn\ \metcyntransk in G333.12--0.56 ALMAe1 with its power law fit following the same procedure as for \methanol\ \mettrans.
The resulting power law index $\alpha$ from an boundless fit is $-0.7\pm0.1$ which is higher than the one obtained from \methanol\ ($\alpha=-0.56\pm0.06$).
Similarly, if the outlier edge points with velocities higher than $-48$\,\kms\ are omitted the power law index is $\alpha=-0.61{\pm}0.07$, still consistent within the uncertainties but expected if \metcyn\ is tracing in part colder regions associated with a rotating and infalling envelope.
On the other hand, the mass within 100\,au from the Keplerian fit to the \metcyn\ emission is $20\pm2$\,\msun\ (with and without outliers), which is roughly 5\,\msun\ higher than that derived from \methanol\ but still within the uncertainties at the $3\sigma$ level.

Figures~\ref{fig:pvcompare}(b) and (c) show a different distribution than that from HNCO for G333.46--0.16 ALMAe1.
In general, HNCO emission is more compact than \metcyn\ and \methanol, even for the sources in NGC 6334I.
The PV maps in Figure~\ref{fig:pvmaps} in the rotation and its orthogonal directions and the power law fit indicate that the HNCO emission is not Keplerian.
The first order moment map in Figure~\ref{fig:mom1} seems to indicate that the kinematics traced by HNCO are more complex.
On the other hand, \metcyn\ and \methanol\ are tracing roughly the same component.
Using the velocity extrema, we obtain the same source mass, 16\,\msun, within a radius of 750 and 780\,au for \methanol\ and \metcyn, respectively.
Therefore, these two sources show that there is no clear cut between the source  properties derived from \methanol\ and \metcyn.
This is also supported by the histogram of power-law indices in Figure~\ref{fig:alpha_hist}.
Detailed radiative transfer modeling of the emission of these molecules in future studies would provide more accurate physical properties.

Finally, c-HCOOH was used for IRAS 18162--2048 ALMAe1 and a relatively good power law fit was obtained.
Although the lack of \methanol\ and \metcyn\ precludes a detailed study of what this molecular line emission is tracing, the central mass derived from the fitting is in line to that estimated from the detailed study performed by \citet{Fernandez-Lopez2023}.

Even though \metcyn\ and \methanol\ seem to be tracing similar kinematics, this may not be true in general across observations mapping different scales. 
For instance, \citet{Law2021} found structured emission of complex organic molecules in G10.62--0.38 at intermediate scales (few hundred to thousand au scales) that trace different kinematics, albeit in one of the more evolved regions in the DIHCA sample.
At smaller scales, \methanol\ abundance may decrease close to the central source due to reactions with H$_3^+$ \citep{Garrod2006}, making them appear weaker and thus hindering their ability to trace the gas motion of disks.
These two issues could be addressed in future works by using principal component analysis in order to determine the best tracers for the same gas components that could be used at different scales \citep[e.g.,][]{Okoda2021} in addition to chemical models.

\subsection{Infall and rotation to form high-mass stars}

In this work we have shown that the disk candidates have Keplerian-like rotation, but also exhibit hints of infall.
Table~\ref{tab:condprops} summarizes the gas motions in the disk candidates.
We propose that high-mass stars form at the center of envelope/disk systems, whereby an unstable envelope channel gas inward from scales orders of magnitude larger than those of the disks.
The unstable envelopes can fragment, allowing the formation of companions, and/or develop preferential channels to feed the disks at high infall rates \citep[e.g.,][]{Olguin2025}.
Anisotropic collapse can also be attained or triggered by interaction with other sources in clustered environments \citep[e.g.,][]{Lu2022}.
The constant streaming of fresh gas allow to maintain a high density in the zone close to the star to avoid feedback effects, like quenching of accretion.
Anisotropic collapse can also help to build up some of the larger and massive, albeit short-lived, Keplerian disks observed \citep[e.g.,][]{Johnston2020}.
As the gas joins the disk, the inflows transition to Keplerian rotation resulting in a region where spectral indices of the power-law PV levels change, and where infall shocks are expected.
These shocks can be studied in future works from some of the shock tracers in the DIHCA spectral coverage (e.g., SO, SiO).

\section{Conclusions} \label{sec:conclusions}

We systematically observed 30 fields forming high-mass stars at 1.3\,mm with ALMA as part of the DIHCA project.
These high-angular resolution observations resolve the kinematics between ${\sim}230$\,au scales.
We search for molecular line emission mainly from the well-known kinematic tracers \methanol\ and \metcyn\ in order to study the circumstellar gas. 
Of the 30 fields, 49 condensation have  line emission over the $5\sigma$ detection level to study their kinematics.

Through PV maps, we explore whether the velocity gradients observed in the first moment maps represent rotation.
We are able to fit a power law distribution to the PV map edges derived from the contours of the low level emission of 32 sources, that we suggest are disk candidates. 
With the analysis of the largest, uniformly sample to date of disk-like structures at ${\sim230}$\,au scales, we find that power-law indices are mostly distributed in the --0.5 to --1.0  range, with a median of --0.7.
This indicates that the gradient can be explained by Keplerian rotation with some contribution of a rotating and infalling envelope.
Future studies performing a detailed modeling of the kinematics would be able to separate the contribution of each component.
We provide constraints for another 14 sources while the kinematics cannot be resolved in the remaining 3 sources.

The PV maps also show that the molecular line emission is highly asymmetric on the opposite sides of the central source.
We argue that this is the result of anisotropic infall, whereby inflows land at different projected radii along the line of sight and perhaps with different angular momentum vectors.
This would result in enhancement of molecular line emission due to density and/or temperature increments.
Given that the mass of the disk candidates are relatively independent of the central source masses, these inflows should play an important role in continuously feeding the central region and eventually the star in order to oppose the feedback as the stellar mass increases.

We conclude that the DIHCA observations reveal the regime at which we begin to separate/distinguish disks from envelopes in high-mass hot cores.
This results in sources that are more compact and less massive than those derived by early ALMA observations.
More sensitive observations will be needed to increase the detection of lines from fainter sources in order to increase the number statistics and bridge the gap to the low-mass proto-stellar population.
At the same time higher angular resolution observations will allow us to determine whether the compact disks are stable and how the inflows feed them.

\begin{acknowledgments}
F.O. and Y.O. acknowledge the support of the NAOJ ALMA Joint Scientific Research Program grant No. 2024-27B.
P.S. was partially supported by a Grant-in-Aid for Scientific Research (KAKENHI Number JP23H01221) of JSPS.
F.O. and H.-R.V.C. acknowledge the support from the National Science and Technology Council (NSTC) of Taiwan grants NSTC 112-2112-M-007-041 and NSTC 112-2811-M-007-048.
Y.O. acknowledges the support by Grant-in-Aids from Ministry of Education, Culture, Sports, Science, and Technologies of Japan (KAKENHI; 21K13954, 25K07367).
R.G.M acknowledges support from UNAM-DGAPA-PAPIIT project IN105225.
QY-L acknowledges the support by JSPS KAKENHI Grant Number JP23K20035.
K.E.I.T. and S.Z. acknowledge support from the NAOJ ALMA Scientific Research Grant Code 2025-29B.
K.E.I.T. acknowledges the support by JSPS KAKENHI Grant Number JP 25K07365.
K.T. is supported by JSPS KAKENHI grant Nos. 21H01142, 24K17096, and 24H00252.
G.G. acknowledges support by the ANID BASAL project FB210003.
X.L. acknowledges support from the Strategic Priority Research Program of the Chinese Academy of Sciences (CAS) Grant No. XDB0800300, the National Key R\&D Program of China (No. 2022YFA1603101), State Key Laboratory of Radio Astronomy and Technology (CAS), the National Natural Science Foundation of China (NSFC) through grant Nos. 12273090 and 12322305, the Natural Science Foundation of Shanghai (No. 23ZR1482100), and the CAS ``Light of West China'' Program No. xbzg-zdsys-202212.
This paper makes use of the following ALMA data: ADS/JAO.ALMA\#2016.1.01036.S and \#2017.1.00237.S. ALMA is a partnership of ESO (representing its member states), NSF (USA) and NINS (Japan), together with NRC (Canada), MOST and ASIAA (Taiwan), and KASI (Republic of Korea), in cooperation with the Republic of Chile. The Joint ALMA Observatory is operated by ESO, AUI/NRAO and NAOJ.
Data analysis was in part carried out on the Multi-wavelength Data Analysis System operated by the Astronomy Data Center (ADC), National Astronomical Observatory of Japan.
The Scientific color maps vik and lipari \citep{crameri_2023_8409685} are used in this work to prevent visual distortion of the data and exclusion of readers with color vision deficiencies \citep{2020NatCo..11.5444C}. 
\end{acknowledgments}

\vspace{5mm}
\facilities{ALMA}

\software{Astropy \citep{astropy:2013, astropy:2018, astropy:2022},  
          CASA \citep{2022PASP..134k4501C},
          Scipy \citep{2020SciPy-NMeth},
          YCLEAN \citep{Contreras18,2018zndo...1216881C,olguin_2025_17197133},
          GoContinuum \citep{2020zndo...4302846O},
          statcont \citep{Sanchez-Monge2018}
          }

\appendix

\section{Auto-masking improvements}

For the cube cleaning, we modified/improved the auto-masking routine YCLEAN \citep{Contreras18,2018zndo...1216881C}.
The algorithm performs several iterations of \texttt{tclean} with an incremental mask calculated from the results of each iterations, and can be divided in three steps: initial (dirty) clean, main cycle, and final clean.
From the initial step parameters are calculated from a dirty run of \texttt{tclean} (\texttt{niter=0}), and include the secondary lobe level from the PSF image (\texttt{secondary\_lobe}), the rms calculated from image data over a primary beam level of 0.2 using the median average deviation over a sample of channels (\texttt{rms}), and residual maximum (\texttt{residual\_max}).
At each iteration of the main cycle the rms and residual maximum are recalculated, and a limit level based on the signal-to-noise ratio (SNR) is calculated as:
\begin{equation}
    \texttt{limit\_level\_snr} = \texttt{secondary\_lobe\_level}\,\frac{\texttt{residual\_max}}{\texttt{rms}} \,.
\end{equation}.
The iterations run until one of the following conditions is achieved:
\begin{itemize}
    \item $\texttt{limit\_level\_snr} > \texttt{min\_limit\_level}$ where the input parameter \texttt{min\_limit\_level} is set to a default value of $1.5$.
    \item The iteration limit is achieved (default value of 10 iterations).
    \item The residual maximum increases.
    \item The corrected residual maximum (see below) is below the threshold of the final iteration (set to $2\texttt{rms}$).
\end{itemize}
For each iteration in the main cycle a mask level and cleaning threshold are calculated.
The mask level (\texttt{mask\_level}) is defined as:
\begin{equation}
    \texttt{mask\_level} = (\texttt{limit\_level\_snr} + 1.5 \exp(-(\texttt{limit\_level\_snr} - 1.5) / 1.5)) \, \texttt{rms}\,.
\end{equation}
For each mask, small clusters of pixels are removed if their areas are smaller than half the beam at a given channel.
The threshold is calculated as: 
\begin{equation}
    \texttt{threshold} = (0.4 + \arctan(\texttt{secondary\_lobe\_level} - 0.2)) \,\texttt{residual\_max}\,.
\end{equation}
This is an improvement on the determination of the threshold from earlier implementations of YCLEAN, where for higher values of the secondary lobe level the threshold did not decrease with iterations.
We implement a corrected \texttt{limit\_level\_snr} and \texttt{residual\_max} when the the new maximum residual value (calculated at the end of the cleaning cycle) is within a tolerance level of the one from the previous iteration in order to keep decreasing the threshold and mask values.
The tolerance is set as default as $0.01\texttt{rms}$ and the corrected values are $0.8^n \texttt{residual\_max}$ and $0.8^n \texttt{limit\_level\_snr}$, where $n$ is the number of times the correction has been applied.
Finally, a mask with a $3\texttt{rms}$ level is produced and a final \texttt{tclean} run with a threshold level of $2\texttt{rms}$ is performed.

\section{Edge parameters}\label{sec:appendix:edge}

In order to determine the PV map edges, we use contour levels based on the rms noise of the PV maps (determined from the rms on a line free region).
This defines regions over an $n\sigma$ level.
We use the largest (or the two largest if the central source is in absorption like in the case of IRAS 18162--2048) contiguous region over the level to extract the edge points as described in \S\ref{subsec:kinematics}.
Since some sources have line forests and/or wider lines that can produce a larger region joining two or more lines, we introduce limits in velocity in the edge extraction.
Similarly, some sources may be connected to emission of a common envelope gas, thus we also introduce limits in offset for these cases.
Table~\ref{tab:edgeparams} lists the rms ($\sigma$), levels over the rms used ($n$), and limits in velocity and offset used in the edge calculations.

To assess the effects of the levels selected, we run the same edge extraction and fitting procedures for levels between $3 - \max(5, n)$ with $n$ in Table~\ref{tab:edgeparams} and in steps of 1.
In general, and as expected, higher contours over the noise level result in lower masses.
Larger variations in mass are obtained for the brightest sources (e.g., G11.92--0.61 ALMAe1, IRAS 18089--1732 ALMAe1) with variations of roughly 8\,\msun\ for levels between 3 to $7-8\sigma$, but in general changes are around 2-3\,\msun\ for levels between 3 and $5\sigma$ (with the exception of sources with erratic results).
Higher levels should produce masses that are more accurate, but since the PV maps are not smooth worse fits may be obtained for higher levels.
This seems to have a larger effect on the values of the power law index, $\alpha$, which do not have a clear trend.
Nevertheless variations of $\alpha$ values are generally within ${\sim}0.3$.

\begin{deluxetable}{lcCCCCCC}
\tablecaption{PV map edge parameters\label{tab:edgeparams}}
\tablewidth{0pt}
\tablehead{
\colhead{Source} & \colhead{ALMAe} & \colhead{$\sigma$} & \colhead{Level} & \multicolumn{2}{c}{Velocity range} & \multicolumn{2}{c}{Offset range} \\
\colhead{} & \colhead{Source} & \colhead{(mJy\,beam$^{-1}$)} & \colhead{} & \multicolumn{2}{c}{(\kms)} & \multicolumn{2}{c}{(arcsec)} 
}
\startdata
G10.62--0.38     & 1  & 1.6 & 4 & -10 & 8                     & -0.5 & 0.5 \\
                 & 3  & 1.6 & 6 & \multicolumn{2}{c}{\nodata} & -0.5 & 0.5 \\
G11.1--0.12      & 1  & 1.3 & 3 & \multicolumn{2}{c}{\nodata} & \multicolumn{2}{c}{\nodata} \\
G11.92--0.61     & 1  & 1.1 & 8 &  20 & 50                    & \multicolumn{2}{c}{\nodata} \\
                 & 4  & 1.1 & 4 & \multicolumn{2}{c}{\nodata} & \multicolumn{2}{c}{\nodata} \\
                 & 6  & 1.1 & 3 & \multicolumn{2}{c}{\nodata} & \multicolumn{2}{c}{\nodata} \\
G14.22--0.50 S   & 3  & 1.3 & 3 & \multicolumn{2}{c}{\nodata} & \multicolumn{2}{c}{\nodata} \\
G29.96--0.02     & 1  & 1.4 & 5 &  85 & 110                   & \multicolumn{2}{c}{\nodata} \\
G333.12--0.56    & 1  & 1.8 & 3 & \multicolumn{2}{c}{\nodata} & \multicolumn{2}{c}{\nodata} \\
                 & 8  & 1.7 & 3 & \multicolumn{2}{c}{\nodata} & \multicolumn{2}{c}{\nodata} \\
G333.23--0.06    & 6  & 1.1 & 5 & \multicolumn{2}{c}{\nodata} & \multicolumn{2}{c}{\nodata} \\
                 & 17 & 1.3 & 3 & \multicolumn{2}{c}{\nodata} &  -0.38 & 0.1 \\ % to update
G333.46--0.16    & 1\tablenotemark{a}  & 1.7 & 4 & -50 & -35                   & \multicolumn{2}{c}{\nodata} \\
                 & 1\tablenotemark{b}  & 1.9 & 3 & \multicolumn{2}{c}{\nodata} & \multicolumn{2}{c}{\nodata} \\
                 & 2  & 1.7 & 3 & \multicolumn{2}{c}{\nodata} & \multicolumn{2}{c}{\nodata} \\
G335.579--0.272  & 1  & 1.4 & 4 & \multicolumn{2}{c}{\nodata} & \multicolumn{2}{c}{\nodata} \\
                 & 4  & 1.4 & 3 & \multicolumn{2}{c}{\nodata} & \multicolumn{2}{c}{\nodata} \\
G335.78+0.17     & 1  & 1.6 & 6 & \multicolumn{2}{c}{\nodata} & \multicolumn{2}{c}{\nodata} \\
                 & 2  & 1.8 & 3 & \multicolumn{2}{c}{\nodata} & \multicolumn{2}{c}{\nodata} \\
G336.01--0.82    & 3  & 1.7 & 4 & \multicolumn{2}{c}{\nodata} & \multicolumn{2}{c}{\nodata} \\
G34.43+0.24 MM1  & 2  & 1.3 & 6 & \multicolumn{2}{c}{\nodata} & -0.5 & 0.5 \\
G343.12--0.06    & 1  & 1.1 & 6 & \multicolumn{2}{c}{\nodata} & \multicolumn{2}{c}{\nodata} \\
G35.03+0.35 A    & 1  & 1.9 & 5 & \multicolumn{2}{c}{\nodata} & \multicolumn{2}{c}{\nodata} \\
G35.13--0.74     & 1  & 2.5 & 5 & \multicolumn{2}{c}{\nodata} & \multicolumn{2}{c}{\nodata} \\
                 & 2  & 2.5 & 6 & \multicolumn{2}{c}{\nodata} & \multicolumn{2}{c}{\nodata} \\
                 & 7  & 2.5 & 5 & \multicolumn{2}{c}{\nodata} & \multicolumn{2}{c}{\nodata} \\
G35.20--0.74 N   & 1  & 2.5 & 6 & \multicolumn{2}{c}{\nodata} & \multicolumn{2}{c}{\nodata} \\
                 & 2  & 2.5 & 3 & \multicolumn{2}{c}{\nodata} & -0.35 & 0.5 \\
G5.89--0.37      & 1  & 1.8 & 6 & \multicolumn{2}{c}{\nodata} & \multicolumn{2}{c}{\nodata} \\
IRAS 16562--3959 & 1  & 1.4 & 6 & \multicolumn{2}{c}{\nodata} & \multicolumn{2}{c}{\nodata} \\
                 & 4  & 1.4 & 6 & \multicolumn{2}{c}{\nodata} & \multicolumn{2}{c}{\nodata} \\
IRAS 18089--1732 & 1  & 1.6 & 7 &  20 & 46                    & -0.7 & 0.7 \\
                 & 2  & 1.6 & 4 & \multicolumn{2}{c}{\nodata} & \multicolumn{2}{c}{\nodata} \\
IRAS 18162--2048 & 1  & 1.2 & 5 & \multicolumn{2}{c}{\nodata} & -0.4 & 0.4 \\
IRAS 18182--1433 & 1  & 1.4 & 5 & \multicolumn{2}{c}{\nodata} & \multicolumn{2}{c}{\nodata} \\
                 & 2  & 1.4 & 3 & \multicolumn{2}{c}{\nodata} & \multicolumn{2}{c}{\nodata} \\
                 & 5  & 1.4 & 4 & \multicolumn{2}{c}{\nodata} & \multicolumn{2}{c}{\nodata} \\
                 & 11 & 1.4 & 6 & \multicolumn{2}{c}{\nodata} & \multicolumn{2}{c}{\nodata} \\
NGC 6334I        & 1  & 1.5 & 5 & \multicolumn{2}{c}{\nodata} & \multicolumn{2}{c}{\nodata} \\
                 & 3  & 1.5 & 3 & \multicolumn{2}{c}{\nodata} & \multicolumn{2}{c}{\nodata} \\
                 & 4  & 1.5 & 4 & \multicolumn{2}{c}{\nodata} & \multicolumn{2}{c}{\nodata} \\
NGC 6334I(N)     & 1  & 1.5 & 3 & \multicolumn{2}{c}{\nodata} & \multicolumn{2}{c}{\nodata} \\
                 & 2  & 1.5 & 3 & \multicolumn{2}{c}{\nodata} & \multicolumn{2}{c}{\nodata} \\
                 & 8  & 1.7 & 3 & \multicolumn{2}{c}{\nodata} & \multicolumn{2}{c}{\nodata} \\
                 & 9  & 1.7 & 3 & \multicolumn{2}{c}{\nodata} & \multicolumn{2}{c}{\nodata} \\
                 & 14 & 1.7 & 3 & \multicolumn{2}{c}{\nodata} & \multicolumn{2}{c}{\nodata} \\
W33A             & 1  & 1.3 & 3 & \multicolumn{2}{c}{\nodata} & -0.5 & 0.5 \\
                 & 4  & 1.3 & 6 & \multicolumn{2}{c}{\nodata} & \multicolumn{2}{c}{\nodata} \\
\enddata
\tablenotetext{a}{For HNCO.}
\tablenotetext{b}{For \methanol\ and \metcyn.}
\end{deluxetable}

\section{Supplementary figures}\label{app:figs}

\begin{figure*}
\figurenum{C1}
\begin{center}
\includegraphics[angle=0,scale=0.3]{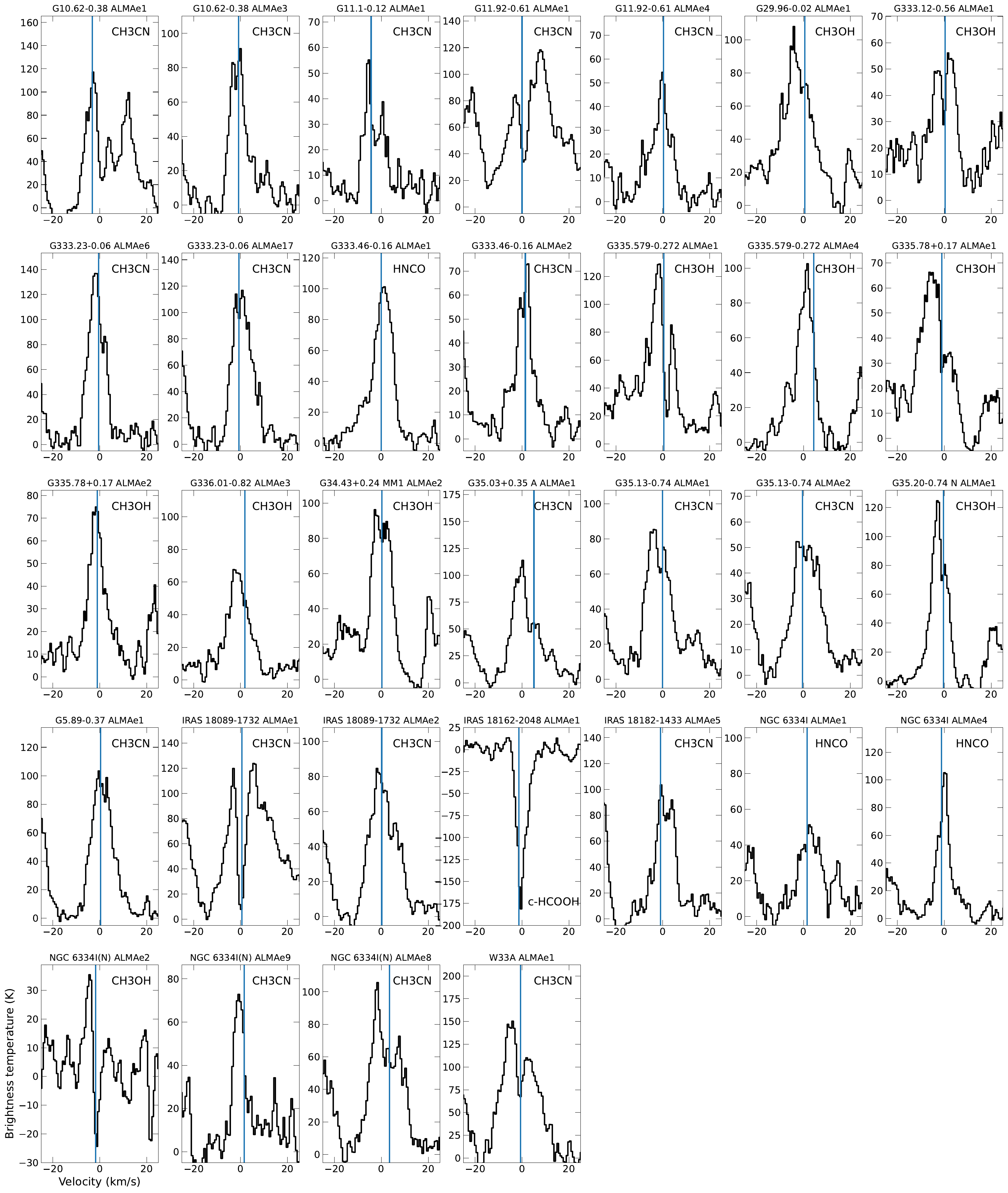}
\end{center}
\caption{Peak spectra calculated as the average with a circle of radius 0\farcs05 centered in the source continuum position.
The blue vertical line shows the velocity offset between the $v_{LSR}$ in Table~\ref{tab:props} (set to zero velocity) and the systemic velocity obtained through the power law fit to the PV map edges in Table~\ref{tab:plfit}.
}\label{fig:peak_spec}
\end{figure*}

\begin{figure}
\figurenum{C2}
\begin{center}
\includegraphics[angle=0,width=\linewidth]{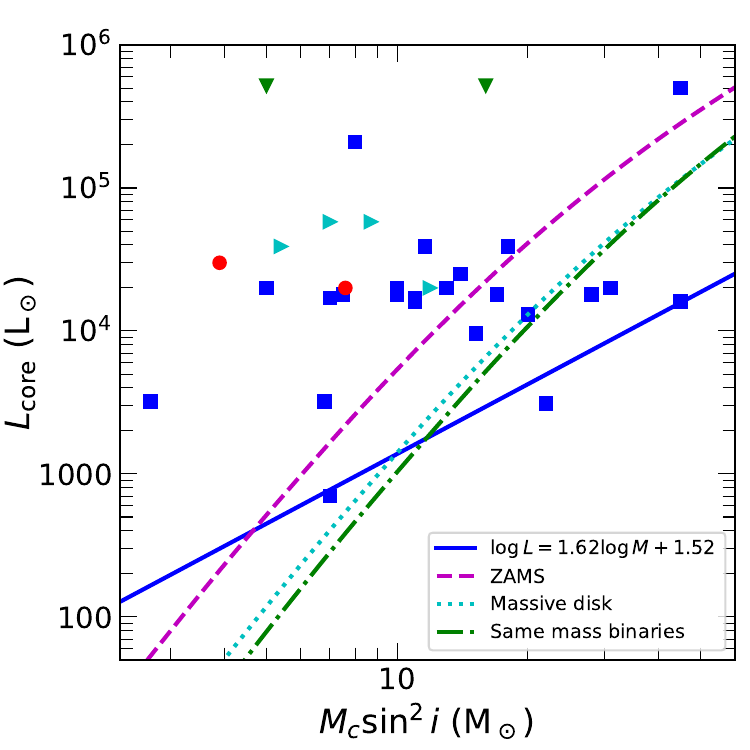}
\end{center}
\caption{Central sources mass-luminosity relation with the luminosity set to that of the clump.
The line and point types are the same as in Figure~\ref{fig:rel}(d).
}\label{fig:rel:aux}
\end{figure}

Figure~\ref{fig:peak_spec} shows the peak spectra of the 32 disk candidates, calculated as the average in a circle of radius 0\farcs05 centered in the source positions in Table~\ref{tab:props}.
Figure~\ref{fig:rel:aux} shows the mass-luminosity relation with the luminosity of each source equal to that of the clump.

\bibliography{manuscript}{}

@ARTICLE{Ahmadi2023,
       author = {{Ahmadi}, A. and {Beuther}, H. and {Bosco}, F. and {Gieser}, C. and {Suri}, S. and {Mottram}, J.~C. and {Kuiper}, R. and {Henning}, T. and {S{\'a}nchez-Monge}, {\'A}. and {Linz}, H. and {Pudritz}, R.~E. and {Semenov}, D. and {Winters}, J.~M. and {M{\"o}ller}, T. and {Beltr{\'a}n}, M.~T. and {Csengeri}, T. and {Galv{\'a}n-Madrid}, R. and {Johnston}, K.~G. and {Keto}, E. and {Klaassen}, P.~D. and {Leurini}, S. and {Longmore}, S.~N. and {Lumsden}, S.~L. and {Maud}, L.~T. and {Moscadelli}, L. and {Palau}, A. and {Peters}, T. and {Ragan}, S.~E. and {Urquhart}, J.~S. and {Zhang}, Q. and {Zinnecker}, H.},
        title = "{Kinematics and stability of high-mass protostellar disk candidates at sub-arcsecond resolution. Insights from the IRAM NOEMA large programme CORE}",
      journal = {\aap},
         year = 2023,
        month = sep,
       volume = {677},
          eid = {A171},
        pages = {A171},
          doi = {10.1051/0004-6361/202245580},
archivePrefix = {arXiv},
       eprint = {2305.00020},
 primaryClass = {astro-ph.GA},
       adsurl = {https://ui.adsabs.harvard.edu/abs/2023A&A...677A.171A}
}

@ARTICLE{Anez-Lopez2020,
       author = {{A{\~n}ez-L{\'o}pez}, N. and {Osorio}, M. and {Busquet}, G. and {Girart}, J.~M. and {Mac{\'\i}as}, E. and {Carrasco-Gonz{\'a}lez}, C. and {Curiel}, S. and {Estalella}, R. and {Fern{\'a}ndez-L{\'o}pez}, M. and {Galv{\'a}n-Madrid}, R. and {Kwon}, J. and {Torrelles}, J.~M.},
        title = "{Modeling the Accretion Disk around the High-mass Protostar GGD 27-MM1}",
      journal = {\apj},
         year = 2020,
        month = jan,
       volume = {888},
       number = {1},
          eid = {41},
        pages = {41},
          doi = {10.3847/1538-4357/ab5dbc},
archivePrefix = {arXiv},
       eprint = {1911.12398},
 primaryClass = {astro-ph.SR},
       adsurl = {https://ui.adsabs.harvard.edu/abs/2020ApJ...888...41A}
}

@ARTICLE{Avison2021,
       author = {{Avison}, A. and {Fuller}, G.~A. and {Peretto}, N. and {Duarte-Cabral}, A. and {Rosen}, A.~L. and {Traficante}, A. and {Pineda}, J.~E. and {G{\"u}sten}, R. and {Cunningham}, N.},
        title = "{Continuity of accretion from clumps to Class 0 high-mass protostars in SDC335}",
      journal = {\aap},
         year = 2021,
        month = jan,
       volume = {645},
          eid = {A142},
        pages = {A142},
          doi = {10.1051/0004-6361/201936043},
archivePrefix = {arXiv},
       eprint = {2012.08948},
 primaryClass = {astro-ph.GA},
       adsurl = {https://ui.adsabs.harvard.edu/abs/2021A&A...645A.142A}
}

@ARTICLE{Aso2015,
       author = {{Aso}, Yusuke and {Ohashi}, Nagayoshi and {Saigo}, Kazuya and {Koyamatsu}, Shin and {Aikawa}, Yuri and {Hayashi}, Masahiko and {Machida}, Masahiro N. and {Saito}, Masao and {Takakuwa}, Shigehisa and {Tomida}, Kengo and {Tomisaka}, Kohji and {Yen}, Hsi-Wei},
        title = "{ALMA Observations of the Transition from Infall Motion to Keplerian Rotation around the Late-phase Protostar TMC-1A}",
      journal = {\apj},
         year = 2015,
        month = oct,
       volume = {812},
       number = {1},
          eid = {27},
        pages = {27},
          doi = {10.1088/0004-637X/812/1/27},
archivePrefix = {arXiv},
       eprint = {1508.07013},
 primaryClass = {astro-ph.SR},
       adsurl = {https://ui.adsabs.harvard.edu/abs/2015ApJ...812...27A}
}

@ARTICLE{Bate2018,
       author = {{Bate}, Matthew R.},
        title = "{On the diversity and statistical properties of protostellar discs}",
      journal = {\mnras},
         year = 2018,
        month = apr,
       volume = {475},
       number = {4},
        pages = {5618-5658},
          doi = {10.1093/mnras/sty169},
archivePrefix = {arXiv},
       eprint = {1801.07721},
 primaryClass = {astro-ph.SR},
       adsurl = {https://ui.adsabs.harvard.edu/abs/2018MNRAS.475.5618B}
}

@ARTICLE{Beltran2014,
       author = {{Beltr{\'a}n}, M.~T. and {S{\'a}nchez-Monge}, {\'A}. and {Cesaroni}, R. and {Kumar}, M.~S.~N. and {Galli}, D. and {Walmsley}, C.~M. and {Etoka}, S. and {Furuya}, R.~S. and {Moscadelli}, L. and {Stanke}, T. and {van der Tak}, F.~F.~S. and {Vig}, S. and {Wang}, K. -S. and {Zinnecker}, H. and {Elia}, D. and {Schisano}, E.},
        title = "{Filamentary structure and Keplerian rotation in the high-mass star-forming region G35.03+0.35 imaged with ALMA}",
      journal = {\aap},
         year = 2014,
        month = nov,
       volume = {571},
          eid = {A52},
        pages = {A52},
          doi = {10.1051/0004-6361/201424031},
       adsurl = {https://ui.adsabs.harvard.edu/abs/2014A&A...571A..52B}
}

@ARTICLE{Beltran2016,
       author = {{Beltr{\'a}n}, M.~T. and {de Wit}, W.~J.},
        title = "{Accretion disks in luminous young stellar objects}",
      journal = {\aapr},
         year = 2016,
        month = jan,
       volume = {24},
          eid = {6},
        pages = {6},
          doi = {10.1007/s00159-015-0089-z},
archivePrefix = {arXiv},
       eprint = {1509.08335},
 primaryClass = {astro-ph.GA},
       adsurl = {https://ui.adsabs.harvard.edu/abs/2016A&ARv..24....6B}
}

@ARTICLE{Beuther2004,
       author = {{Beuther}, H. and {Hunter}, T.~R. and {Zhang}, Q. and {Sridharan}, T.~K. and {Zhao}, J. -H. and {Sollins}, P. and {Ho}, P.~T.~P. and {Ohashi}, N. and {Su}, Y.~N. and {Lim}, J. and {Liu}, S. -Y.},
        title = "{Submillimeter Array Outflow/Disk Studies in the Massive Star-forming Region IRAS 18089-1732}",
      journal = {\apjl},
         year = 2004,
        month = nov,
       volume = {616},
       number = {1},
        pages = {L23-L26},
          doi = {10.1086/383570},
archivePrefix = {arXiv},
       eprint = {astro-ph/0402505},
 primaryClass = {astro-ph},
       adsurl = {https://ui.adsabs.harvard.edu/abs/2004ApJ...616L..23B}
}

@ARTICLE{Beuther2018,
       author = {{Beuther}, H. and {Mottram}, J.~C. and {Ahmadi}, A. and {Bosco}, F. and {Linz}, H. and {Henning}, Th. and {Klaassen}, P. and {Winters}, J.~M. and {Maud}, L.~T. and {Kuiper}, R. and {Semenov}, D. and {Gieser}, C. and {Peters}, T. and {Urquhart}, J.~S. and {Pudritz}, R. and {Ragan}, S.~E. and {Feng}, S. and {Keto}, E. and {Leurini}, S. and {Cesaroni}, R. and {Beltran}, M. and {Palau}, A. and {S{\'a}nchez-Monge}, {\'A}. and {Galvan-Madrid}, R. and {Zhang}, Q. and {Schilke}, P. and {Wyrowski}, F. and {Johnston}, K.~G. and {Longmore}, S.~N. and {Lumsden}, S. and {Hoare}, M. and {Menten}, K.~M. and {Csengeri}, T.},
        title = "{Fragmentation and disk formation during high-mass star formation. IRAM NOEMA (Northern Extended Millimeter Array) large program CORE}",
      journal = {\aap},
         year = 2018,
        month = sep,
       volume = {617},
          eid = {A100},
        pages = {A100},
          doi = {10.1051/0004-6361/201833021},
archivePrefix = {arXiv},
       eprint = {1805.01191},
 primaryClass = {astro-ph.GA},
       adsurl = {https://ui.adsabs.harvard.edu/abs/2018A&A...617A.100B}
}

@ARTICLE{Beuther2017,
       author = {{Beuther}, H. and {Walsh}, A.~J. and {Johnston}, K.~G. and {Henning}, Th. and {Kuiper}, R. and {Longmore}, S.~N. and {Walmsley}, C.~M.},
        title = "{Fragmentation and disk formation in high-mass star formation: The ALMA view of G351.77-0.54 at 0.06'' resolution}",
      journal = {\aap},
         year = 2017,
        month = jun,
       volume = {603},
          eid = {A10},
        pages = {A10},
          doi = {10.1051/0004-6361/201630126},
archivePrefix = {arXiv},
       eprint = {1703.07235},
 primaryClass = {astro-ph.GA},
       adsurl = {https://ui.adsabs.harvard.edu/abs/2017A&A...603A..10B}
}

@ARTICLE{Beuther2019,
       author = {{Beuther}, H. and {Ahmadi}, A. and {Mottram}, J.~C. and {Linz}, H. and {Maud}, L.~T. and {Henning}, Th. and {Kuiper}, R. and {Walsh}, A.~J. and {Johnston}, K.~G. and {Longmore}, S.~N.},
        title = "{High-mass star formation at sub-50 au scales}",
      journal = {\aap},
         year = 2019,
        month = jan,
       volume = {621},
          eid = {A122},
        pages = {A122},
          doi = {10.1051/0004-6361/201834064},
archivePrefix = {arXiv},
       eprint = {1811.10245},
 primaryClass = {astro-ph.SR},
       adsurl = {https://ui.adsabs.harvard.edu/abs/2019A&A...621A.122B}
}

@ARTICLE{Beuther2025,
       author = {{Beuther}, H. and {Olguin}, F.~A. and {Sanhueza}, P. and {Cunningham}, N. and {Ginsburg}, A.},
        title = "{Hierarchical accretion flow from the G351 infrared dark filament to its central cores}",
      journal = {\aap},
         year = 2025,
        month = mar,
       volume = {695},
          eid = {A51},
        pages = {A51},
          doi = {10.1051/0004-6361/202452754},
archivePrefix = {arXiv},
       eprint = {2502.13866},
 primaryClass = {astro-ph.GA},
       adsurl = {https://ui.adsabs.harvard.edu/abs/2025A&A...695A..51B}
}

@ARTICLE{Beuther2025rev,
       author = {{Beuther}, H. and {Kuiper}, R. and {Tafalla}, M.},
        title = "{Star Formation from Low to High Mass: A Comparative View}",
      journal = {\araa},
         year = 2025,
        month = aug,
       volume = {63},
       number = {1},
        pages = {1-44},
          doi = {10.1146/annurev-astro-013125-122023},
archivePrefix = {arXiv},
       eprint = {2501.16866},
 primaryClass = {astro-ph.GA},
       adsurl = {https://ui.adsabs.harvard.edu/abs/2025ARA&A..63....1B}
}

@ARTICLE{Birnstiel2018,
       author = {{Birnstiel}, Tilman and {Dullemond}, Cornelis P. and {Zhu}, Zhaohuan and {Andrews}, Sean M. and {Bai}, Xue-Ning and {Wilner}, David J. and {Carpenter}, John M. and {Huang}, Jane and {Isella}, Andrea and {Benisty}, Myriam and {P{\'e}rez}, Laura M. and {Zhang}, Shangjia},
        title = "{The Disk Substructures at High Angular Resolution Project (DSHARP). V. Interpreting ALMA Maps of Protoplanetary Disks in Terms of a Dust Model}",
      journal = {\apjl},
         year = 2018,
        month = dec,
       volume = {869},
       number = {2},
          eid = {L45},
        pages = {L45},
          doi = {10.3847/2041-8213/aaf743},
archivePrefix = {arXiv},
       eprint = {1812.04043},
 primaryClass = {astro-ph.SR},
       adsurl = {https://ui.adsabs.harvard.edu/abs/2018ApJ...869L..45B}
}

@ARTICLE{Carrasco-Gonzalez2012,
       author = {{Carrasco-Gonz{\'a}lez}, Carlos and {Galv{\'a}n-Madrid}, Roberto and {Anglada}, Guillem and {Osorio}, Mayra and {D'Alessio}, Paola and {Hofner}, Peter and {Rodr{\'\i}guez}, Luis F. and {Linz}, Hendrik and {Araya}, Esteban D.},
        title = "{Resolving the Circumstellar Disk around the Massive Protostar Driving the HH 80-81 Jet}",
      journal = {\apjl},
         year = 2012,
        month = jun,
       volume = {752},
       number = {2},
          eid = {L29},
        pages = {L29},
          doi = {10.1088/2041-8205/752/2/L29},
archivePrefix = {arXiv},
       eprint = {1205.3302},
 primaryClass = {astro-ph.GA},
       adsurl = {https://ui.adsabs.harvard.edu/abs/2012ApJ...752L..29C}
}

@ARTICLE{Cesaroni2014,
       author = {{Cesaroni}, R. and {Galli}, D. and {Neri}, R. and {Walmsley}, C.~M.},
        title = "{Imaging the disk around IRAS 20126+4104 at subarcsecond resolution}",
      journal = {\aap},
         year = 2014,
        month = jun,
       volume = {566},
          eid = {A73},
        pages = {A73},
          doi = {10.1051/0004-6361/201323065},
       adsurl = {https://ui.adsabs.harvard.edu/abs/2014A&A...566A..73C}
}

@ARTICLE{Cesaroni2017,
       author = {{Cesaroni}, R. and {S{\'a}nchez-Monge}, {\'A}. and {Beltr{\'a}n}, M.~T. and {Johnston}, K.~G. and {Maud}, L.~T. and {Moscadelli}, L. and {Mottram}, J.~C. and {Ahmadi}, A. and {Allen}, V. and {Beuther}, H. and {Csengeri}, T. and {Etoka}, S. and {Fuller}, G.~A. and {Galli}, D. and {Galv{\'a}n-Madrid}, R. and {Goddi}, C. and {Henning}, T. and {Hoare}, M.~G. and {Klaassen}, P.~D. and {Kuiper}, R. and {Kumar}, M.~S.~N. and {Lumsden}, S. and {Peters}, T. and {Rivilla}, V.~M. and {Schilke}, P. and {Testi}, L. and {van der Tak}, F. and {Vig}, S. and {Walmsley}, C.~M. and {Zinnecker}, H.},
        title = "{Chasing discs around O-type (proto)stars: Evidence from ALMA observations}",
      journal = {\aap},
         year = 2017,
        month = jun,
       volume = {602},
          eid = {A59},
        pages = {A59},
          doi = {10.1051/0004-6361/201630184},
       adsurl = {https://ui.adsabs.harvard.edu/abs/2017A&A...602A..59C}
}

@ARTICLE{Chen2016,
       author = {{Chen}, Huei-Ru Vivien and {Keto}, Eric and {Zhang}, Qizhou and {Sridharan}, T.~K. and {Liu}, Sheng-Yuan and {Su}, Yu-Nung},
        title = "{A Hot and Massive Accretion Disk around the High-mass Protostar IRAS 20126+4104}",
      journal = {\apj},
         year = 2016,
        month = jun,
       volume = {823},
       number = {2},
          eid = {125},
        pages = {125},
          doi = {10.3847/0004-637X/823/2/125},
archivePrefix = {arXiv},
       eprint = {1604.00523},
 primaryClass = {astro-ph.GA},
       adsurl = {https://ui.adsabs.harvard.edu/abs/2016ApJ...823..125C}
}

@ARTICLE{Commercon2022,
       author = {{Commer{\c{c}}on}, B. and {Gonz{\'a}lez}, M. and {Mignon-Risse}, R. and {Hennebelle}, P. and {Vaytet}, N.},
        title = "{Discs and outflows in the early phases of massive star formation: Influence of magnetic fields and ambipolar diffusion}",
      journal = {\aap},
         year = 2022,
        month = feb,
       volume = {658},
          eid = {A52},
        pages = {A52},
          doi = {10.1051/0004-6361/202037479},
archivePrefix = {arXiv},
       eprint = {2109.10580},
 primaryClass = {astro-ph.SR},
       adsurl = {https://ui.adsabs.harvard.edu/abs/2022A&A...658A..52C}
}

@ARTICLE{Contreras18,
       author = {{Contreras}, Yanett and {Sanhueza}, Patricio and {Jackson}, James M. and {Guzm{\'a}n}, Andr{\'e}s E. and {Longmore}, Steven and {Garay}, Guido and {Zhang}, Qizhou and {Nguyễn-Lu'o'ng}, Quang and {Tatematsu}, Ken'ichi and {Nakamura}, Fumitaka and {Sakai}, Takeshi and {Ohashi}, Satoshi and {Liu}, Tie and {Saito}, Masao and {Gomez}, Laura and {Rathborne}, Jill and {Whitaker}, Scott},
        title = "{Infall Signatures in a Prestellar Core Embedded in the High-mass 70 {\ensuremath{\mu}}m Dark IRDC G331.372-00.116}",
      journal = {\apj},
         year = 2018,
        month = jul,
       volume = {861},
       number = {1},
          eid = {14},
        pages = {14},
          doi = {10.3847/1538-4357/aac2ec},
archivePrefix = {arXiv},
       eprint = {1805.01802},
 primaryClass = {astro-ph.GA},
       adsurl = {https://ui.adsabs.harvard.edu/abs/2018ApJ...861...14C}
}

@ARTICLE{Cortes2021,
       author = {{Cort{\'e}s}, Paulo C. and {Sanhueza}, Patricio and {Houde}, Martin and {Mart{\'\i}n}, Sergio and {Hull}, Charles L.~H. and {Girart}, Josep M. and {Zhang}, Qizhou and {Fernandez-Lopez}, Manuel and {Zapata}, Luis A. and {Stephens}, Ian W. and {Li}, Hua-bai and {Wu}, Benjamin and {Olguin}, Fernando and {Lu}, Xing and {Guzm{\'a}n}, Andres E. and {Nakamura}, Fumitaka},
        title = "{Magnetic Fields in Massive Star-forming Regions (MagMaR). II. Tomography through Dust and Molecular Line Polarization in NGC 6334I(N)}",
      journal = {\apj},
         year = 2021,
        month = dec,
       volume = {923},
       number = {2},
          eid = {204},
        pages = {204},
          doi = {10.3847/1538-4357/ac28a1},
archivePrefix = {arXiv},
       eprint = {2109.09270},
 primaryClass = {astro-ph.GA},
       adsurl = {https://ui.adsabs.harvard.edu/abs/2021ApJ...923..204C}
}

@ARTICLE{Cortes2024,
       author = {{Cort{\'e}s}, Paulo C. and {Girart}, Josep M. and {Sanhueza}, Patricio and {Liu}, Junhao and {Mart{\'\i}n}, Sergio and {Stephens}, Ian W. and {Beuther}, Henrik and {Koch}, Patrick M. and {Fern{\'a}ndez-L{\'o}pez}, M. and {S{\'a}nchez-Monge}, {\'A}lvaro and {Wang}, Jia-Wei and {Morii}, Kaho and {Li}, Shanghuo and {Saha}, Piyali and {Zhang}, Qizhou and {Rebolledo}, David and {Zapata}, Luis A. and {Kang}, Ji-hyun and {Jiao}, Wenyu and {Kim}, Jongsoo and {Cheng}, Yu and {Hwang}, Jihye and {Chung}, Eun Jung and {Choudhury}, Spandan and {Lyo}, A. -Ran and {Olguin}, Fernando},
        title = "{MagMaR III{\textemdash}Resisting the Pressure, Is the Magnetic Field Overwhelmed in NGC6334I?}",
      journal = {\apj},
         year = 2024,
        month = sep,
       volume = {972},
       number = {1},
          eid = {115},
        pages = {115},
          doi = {10.3847/1538-4357/ad59a7},
archivePrefix = {arXiv},
       eprint = {2406.14663},
 primaryClass = {astro-ph.GA},
       adsurl = {https://ui.adsabs.harvard.edu/abs/2024ApJ...972..115C}
}

@ARTICLE{Cunningham2011,
       author = {{Cunningham}, Andrew J. and {Klein}, Richard I. and {Krumholz}, Mark R. and {McKee}, Christopher F.},
        title = "{Radiation-hydrodynamic Simulations of Massive Star Formation with Protostellar Outflows}",
      journal = {\apj},
         year = 2011,
        month = oct,
       volume = {740},
       number = {2},
          eid = {107},
        pages = {107},
          doi = {10.1088/0004-637X/740/2/107},
archivePrefix = {arXiv},
       eprint = {1104.1218},
 primaryClass = {astro-ph.SR},
       adsurl = {https://ui.adsabs.harvard.edu/abs/2011ApJ...740..107C}
}

@ARTICLE{Cyganowski2011,
       author = {{Cyganowski}, C.~J. and {Brogan}, C.~L. and {Hunter}, T.~R. and {Churchwell}, E. and {Zhang}, Q.},
        title = "{Bipolar Molecular Outflows and Hot Cores in Glimpse Extended Green Objects (EGOs)}",
      journal = {\apj},
         year = 2011,
        month = mar,
       volume = {729},
       number = {2},
          eid = {124},
        pages = {124},
          doi = {10.1088/0004-637X/729/2/124},
archivePrefix = {arXiv},
       eprint = {1012.0851},
 primaryClass = {astro-ph.SR},
       adsurl = {https://ui.adsabs.harvard.edu/abs/2011ApJ...729..124C}
}

@ARTICLE{Cyganowski2017,
       author = {{Cyganowski}, C.~J. and {Brogan}, C.~L. and {Hunter}, T.~R. and {Smith}, R. and {Kruijssen}, J.~M.~D. and {Bonnell}, I.~A. and {Zhang}, Q.},
        title = "{Simultaneous low- and high-mass star formation in a massive protocluster: ALMA observations of G11.92-0.61$^{★}$}",
      journal = {\mnras},
         year = 2017,
        month = jul,
       volume = {468},
       number = {3},
        pages = {3694-3708},
          doi = {10.1093/mnras/stx043},
archivePrefix = {arXiv},
       eprint = {1701.02802},
 primaryClass = {astro-ph.GA},
       adsurl = {https://ui.adsabs.harvard.edu/abs/2017MNRAS.468.3694C}
}

@ARTICLE{Cyganowski2022,
       author = {{Cyganowski}, C.~J. and {Ilee}, J.~D. and {Brogan}, C.~L. and {Hunter}, T.~R. and {Zhang}, S. and {Harries}, T.~J. and {Haworth}, T.~J.},
        title = "{Discovery of a 500 au Protobinary in the Massive Prestellar Core G11.92-0.61 MM2}",
      journal = {\apjl},
         year = 2022,
        month = jun,
       volume = {931},
       number = {2},
          eid = {L31},
        pages = {L31},
          doi = {10.3847/2041-8213/ac69ca},
archivePrefix = {arXiv},
       eprint = {2204.09163},
 primaryClass = {astro-ph.GA},
       adsurl = {https://ui.adsabs.harvard.edu/abs/2022ApJ...931L..31C}
}

@software{Dominik2021,
       author = {{Dominik}, Carsten and {Min}, Michiel and {Tazaki}, Ryo},
        title = "{OpTool: Command-line driven tool for creating complex dust opacities}",
 howpublished = {Astrophysics Source Code Library, record ascl:2104.010},
         year = 2021,
        month = apr,
          eid = {ascl:2104.010},
       adsurl = {https://ui.adsabs.harvard.edu/abs/2021ascl.soft04010D}
}

@ARTICLE{Dullemond2019,
       author = {{Dullemond}, C.~P. and {K{\"u}ffmeier}, M. and {Goicovic}, F. and {Fukagawa}, M. and {Oehl}, V. and {Kramer}, M.},
        title = "{Cloudlet capture by transitional disk and FU Orionis stars}",
      journal = {\aap},
         year = 2019,
        month = aug,
       volume = {628},
          eid = {A20},
        pages = {A20},
          doi = {10.1051/0004-6361/201832632},
archivePrefix = {arXiv},
       eprint = {1911.05158},
 primaryClass = {astro-ph.EP},
       adsurl = {https://ui.adsabs.harvard.edu/abs/2019A&A...628A..20D}
}

@ARTICLE{Ekstrom2012,
       author = {{Ekstr{\"o}m}, S. and {Georgy}, C. and {Eggenberger}, P. and {Meynet}, G. and {Mowlavi}, N. and {Wyttenbach}, A. and {Granada}, A. and {Decressin}, T. and {Hirschi}, R. and {Frischknecht}, U. and {Charbonnel}, C. and {Maeder}, A.},
        title = "{Grids of stellar models with rotation. I. Models from 0.8 to 120 M$_{{\ensuremath{\odot}}}$ at solar metallicity (Z = 0.014)}",
      journal = {\aap},
         year = 2012,
        month = jan,
       volume = {537},
          eid = {A146},
        pages = {A146},
          doi = {10.1051/0004-6361/201117751},
archivePrefix = {arXiv},
       eprint = {1110.5049},
 primaryClass = {astro-ph.SR},
       adsurl = {https://ui.adsabs.harvard.edu/abs/2012A&A...537A.146E}
}

@ARTICLE{Fernandez-Lopez2021,
       author = {{Fern{\'a}ndez-L{\'o}pez}, M. and {Sanhueza}, P. and {Zapata}, L.~A. and {Stephens}, I. and {Hull}, C. and {Zhang}, Q. and {Girart}, J.~M. and {Koch}, P.~M. and {Cort{\'e}s}, P. and {Silva}, A. and {Tatematsu}, K. and {Nakamura}, F. and {Guzm{\'a}n}, A.~E. and {Nguyen Luong}, Q. and {Guzm{\'a}n Ccolque}, E. and {Tang}, Y. -W. and {Chen}, H. -R.~V.},
        title = "{Magnetic Fields in Massive Star-forming Regions (MagMaR). I. Linear Polarized Imaging of the Ultracompact H II Region G5.89-0.39}",
      journal = {\apj},
         year = 2021,
        month = may,
       volume = {913},
       number = {1},
          eid = {29},
        pages = {29},
          doi = {10.3847/1538-4357/abf2b6},
archivePrefix = {arXiv},
       eprint = {2104.03331},
 primaryClass = {astro-ph.GA},
       adsurl = {https://ui.adsabs.harvard.edu/abs/2021ApJ...913...29F}
}

@ARTICLE{Fernandez-Lopez2023,
       author = {{Fern{\'a}ndez-L{\'o}pez}, M. and {Girart}, J.~M. and {L{\'o}pez-V{\'a}zquez}, J.~A. and {Estalella}, R. and {Busquet}, G. and {Curiel}, S. and {A{\~n}ez-L{\'o}pez}, N.},
        title = "{Disk and Envelope Streamers of the GGD 27-MM1 Massive Protostar}",
      journal = {\apj},
         year = 2023,
        month = oct,
       volume = {956},
       number = {2},
          eid = {82},
        pages = {82},
          doi = {10.3847/1538-4357/ace786},
archivePrefix = {arXiv},
       eprint = {2307.06178},
 primaryClass = {astro-ph.GA},
       adsurl = {https://ui.adsabs.harvard.edu/abs/2023ApJ...956...82F}
}

@ARTICLE{Garrod2006,
       author = {{Garrod}, R.~T. and {Herbst}, E.},
        title = "{Formation of methyl formate and other organic species in the warm-up phase of hot molecular cores}",
      journal = {\aap},
         year = 2006,
        month = oct,
       volume = {457},
       number = {3},
        pages = {927-936},
          doi = {10.1051/0004-6361:20065560},
archivePrefix = {arXiv},
       eprint = {astro-ph/0607560},
 primaryClass = {astro-ph},
       adsurl = {https://ui.adsabs.harvard.edu/abs/2006A&A...457..927G}
}

@ARTICLE{Ginsburg2023,
       author = {{Ginsburg}, Adam and {McGuire}, Brett A. and {Sanhueza}, Patricio and {Olguin}, Fernando and {Maud}, Luke T. and {Tanaka}, Kei E.~I. and {Zhang}, Yichen and {Beuther}, Henrik and {Indriolo}, Nick},
        title = "{Salt-bearing Disk Candidates around High-mass Young Stellar Objects}",
      journal = {\apj},
         year = 2023,
        month = jan,
       volume = {942},
       number = {2},
          eid = {66},
        pages = {66},
          doi = {10.3847/1538-4357/ac9f4a},
archivePrefix = {arXiv},
       eprint = {2211.02502},
 primaryClass = {astro-ph.GA},
       adsurl = {https://ui.adsabs.harvard.edu/abs/2023ApJ...942...66G}
}

@ARTICLE{Goddi2020,
       author = {{Goddi}, C. and {Ginsburg}, A. and {Maud}, L.~T. and {Zhang}, Q. and {Zapata}, Luis A.},
        title = "{Multidirectional Mass Accretion and Collimated Outflows on Scales of 100-2000 au in Early Stages of High-mass Protostars}",
      journal = {\apj},
         year = 2020,
        month = dec,
       volume = {905},
       number = {1},
          eid = {25},
        pages = {25},
          doi = {10.3847/1538-4357/abc88e},
archivePrefix = {arXiv},
       eprint = {1805.05364},
 primaryClass = {astro-ph.GA},
       adsurl = {https://ui.adsabs.harvard.edu/abs/2020ApJ...905...25G}
}

@ARTICLE{Guzman2020,
       author = {{Guzm{\'a}n}, Andr{\'e}s E. and {Sanhueza}, Patricio and {Zapata}, Luis and {Garay}, Guido and {Rodr{\'\i}guez}, Luis Felipe},
        title = "{A Photoionized Accretion Disk around a Young High-mass Star}",
      journal = {\apj},
         year = 2020,
        month = nov,
       volume = {904},
       number = {1},
          eid = {77},
        pages = {77},
          doi = {10.3847/1538-4357/abbe09},
archivePrefix = {arXiv},
       eprint = {2010.00244},
 primaryClass = {astro-ph.SR},
       adsurl = {https://ui.adsabs.harvard.edu/abs/2020ApJ...904...77G}
}

@ARTICLE{Ilee2018,
       author = {{Ilee}, J.~D. and {Cyganowski}, C.~J. and {Brogan}, C.~L. and {Hunter}, T.~R. and {Forgan}, D.~H. and {Haworth}, T.~J. and {Clarke}, C.~J. and {Harries}, T.~J.},
        title = "{G11.92-0.61 MM 1: A Fragmented Keplerian Disk Surrounding a Proto-O Star}",
      journal = {\apjl},
         year = 2018,
        month = dec,
       volume = {869},
       number = {2},
          eid = {L24},
        pages = {L24},
          doi = {10.3847/2041-8213/aaeffc},
archivePrefix = {arXiv},
       eprint = {1811.05267},
 primaryClass = {astro-ph.SR},
       adsurl = {https://ui.adsabs.harvard.edu/abs/2018ApJ...869L..24I}
}

@ARTICLE{Ishihara2024,
       author = {{Ishihara}, Kousuke and {Sanhueza}, Patricio and {Nakamura}, Fumitaka and {Saito}, Masao and {Chen}, Huei-Ru Vivien and {Li}, Shanghuo and {Olguin}, Fernando and {Taniguchi}, Kotomi and {Morii}, Kaho and {Lu}, Xing and {Luo}, Qiu-yi and {Sakai}, Takeshi and {Zhang}, Qizhou},
        title = "{Digging into the Interior of Hot Cores with ALMA (DIHCA). IV. Fragmentation in High-mass Star-forming Clumps}",
      journal = {\apj},
         year = 2024,
        month = oct,
       volume = {974},
       number = {1},
          eid = {95},
        pages = {95},
          doi = {10.3847/1538-4357/ad630f},
archivePrefix = {arXiv},
       eprint = {2407.06845},
 primaryClass = {astro-ph.GA},
       adsurl = {https://ui.adsabs.harvard.edu/abs/2024ApJ...974...95I}
}

@ARTICLE{Izquierdo2018,
       author = {{Izquierdo}, Andr{\'e}s F. and {Galv{\'a}n-Madrid}, Roberto and {Maud}, Luke T. and {Hoare}, Melvin G. and {Johnston}, Katharine G. and {Keto}, Eric R. and {Zhang}, Qizhou and {de Wit}, Willem-Jan},
        title = "{Radiative transfer modelling of W33A MM1: 3D structure and dynamics of a complex massive star-forming region}",
      journal = {\mnras},
         year = 2018,
        month = aug,
       volume = {478},
       number = {2},
        pages = {2505-2525},
          doi = {10.1093/mnras/sty1096},
archivePrefix = {arXiv},
       eprint = {1804.09204},
 primaryClass = {astro-ph.GA},
       adsurl = {https://ui.adsabs.harvard.edu/abs/2018MNRAS.478.2505I}
}

@ARTICLE{Johnston2015,
       author = {{Johnston}, Katharine G. and {Robitaille}, Thomas P. and {Beuther}, Henrik and {Linz}, Hendrik and {Boley}, Paul and {Kuiper}, Rolf and {Keto}, Eric and {Hoare}, Melvin G. and {van Boekel}, Roy},
        title = "{A Keplerian-like Disk around the Forming O-type Star AFGL 4176}",
      journal = {\apjl},
         year = 2015,
        month = nov,
       volume = {813},
       number = {1},
          eid = {L19},
        pages = {L19},
          doi = {10.1088/2041-8205/813/1/L19},
archivePrefix = {arXiv},
       eprint = {1509.08469},
 primaryClass = {astro-ph.SR},
       adsurl = {https://ui.adsabs.harvard.edu/abs/2015ApJ...813L..19J}
}

@ARTICLE{Johnston2020,
       author = {{Johnston}, Katharine G. and {Hoare}, Melvin G. and {Beuther}, Henrik and {Kuiper}, Rolf and {Kee}, Nathaniel Dylan and {Linz}, Hendrik and {Boley}, Paul and {Maud}, Luke T. and {Ahmadi}, Aida and {Robitaille}, Thomas P.},
        title = "{Spiral arms and instability within the AFGL 4176 mm1 disc}",
      journal = {\aap},
         year = 2020,
        month = feb,
       volume = {634},
          eid = {L11},
        pages = {L11},
          doi = {10.1051/0004-6361/201937154},
archivePrefix = {arXiv},
       eprint = {1911.09692},
 primaryClass = {astro-ph.SR},
       adsurl = {https://ui.adsabs.harvard.edu/abs/2020A&A...634L..11J}
}

@ARTICLE{Krumholz2009,
       author = {{Krumholz}, Mark R. and {Klein}, Richard I. and {McKee}, Christopher F. and {Offner}, Stella S.~R. and {Cunningham}, Andrew J.},
        title = "{The Formation of Massive Star Systems by Accretion}",
      journal = {Science},
         year = 2009,
        month = feb,
       volume = {323},
       number = {5915},
        pages = {754},
          doi = {10.1126/science.1165857},
archivePrefix = {arXiv},
       eprint = {0901.3157},
 primaryClass = {astro-ph.SR},
       adsurl = {https://ui.adsabs.harvard.edu/abs/2009Sci...323..754K}
}

@ARTICLE{2007ApJ...656..959K,
       author = {{Krumholz}, Mark R. and {Klein}, Richard I. and {McKee}, Christopher F.},
        title = "{Radiation-Hydrodynamic Simulations of Collapse and Fragmentation in Massive Protostellar Cores}",
      journal = {\apj},
         year = 2007,
        month = feb,
       volume = {656},
       number = {2},
        pages = {959-979},
          doi = {10.1086/510664},
archivePrefix = {arXiv},
       eprint = {astro-ph/0609798},
 primaryClass = {astro-ph},
       adsurl = {https://ui.adsabs.harvard.edu/abs/2007ApJ...656..959K}
}

@ARTICLE{2007ApJ...665..478K,
       author = {{Krumholz}, Mark R. and {Klein}, Richard I. and {McKee}, Christopher F.},
        title = "{Molecular Line Emission from Massive Protostellar Disks: Predictions for ALMA and EVLA}",
      journal = {\apj},
         year = 2007,
        month = aug,
       volume = {665},
       number = {1},
        pages = {478-491},
          doi = {10.1086/519305},
archivePrefix = {arXiv},
       eprint = {0705.0536},
 primaryClass = {astro-ph},
       adsurl = {https://ui.adsabs.harvard.edu/abs/2007ApJ...665..478K}
}

@ARTICLE{Kuiper2018,
       author = {{Kuiper}, R. and {Hosokawa}, T.},
        title = "{First hydrodynamics simulations of radiation forces and photoionization feedback in massive star formation}",
      journal = {\aap},
         year = 2018,
        month = aug,
       volume = {616},
          eid = {A101},
        pages = {A101},
          doi = {10.1051/0004-6361/201832638},
archivePrefix = {arXiv},
       eprint = {1804.10211},
 primaryClass = {astro-ph.GA},
       adsurl = {https://ui.adsabs.harvard.edu/abs/2018A&A...616A.101K}
}

@ARTICLE{Law2021,
       author = {{Law}, Charles J. and {Zhang}, Qizhou and {{\"O}berg}, Karin I. and {Galv{\'a}n-Madrid}, Roberto and {Keto}, Eric and {Liu}, Hauyu Baobab and {Ho}, Paul T.~P.},
        title = "{Subarcsecond Imaging of the Complex Organic Chemistry in Massive Star-forming Region G10.6-0.4}",
      journal = {\apj},
         year = 2021,
        month = mar,
       volume = {909},
       number = {2},
          eid = {214},
        pages = {214},
          doi = {10.3847/1538-4357/abdeb8},
archivePrefix = {arXiv},
       eprint = {2101.07801},
 primaryClass = {astro-ph.GA},
       adsurl = {https://ui.adsabs.harvard.edu/abs/2021ApJ...909..214L}
}

@ARTICLE{Li2024,
       author = {{Li}, Shanghuo and {Sanhueza}, Patricio and {Beuther}, Henrik and {Chen}, Huei-Ru Vivien and {Kuiper}, Rolf and {Olguin}, Fernando A. and {Pudritz}, Ralph E. and {Stephens}, Ian W. and {Zhang}, Qizhou and {Nakamura}, Fumitaka and {Lu}, Xing and {Kuruwita}, Rajika L. and {Sakai}, Takeshi and {Henning}, Thomas and {Taniguchi}, Kotomi and {Li}, Fei},
        title = "{Observations of high-order multiplicity in a high-mass stellar protocluster}",
      journal = {Nature Astronomy},
         year = 2024,
        month = apr,
       volume = {8},
        pages = {472-481},
          doi = {10.1038/s41550-023-02181-9},
archivePrefix = {arXiv},
       eprint = {2401.06545},
 primaryClass = {astro-ph.GA},
       adsurl = {https://ui.adsabs.harvard.edu/abs/2024NatAs...8..472L}
}

@ARTICLE{Li2025,
       author = {{Li}, Shanghuo and {Beuther}, Henrik and {Oliva}, Andr{\'e} and {Elbakyan}, Vardan G. and {Offner}, Stella S.~R. and {Kuiper}, Rolf and {Qiu}, Keping and {Lu}, Xing and {Sanhueza}, Patricio and {Chen}, Huei-Ru Vivien and {Zhang}, Qizhou and {Olguin}, Fernando A. and {Lee}, Chang Won and {Pudritz}, Ralph E. and {Kong}, Shuo and {Kuruwita}, Rajika L. and {Luo}, Qiuyi and {Liu}, Junhao},
        title = "{Detection of a septuple stellar system in formation via disk fragmentation}",
      journal = {Nature Astronomy},
         year = 2025,
        month = dec,
       volume = {9},
        pages = {1833-1844},
          doi = {10.1038/s41550-025-02682-9},
archivePrefix = {arXiv},
       eprint = {2509.06787},
 primaryClass = {astro-ph.GA},
       adsurl = {https://ui.adsabs.harvard.edu/abs/2025NatAs...9.1833L}
}

@ARTICLE{Lu2022,
       author = {{Lu}, Xing and {Li}, Guang-Xing and {Zhang}, Qizhou and {Lin}, Yuxin},
        title = "{A massive Keplerian protostellar disk with flyby-induced spirals in the Central Molecular Zone}",
      journal = {Nature Astronomy},
         year = 2022,
        month = may,
       volume = {6},
        pages = {837-843},
          doi = {10.1038/s41550-022-01681-4},
archivePrefix = {arXiv},
       eprint = {2206.00202},
 primaryClass = {astro-ph.GA},
       adsurl = {https://ui.adsabs.harvard.edu/abs/2022NatAs...6..837L}
}

@misc{Luo2026,
      title={Digging into the Interior of Hot Cores with ALMA. VI. The Formation of Low-mass Multiple Systems in High-mass Cluster-forming Regions}, 
      author={Qiuyi Luo and Patricio Sanhueza and Stella S. R. Offner and Fernando Olguin and Adam Ginsburg and Fumitaka Nakamura and Kaho Morii and Yu Cheng and Kei Tanaka and Junhao Liu and Tie Liu and Xing Lu and Qizhou Zhang and Kotomi Taniguchi and Piyali Saha and Shanghuo Li and Xiaofeng Mai},
      year={2026},
      eprint={2601.08904},
      archivePrefix={arXiv},
      primaryClass={astro-ph.GA},
      url={https://arxiv.org/abs/2601.08904}, 
}

@ARTICLE{Mai2025,
       author = {{Mai}, Xiaofeng and {Liu}, Tie and {Liu}, Xunchuan and {Zhang}, Bo and {Goldsmith}, Paul F. and {Evans}, II, Neal J. and {Zhang}, Qizhou and {Kim}, Kee-Tae and {Yang}, Dongting and {Juvela}, Mika and {Xu}, Fengwei and {Jiao}, Wenyu and {Liu}, Hongli and {Sanhueza}, Patricio and {Garay}, Guido and {Chen}, Xi and {Qin}, Shengli and {Vorster}, Jakobus M. and {Tej}, Anandmayee and {Ren}, Zhiyuan and {Dib}, Sami and {Li}, Shanghuo and {Luo}, Qiuyi and {Hwang}, Jihye and {Gorai}, Prasanta and {Hoque}, Ariful and {Zhang}, Yichen and {Lee}, Jeong-Eun and {Zhang}, Siju and {Mannfors}, Emma and {Tharakkal}, Devika and {Dewangan}, Lokesh and {Bronfman}, Leonardo and {Garc{\'\i}a}, Pablo and {Tang}, Xindi and {Das}, Swagat R. and {Wu}, Gang and {Lee}, Chang-Won and {Chibueze}, James O. and {Zhang}, Yankun and {Gu}, Qilao and {Tatematsu}, Ken'ichi and {Wang}, Guangli and {Zhu}, Lei and {Shen}, Zhiqiang},
        title = "{A misaligned protostellar disk fed by gas streamers in a barred spiral-like massive dense core}",
      journal = {Science Advances},
         year = 2025,
        month = sep,
       volume = {11},
       number = {38},
          eid = {eady6953},
        pages = {eady6953},
          doi = {10.1126/sciadv.ady6953},
       adsurl = {https://ui.adsabs.harvard.edu/abs/2025SciA...11.6953M}
}

@ARTICLE{Maud2017,
       author = {{Maud}, L.~T. and {Hoare}, M.~G. and {Galv{\'a}n-Madrid}, R. and {Zhang}, Q. and {de Wit}, W.~J. and {Keto}, E. and {Johnston}, K.~G. and {Pineda}, J.~E.},
        title = "{The ALMA view of W33A: a spiral filament feeding the candidate disc in MM1-Main}",
      journal = {\mnras},
         year = 2017,
        month = may,
       volume = {467},
       number = {1},
        pages = {L120-L124},
          doi = {10.1093/mnrasl/slx010},
archivePrefix = {arXiv},
       eprint = {1701.06958},
 primaryClass = {astro-ph.SR},
       adsurl = {https://ui.adsabs.harvard.edu/abs/2017MNRAS.467L.120M}
}

@ARTICLE{Maud2019,
       author = {{Maud}, L.~T. and {Cesaroni}, R. and {Kumar}, M.~S.~N. and {Rivilla}, V.~M. and {Ginsburg}, A. and {Klaassen}, P.~D. and {Harsono}, D. and {S{\'a}nchez-Monge}, {\'A}. and {Ahmadi}, A. and {Allen}, V. and {Beltr{\'a}n}, M.~T. and {Beuther}, H. and {Galv{\'a}n-Madrid}, R. and {Goddi}, C. and {Hoare}, M.~G. and {Hogerheijde}, M.~R. and {Johnston}, K.~G. and {Kuiper}, R. and {Moscadelli}, L. and {Peters}, T. and {Testi}, L. and {van der Tak}, F.~F.~S. and {de Wit}, W.~J.},
        title = "{Substructures in the Keplerian disc around the O-type (proto-)star G17.64+0.16}",
      journal = {\aap},
         year = 2019,
        month = jul,
       volume = {627},
          eid = {L6},
        pages = {L6},
          doi = {10.1051/0004-6361/201935633},
archivePrefix = {arXiv},
       eprint = {1906.06548},
 primaryClass = {astro-ph.SR},
       adsurl = {https://ui.adsabs.harvard.edu/abs/2019A&A...627L...6M}
}

@ARTICLE{Meyer2017,
       author = {{Meyer}, D.~M. -A. and {Vorobyov}, E.~I. and {Kuiper}, R. and {Kley}, W.},
        title = "{On the existence of accretion-driven bursts in massive star formation}",
      journal = {\mnras},
         year = 2017,
        month = jan,
       volume = {464},
       number = {1},
        pages = {L90-L94},
          doi = {10.1093/mnrasl/slw187},
archivePrefix = {arXiv},
       eprint = {1609.03402},
 primaryClass = {astro-ph.SR},
       adsurl = {https://ui.adsabs.harvard.edu/abs/2017MNRAS.464L..90M}
}

@ARTICLE{Momose1998,
       author = {{Momose}, Munetake and {Ohashi}, Nagayoshi and {Kawabe}, Ryohei and {Nakano}, Takenori and {Hayashi}, Masahiko},
        title = "{Aperture Synthesis C$^{18}$O J = 1-0 Observations of L1551 IRS 5: Detailed Structure of the Infalling Envelope}",
      journal = {\apj},
         year = 1998,
        month = sep,
       volume = {504},
       number = {1},
        pages = {314-333},
          doi = {10.1086/306061},
       adsurl = {https://ui.adsabs.harvard.edu/abs/1998ApJ...504..314M}
}

@ARTICLE{Mori2024,
       author = {{Mori}, Shoji and {Aikawa}, Yuri and {Oya}, Yoko and {Yamamoto}, Satoshi and {Sakai}, Nami},
        title = "{Synthetic Observations of the Infalling Rotating Envelope: Links between the Physical Structure and Observational Features}",
      journal = {\apj},
         year = 2024,
        month = jan,
       volume = {961},
       number = {1},
          eid = {31},
        pages = {31},
          doi = {10.3847/1538-4357/ad0634},
archivePrefix = {arXiv},
       eprint = {2401.06213},
 primaryClass = {astro-ph.SR},
       adsurl = {https://ui.adsabs.harvard.edu/abs/2024ApJ...961...31M}
}

@ARTICLE{Okoda2021,
       author = {{Okoda}, Yuki and {Oya}, Yoko and {Abe}, Shotaro and {Komaki}, Ayano and {Watanabe}, Yoshimasa and {Yamamoto}, Satoshi},
        title = "{Molecular Distributions of the Disk/Envelope System of L483: Principal Component Analysis for the Image Cube Data}",
      journal = {\apj},
         year = 2021,
        month = dec,
       volume = {923},
       number = {2},
          eid = {168},
        pages = {168},
          doi = {10.3847/1538-4357/ac2c6c},
archivePrefix = {arXiv},
       eprint = {2110.00150},
 primaryClass = {astro-ph.GA},
       adsurl = {https://ui.adsabs.harvard.edu/abs/2021ApJ...923..168O}
}

@ARTICLE{Olguin2021,
       author = {{Olguin}, Fernando A. and {Sanhueza}, Patricio and {Guzm{\'a}n}, Andr{\'e}s E. and {Lu}, Xing and {Saigo}, Kazuya and {Zhang}, Qizhou and {Silva}, Andrea and {Chen}, Huei-Ru Vivien and {Li}, Shanghuo and {Ohashi}, Satoshi and {Nakamura}, Fumitaka and {Sakai}, Takeshi and {Wu}, Benjamin},
        title = "{Digging into the Interior of Hot Cores with ALMA (DIHCA). I. Dissecting the High-mass Star-forming Core G335.579-0.292 MM1}",
      journal = {\apj},
         year = 2021,
        month = mar,
       volume = {909},
       number = {2},
          eid = {199},
        pages = {199},
          doi = {10.3847/1538-4357/abde3f},
archivePrefix = {arXiv},
       eprint = {2101.08284},
 primaryClass = {astro-ph.GA},
       adsurl = {https://ui.adsabs.harvard.edu/abs/2021ApJ...909..199O}
}

@ARTICLE{Olguin2022,
       author = {{Olguin}, Fernando A. and {Sanhueza}, Patricio and {Ginsburg}, Adam and {Chen}, Huei-Ru Vivien and {Zhang}, Qizhou and {Li}, Shanghuo and {Lu}, Xing and {Sakai}, Takeshi},
        title = "{Digging into the Interior of Hot Cores with ALMA (DIHCA). II. Exploring the Inner Binary (Multiple) System Embedded in G335 MM1 ALMA1}",
      journal = {\apj},
         year = 2022,
        month = apr,
       volume = {929},
       number = {1},
          eid = {68},
        pages = {68},
          doi = {10.3847/1538-4357/ac5bd8},
archivePrefix = {arXiv},
       eprint = {2203.04333},
 primaryClass = {astro-ph.GA},
       adsurl = {https://ui.adsabs.harvard.edu/abs/2022ApJ...929...68O}
}

@ARTICLE{Olguin2023,
       author = {{Olguin}, Fernando A. and {Sanhueza}, Patricio and {Chen}, Huei-Ru Vivien and {Lu}, Xing and {Oya}, Yoko and {Zhang}, Qizhou and {Ginsburg}, Adam and {Taniguchi}, Kotomi and {Li}, Shanghuo and {Morii}, Kaho and {Sakai}, Takeshi and {Nakamura}, Fumitaka},
        title = "{Digging into the Interior of Hot Cores with ALMA: Spiral Accretion into the High-mass Protostellar Core G336.01-0.82}",
      journal = {\apjl},
         year = 2023,
        month = dec,
       volume = {959},
       number = {2},
          eid = {L31},
        pages = {L31},
          doi = {10.3847/2041-8213/ad1100},
archivePrefix = {arXiv},
       eprint = {2311.18006},
 primaryClass = {astro-ph.GA},
       adsurl = {https://ui.adsabs.harvard.edu/abs/2023ApJ...959L..31O}
}

@ARTICLE{Olguin2025,
       author = {{Olguin}, Fernando A. and {Sanhueza}, Patricio and {Ginsburg}, Adam and {Chen}, Huei-Ru Vivien and {Tanaka}, Kei E.~I. and {Lu}, Xing and {Morii}, Kaho and {Nakamura}, Fumitaka and {Li}, Shanghuo and {Cheng}, Yu and {Zhang}, Qizhou and {Luo}, Qiuyi and {Oya}, Yoko and {Sakai}, Takeshi and {Saito}, Masao and {Guzm{\'a}n}, Andr{\'e}s E.},
        title = "{Massive extended streamers feed high-mass young stars}",
      journal = {Science Advances},
         year = 2025,
        month = aug,
       volume = {11},
       number = {34},
          eid = {eadw4512},
        pages = {eadw4512},
          doi = {10.1126/sciadv.adw4512},
       adsurl = {https://ui.adsabs.harvard.edu/abs/2025SciA...11.4512O}
}

@ARTICLE{Ossenkopf1994,
       author = {{Ossenkopf}, V. and {Henning}, Th.},
        title = "{Dust opacities for protostellar cores.}",
      journal = {\aap},
         year = 1994,
        month = nov,
       volume = {291},
        pages = {943-959},
       adsurl = {https://ui.adsabs.harvard.edu/abs/1994A&A...291..943O}
}

@ARTICLE{Oya2014,
       author = {{Oya}, Yoko and {Sakai}, Nami and {Sakai}, Takeshi and {Watanabe}, Yoshimasa and {Hirota}, Tomoya and {Lindberg}, Johan E. and {Bisschop}, Suzanne E. and {J{\o}rgensen}, Jes K. and {van Dishoeck}, Ewine F. and {Yamamoto}, Satoshi},
        title = "{A Substellar-mass Protostar and its Outflow of IRAS 15398-3359 Revealed by Subarcsecond-resolution Observations of H$_{2}$CO and CCH}",
      journal = {\apj},
         year = 2014,
        month = nov,
       volume = {795},
       number = {2},
          eid = {152},
        pages = {152},
          doi = {10.1088/0004-637X/795/2/152},
archivePrefix = {arXiv},
       eprint = {1410.5945},
 primaryClass = {astro-ph.SR},
       adsurl = {https://ui.adsabs.harvard.edu/abs/2014ApJ...795..152O}
}

@ARTICLE{Oya2020,
       author = {{Oya}, Yoko and {Yamamoto}, Satoshi},
        title = "{Substructures in the Disk-forming Region of the Class 0 Low-mass Protostellar Source IRAS 16293-2422 Source A on a 10 au Scale}",
      journal = {\apj},
         year = 2020,
        month = dec,
       volume = {904},
       number = {2},
          eid = {185},
        pages = {185},
          doi = {10.3847/1538-4357/abbe14},
archivePrefix = {arXiv},
       eprint = {2010.01273},
 primaryClass = {astro-ph.SR},
       adsurl = {https://ui.adsabs.harvard.edu/abs/2020ApJ...904..185O}
}

@ARTICLE{Oya2022,
       author = {{Oya}, Yoko and {Kibukawa}, Hirofumi and {Miyake}, Shota and {Yamamoto}, Satoshi},
        title = "{FERIA: Flat Envelope Model with Rotation and Infall under Angular Momentum Conservation}",
      journal = {\pasp},
         year = 2022,
        month = sep,
       volume = {134},
       number = {1039},
          eid = {094301},
        pages = {094301},
          doi = {10.1088/1538-3873/ac8839},
archivePrefix = {arXiv},
       eprint = {2208.04581},
 primaryClass = {astro-ph.IM},
       adsurl = {https://ui.adsabs.harvard.edu/abs/2022PASP..134i4301O}
}

@ARTICLE{Pan2025,
       author = {{Pan}, Xing and {Qiu}, Keping and {Zhang}, Qizhou},
        title = "{Surveys of clumps, cores, and condensations in Cygnus-X: Searching for Keplerian disks on the scale of 500 au}",
      journal = {\aap},
         year = 2025,
        month = apr,
       volume = {696},
          eid = {A195},
        pages = {A195},
          doi = {10.1051/0004-6361/202452944},
archivePrefix = {arXiv},
       eprint = {2503.16712},
 primaryClass = {astro-ph.GA},
       adsurl = {https://ui.adsabs.harvard.edu/abs/2025A&A...696A.195P}
}

@ARTICLE{Pouteau2022,
       author = {{Pouteau}, Y. and {Motte}, F. and {Nony}, T. and {Galv{\'a}n-Madrid}, R. and {Men'shchikov}, A. and {Bontemps}, S. and {Robitaille}, J. -F. and {Louvet}, F. and {Ginsburg}, A. and {Herpin}, F. and {L{\'o}pez-Sepulcre}, A. and {Dell'Ova}, P. and {Gusdorf}, A. and {Sanhueza}, P. and {Stutz}, A.~M. and {Brouillet}, N. and {Thomasson}, B. and {Armante}, M. and {Baug}, T. and {Bonfand}, M. and {Busquet}, G. and {Csengeri}, T. and {Cunningham}, N. and {Fern{\'a}ndez-L{\'o}pez}, M. and {Liu}, H. -L. and {Olguin}, F. and {Towner}, A.~P.~M. and {Bally}, J. and {Braine}, J. and {Bronfman}, L. and {Joncour}, I. and {Gonz{\'a}lez}, M. and {Hennebelle}, P. and {Lu}, X. and {Menten}, K.~M. and {Moraux}, E. and {Tatematsu}, K. and {Walker}, D. and {Whitworth}, A.~P.},
        title = "{ALMA-IMF. III. Investigating the origin of stellar masses: top-heavy core mass function in the W43-MM2\&MM3 mini-starburst}",
      journal = {\aap},
         year = 2022,
        month = aug,
       volume = {664},
          eid = {A26},
        pages = {A26},
          doi = {10.1051/0004-6361/202142951},
archivePrefix = {arXiv},
       eprint = {2203.03276},
 primaryClass = {astro-ph.GA},
       adsurl = {https://ui.adsabs.harvard.edu/abs/2022A&A...664A..26P}
}

@ARTICLE{Rosen2022,
       author = {{Rosen}, Anna L.},
        title = "{A Massive Star Is Born: How Feedback from Stellar Winds, Radiation Pressure, and Collimated Outflows Limits Accretion onto Massive Stars}",
      journal = {\apj},
         year = 2022,
        month = dec,
       volume = {941},
       number = {2},
          eid = {202},
        pages = {202},
          doi = {10.3847/1538-4357/ac9f3d},
archivePrefix = {arXiv},
       eprint = {2204.09700},
 primaryClass = {astro-ph.SR},
       adsurl = {https://ui.adsabs.harvard.edu/abs/2022ApJ...941..202R}
}

@ARTICLE{Sai2020,
       author = {{Sai}, Jinshi and {Ohashi}, Nagayoshi and {Saigo}, Kazuya and {Matsumoto}, Tomoaki and {Aso}, Yusuke and {Takakuwa}, Shigehisa and {Aikawa}, Yuri and {Kurose}, Ippei and {Yen}, Hsi-Wei and {Tomisaka}, Kohji and {Tomida}, Kengo and {Machida}, Masahiro N.},
        title = "{Disk Structure around the Class I Protostar L1489 IRS Revealed by ALMA: A Warped-disk System}",
      journal = {\apj},
         year = 2020,
        month = apr,
       volume = {893},
       number = {1},
          eid = {51},
        pages = {51},
          doi = {10.3847/1538-4357/ab8065},
archivePrefix = {arXiv},
       eprint = {2003.07067},
 primaryClass = {astro-ph.SR},
       adsurl = {https://ui.adsabs.harvard.edu/abs/2020ApJ...893...51S}
}

@ARTICLE{Saha2024,
       author = {{Saha}, Piyali and {Sanhueza}, Patricio and {Padovani}, Marco and {Girart}, Josep M. and {Cort{\'e}s}, Paulo C. and {Morii}, Kaho and {Liu}, Junhao and {S{\'a}nchez-Monge}, {\'A}. and {Galli}, Daniele and {Basu}, Shantanu and {Koch}, Patrick M. and {Beltr{\'a}n}, Maria T. and {Li}, Shanghuo and {Beuther}, Henrik and {Stephens}, Ian W. and {Nakamura}, Fumitaka and {Zhang}, Qizhou and {Jiao}, Wenyu and {Fern{\'a}ndez-L{\'o}pez}, M. and {Hwang}, Jihye and {Chung}, Eun Jung and {Pattle}, Kate and {Zapata}, Luis A. and {Xu}, Fengwei and {Olguin}, Fernando A. and {Kang}, Ji-hyun and {Karoly}, Janik and {Law}, Chi-Yan and {Wang}, Jia-Wei and {Csengeri}, Timea and {Lu}, Xing and {Cheng}, Yu and {Kim}, Jongsoo and {Choudhury}, Spandan and {Chen}, Huei-Ru Vivien and {Hull}, Charles L.~H.},
        title = "{Magnetic Fields in Massive Star-forming Regions (MagMaR): Unveiling an Hourglass Magnetic Field in G333.46{\textendash}0.16 Using ALMA}",
      journal = {\apjl},
         year = 2024,
        month = sep,
       volume = {972},
       number = {1},
          eid = {L6},
        pages = {L6},
          doi = {10.3847/2041-8213/ad660c},
archivePrefix = {arXiv},
       eprint = {2407.16654},
 primaryClass = {astro-ph.GA},
       adsurl = {https://ui.adsabs.harvard.edu/abs/2024ApJ...972L...6S}
}

@ARTICLE{Sakai2025,
       author = {{Sakai}, Takeshi and {Shiomura}, Nobuhito and {Sanhueza}, Patricio and {Furuya}, Kenji and {Olguin}, Fernando A. and {Tatematsu}, Ken'ichi and {Aikawa}, Yuri and {Taniguchi}, Kotomi and {Chen}, Huei-Ru Vivien and {Morii}, Kaho and {Nakamura}, Fumitaka and {Li}, Shanghuo and {Lu}, Xing and {Zhang}, Qizhou and {Hirota}, Tomoya and {Ishihara}, Kousuke and {Ke}, Hongda and {Sakai}, Nami and {Yamamoto}, Satoshi},
        title = "{Digging Into the Interior of Hot Cores with ALMA (DIHCA). V. Deuterium Fractionation of Methanol}",
      journal = {\apj},
         year = 2025,
        month = apr,
       volume = {983},
       number = {1},
          eid = {37},
        pages = {37},
          doi = {10.3847/1538-4357/adba5a},
archivePrefix = {arXiv},
       eprint = {2503.05094},
 primaryClass = {astro-ph.GA},
       adsurl = {https://ui.adsabs.harvard.edu/abs/2025ApJ...983...37S}
}

@ARTICLE{Sana2012,
       author = {{Sana}, H. and {de Mink}, S.~E. and {de Koter}, A. and {Langer}, N. and {Evans}, C.~J. and {Gieles}, M. and {Gosset}, E. and {Izzard}, R.~G. and {Le Bouquin}, J. -B. and {Schneider}, F.~R.~N.},
        title = "{Binary Interaction Dominates the Evolution of Massive Stars}",
      journal = {Science},
         year = 2012,
        month = jul,
       volume = {337},
       number = {6093},
        pages = {444},
          doi = {10.1126/science.1223344},
archivePrefix = {arXiv},
       eprint = {1207.6397},
 primaryClass = {astro-ph.SR},
       adsurl = {https://ui.adsabs.harvard.edu/abs/2012Sci...337..444S}
}

@ARTICLE{Sanchez-Monge2013,
       author = {{S{\'a}nchez-Monge}, {\'A}. and {Cesaroni}, R. and {Beltr{\'a}n}, M.~T. and {Kumar}, M.~S.~N. and {Stanke}, T. and {Zinnecker}, H. and {Etoka}, S. and {Galli}, D. and {Hummel}, C.~A. and {Moscadelli}, L. and {Preibisch}, T. and {Ratzka}, T. and {van der Tak}, F.~F.~S. and {Vig}, S. and {Walmsley}, C.~M. and {Wang}, K. -S.},
        title = "{A candidate circumbinary Keplerian disk in G35.20-0.74 N: A study with ALMA}",
      journal = {\aap},
         year = 2013,
        month = apr,
       volume = {552},
          eid = {L10},
        pages = {L10},
          doi = {10.1051/0004-6361/201321134},
archivePrefix = {arXiv},
       eprint = {1303.4242},
 primaryClass = {astro-ph.GA},
       adsurl = {https://ui.adsabs.harvard.edu/abs/2013A&A...552L..10S}
}

@ARTICLE{Sanchez-Monge2014,
       author = {{S{\'a}nchez-Monge}, {\'A}. and {Beltr{\'a}n}, M.~T. and {Cesaroni}, R. and {Etoka}, S. and {Galli}, D. and {Kumar}, M.~S.~N. and {Moscadelli}, L. and {Stanke}, T. and {van der Tak}, F.~F.~S. and {Vig}, S. and {Walmsley}, C.~M. and {Wang}, K. -S. and {Zinnecker}, H. and {Elia}, D. and {Molinari}, S. and {Schisano}, E.},
        title = "{A necklace of dense cores in the high-mass star forming region G35.20-0.74 N: ALMA observations}",
      journal = {\aap},
         year = 2014,
        month = sep,
       volume = {569},
          eid = {A11},
        pages = {A11},
          doi = {10.1051/0004-6361/201424032},
archivePrefix = {arXiv},
       eprint = {1406.4081},
 primaryClass = {astro-ph.GA},
       adsurl = {https://ui.adsabs.harvard.edu/abs/2014A&A...569A..11S}
}

@ARTICLE{Sanchez-Monge2018,
       author = {{S{\'a}nchez-Monge}, {\'A}. and {Schilke}, P. and {Ginsburg}, A. and {Cesaroni}, R. and {Schmiedeke}, A.},
        title = "{STATCONT: A statistical continuum level determination method for line-rich sources}",
      journal = {\aap},
         year = 2018,
        month = jan,
       volume = {609},
          eid = {A101},
        pages = {A101},
          doi = {10.1051/0004-6361/201730425},
archivePrefix = {arXiv},
       eprint = {1710.02419},
 primaryClass = {astro-ph.IM},
       adsurl = {https://ui.adsabs.harvard.edu/abs/2018A&A...609A.101S}
}

@INPROCEEDINGS{Sanchez-Monge2019,
       author = {{Sanchez-Monge}, Alvaro},
        title = "{Caught in the act: disruption of a high-mass disk by anisotropic accretion}",
    booktitle = {ALMA2019: Science Results and Cross-Facility Synergies},
         year = 2019,
        month = dec,
          eid = {122},
        pages = {122},
          doi = {10.5281/zenodo.3585445},
       adsurl = {https://ui.adsabs.harvard.edu/abs/2019asrc.confE.122S}
}

@ARTICLE{Sanhueza2021,
       author = {{Sanhueza}, Patricio and {Girart}, Josep Miquel and {Padovani}, Marco and {Galli}, Daniele and {Hull}, Charles L.~H. and {Zhang}, Qizhou and {Cortes}, Paulo and {Stephens}, Ian W. and {Fern{\'a}ndez-L{\'o}pez}, Manuel and {Jackson}, James M. and {Frau}, Pau and {Kock}, Patrick M. and {Wu}, Benjamin and {Zapata}, Luis A. and {Olguin}, Fernando and {Lu}, Xing and {Silva}, Andrea and {Tang}, Ya-Wen and {Sakai}, Takeshi and {Guzm{\'a}n}, Andr{\'e}s E. and {Tatematsu}, Ken'ichi and {Nakamura}, Fumitaka and {Chen}, Huei-Ru Vivien},
        title = "{Gravity-driven Magnetic Field at  1000 au Scales in High-mass Star Formation}",
      journal = {\apjl},
         year = 2021,
        month = jul,
       volume = {915},
       number = {1},
          eid = {L10},
        pages = {L10},
          doi = {10.3847/2041-8213/ac081c},
archivePrefix = {arXiv},
       eprint = {2106.03866},
 primaryClass = {astro-ph.GA},
       adsurl = {https://ui.adsabs.harvard.edu/abs/2021ApJ...915L..10S}
}

@ARTICLE{Sanhueza2025,
       author = {{Sanhueza}, Patricio and {Liu}, Junhao and {Morii}, Kaho and {Girart}, Josep Miquel and {Zhang}, Qizhou and {Stephens}, Ian W. and {Jackson}, James M. and {Cort{\'e}s}, Paulo C. and {Koch}, Patrick M. and {Cyganowski}, Claudia J. and {Saha}, Piyali and {Beuther}, Henrik and {Zhang}, Suinan and {Beltr{\'a}n}, Maria T. and {Cheng}, Yu and {Olguin}, Fernando A. and {Lu}, Xing and {Choudhury}, Spandan and {Pattle}, Kate and {Fern{\'a}ndez-L{\'o}pez}, Manuel and {Hwang}, Jihye and {Kang}, Ji-hyun and {Karoly}, Janik and {Ginsburg}, Adam and {Lyo}, A. -Ran and {Taniguchi}, Kotomi and {Jiao}, Wenyu and {Eswaraiah}, Chakali and {Luo}, Qiu-yi and {Wang}, Jia-Wei and {Commer{\c{c}}on}, Beno{\^\i}t and {Li}, Shanghuo and {Xu}, Fengwei and {Chen}, Huei-Ru Vivien and {Zapata}, Luis A. and {Chung}, Eun Jung and {Nakamura}, Fumitaka and {Panigrahy}, Sandhyarani and {Sakai}, Takeshi},
        title = "{Magnetic Fields in Massive Star-forming Regions (MagMaR). V. The Magnetic Field at the Onset of High-mass Star Formation}",
      journal = {\apj},
         year = 2025,
        month = feb,
       volume = {980},
       number = {1},
          eid = {87},
        pages = {87},
          doi = {10.3847/1538-4357/ad9d40},
archivePrefix = {arXiv},
       eprint = {2412.08790},
 primaryClass = {astro-ph.GA},
       adsurl = {https://ui.adsabs.harvard.edu/abs/2025ApJ...980...87S}
}

@ARTICLE{Seifried2015,
       author = {{Seifried}, D. and {Banerjee}, R. and {Pudritz}, R.~E. and {Klessen}, R.~S.},
        title = "{Accretion and magnetic field morphology around Class 0 stage protostellar discs}",
      journal = {\mnras},
         year = 2015,
        month = jan,
       volume = {446},
       number = {3},
        pages = {2776-2788},
          doi = {10.1093/mnras/stu2282},
archivePrefix = {arXiv},
       eprint = {1408.2989},
 primaryClass = {astro-ph.GA},
       adsurl = {https://ui.adsabs.harvard.edu/abs/2015MNRAS.446.2776S}
}

@ARTICLE{Seifried2016,
       author = {{Seifried}, D. and {S{\'a}nchez-Monge}, {\'A}. and {Walch}, S. and {Banerjee}, R.},
        title = "{Revealing the dynamics of Class 0 protostellar discs with ALMA}",
      journal = {\mnras},
         year = 2016,
        month = jun,
       volume = {459},
       number = {2},
        pages = {1892-1906},
          doi = {10.1093/mnras/stw785},
archivePrefix = {arXiv},
       eprint = {1601.02384},
 primaryClass = {astro-ph.SR},
       adsurl = {https://ui.adsabs.harvard.edu/abs/2016MNRAS.459.1892S}
}

@ARTICLE{Tanaka2017,
       author = {{Tanaka}, Kei E.~I. and {Tan}, Jonathan C. and {Zhang}, Yichen},
        title = "{The Impact of Feedback During Massive Star Formation by Core Accretion}",
      journal = {\apj},
         year = 2017,
        month = jan,
       volume = {835},
       number = {1},
          eid = {32},
        pages = {32},
          doi = {10.3847/1538-4357/835/1/32},
archivePrefix = {arXiv},
       eprint = {1610.08856},
 primaryClass = {astro-ph.SR},
       adsurl = {https://ui.adsabs.harvard.edu/abs/2017ApJ...835...32T}
}

@ARTICLE{Tanaka2020,
       author = {{Tanaka}, Kei E.~I. and {Zhang}, Yichen and {Hirota}, Tomoya and {Sakai}, Nami and {Motogi}, Kazuhito and {Tomida}, Kengo and {Tan}, Jonathan C. and {Rosero}, Viviana and {Higuchi}, Aya E. and {Ohashi}, Satoshi and {Liu}, Mengyao and {Sugiyama}, Koichiro},
        title = "{Salt, Hot Water, and Silicon Compounds Tracing Massive Twin Disks}",
      journal = {\apjl},
         year = 2020,
        month = sep,
       volume = {900},
       number = {1},
          eid = {L2},
        pages = {L2},
          doi = {10.3847/2041-8213/abadfc},
archivePrefix = {arXiv},
       eprint = {2007.02962},
 primaryClass = {astro-ph.SR},
       adsurl = {https://ui.adsabs.harvard.edu/abs/2020ApJ...900L...2T}
}

@ARTICLE{Taniguchi2023,
       author = {{Taniguchi}, Kotomi and {Sanhueza}, Patricio and {Olguin}, Fernando A. and {Gorai}, Prasanta and {Das}, Ankan and {Nakamura}, Fumitaka and {Saito}, Masao and {Zhang}, Qizhou and {Lu}, Xing and {Li}, Shanghuo and {Chen}, Huei-Ru Vivien},
        title = "{Digging into the Interior of Hot Cores with the ALMA (DIHCA). III. The Chemical Link between NH$_{2}$CHO, HNCO, and H$_{2}$CO}",
      journal = {\apj},
         year = 2023,
        month = jun,
       volume = {950},
       number = {1},
          eid = {57},
        pages = {57},
          doi = {10.3847/1538-4357/acca1d},
archivePrefix = {arXiv},
       eprint = {2304.00267},
 primaryClass = {astro-ph.GA},
       adsurl = {https://ui.adsabs.harvard.edu/abs/2023ApJ...950...57T}
}

@ARTICLE{Urquhart2018,
       author = {{Urquhart}, J.~S. and {K{\"o}nig}, C. and {Giannetti}, A. and {Leurini}, S. and {Moore}, T.~J.~T. and {Eden}, D.~J. and {Pillai}, T. and {Thompson}, M.~A. and {Braiding}, C. and {Burton}, M.~G. and {Csengeri}, T. and {Dempsey}, J.~T. and {Figura}, C. and {Froebrich}, D. and {Menten}, K.~M. and {Schuller}, F. and {Smith}, M.~D. and {Wyrowski}, F.},
        title = "{ATLASGAL - properties of a complete sample of Galactic clumps}",
      journal = {\mnras},
         year = 2018,
        month = jan,
       volume = {473},
       number = {1},
        pages = {1059-1102},
          doi = {10.1093/mnras/stx2258},
archivePrefix = {arXiv},
       eprint = {1709.00392},
 primaryClass = {astro-ph.GA},
       adsurl = {https://ui.adsabs.harvard.edu/abs/2018MNRAS.473.1059U}
}

@ARTICLE{Yamamuro2025,
       author = {{Yamamuro}, Ryota and {Tanaka}, Kei E.~I. and {Okuzumi}, Satoshi},
        title = "{The Impact of Silicate Grain Coagulation on Millimeter Emission from Massive Protostellar Disks}",
      journal = {\apj},
         year = 2025,
        month = sep,
       volume = {990},
       number = {1},
          eid = {59},
        pages = {59},
          doi = {10.3847/1538-4357/adf49c},
archivePrefix = {arXiv},
       eprint = {2508.17727},
 primaryClass = {astro-ph.SR},
       adsurl = {https://ui.adsabs.harvard.edu/abs/2025ApJ...990...59Y}
}

@ARTICLE{Yano2024,
       author = {{Yano}, Yuta and {Nakamura}, Fumitaka and {Kinoshita}, Shinichi. W.},
        title = "{Dense Core Collisions in Molecular Clouds: Formation of Streamers and Binary Stars}",
      journal = {\apj},
         year = 2024,
        month = apr,
       volume = {964},
       number = {2},
          eid = {119},
        pages = {119},
          doi = {10.3847/1538-4357/ad2a54},
archivePrefix = {arXiv},
       eprint = {2402.11147},
 primaryClass = {astro-ph.GA},
       adsurl = {https://ui.adsabs.harvard.edu/abs/2024ApJ...964..119Y}
}

@ARTICLE{Yorke1999,
       author = {{Yorke}, Harold W. and {Bodenheimer}, Peter},
        title = "{The Formation of Protostellar Disks. III. The Influence of Gravitationally Induced Angular Momentum Transport on Disk Structure and Appearance}",
      journal = {\apj},
         year = 1999,
        month = nov,
       volume = {525},
       number = {1},
        pages = {330-342},
          doi = {10.1086/307867},
       adsurl = {https://ui.adsabs.harvard.edu/abs/1999ApJ...525..330Y}
}

@ARTICLE{Zapata2019,
       author = {{Zapata}, Luis A. and {Garay}, Guido and {Palau}, Aina and {Rodr{\'\i}guez}, Luis F. and {Fern{\'a}ndez-L{\'o}pez}, Manuel and {Estalella}, Robert and {Guzm{\'a}n}, Andres},
        title = "{An Asymmetric Keplerian Disk Surrounding the O-type Protostar IRAS 16547-4247}",
      journal = {\apj},
         year = 2019,
        month = feb,
       volume = {872},
       number = {2},
          eid = {176},
        pages = {176},
          doi = {10.3847/1538-4357/aafedf},
archivePrefix = {arXiv},
       eprint = {1901.04896},
 primaryClass = {astro-ph.SR},
       adsurl = {https://ui.adsabs.harvard.edu/abs/2019ApJ...872..176Z}
}

@ARTICLE{Zapata2020,
       author = {{Zapata}, Luis A. and {Ho}, Paul T.~P. and {Fern{\'a}ndez-L{\'o}pez}, Manuel and {Ccolque}, Estrella Guzm{\'a}n and {Rodr{\'\i}guez}, Luis F. and {Reyes-Vald{\'e}s}, Jos{\'e} and {Bally}, John and {Palau}, Aina and {Saito}, Masao and {Sanhueza}, Patricio and {Rivera-Ortiz}, P.~R. and {Rodr{\'\i}guez-Gonz{\'a}lez}, A.},
        title = "{Confirming the Explosive Outflow in G5.89 with ALMA}",
      journal = {\apjl},
         year = 2020,
        month = oct,
       volume = {902},
       number = {2},
          eid = {L47},
        pages = {L47},
          doi = {10.3847/2041-8213/abbd3f},
archivePrefix = {arXiv},
       eprint = {2010.13835},
 primaryClass = {astro-ph.GA},
       adsurl = {https://ui.adsabs.harvard.edu/abs/2020ApJ...902L..47Z}
}

@ARTICLE{Zapata2024,
       author = {{Zapata}, Luis A. and {Fern{\'a}ndez-L{\'o}pez}, Manuel and {Sanhueza}, Patricio and {Girart}, Josep M. and {Rodr{\'\i}guez}, Luis F. and {Cort{\'e}s}, Paulo and {Koch}, Patrick and {Beltr{\'a}n}, Maria T. and {Pattle}, Kate and {Beuther}, Henrik and {Saha}, Piyali and {Jiao}, Wenyu and {Xu}, Fengwei and {Lu}, Xing Walker and {Olguin}, Fernando and {Li}, Shanghuo and {Stephens}, Ian W. and {Kang}, Ji-hyun and {Cheng}, Yu and {Choudhury}, Spandan and {Morii}, Kaho and {Chung}, Eun Jung and {Wang}, Jia-Wei and {Hwang}, Jihye and {Lyo}, A. -Ran and {Zhang}, Q. and {Chen}, Huei-Ru Vivien},
        title = "{Magnetic Fields in Massive Star-forming Regions (MagMaR). IV. Tracing the Magnetic Fields in the O-type Protostellar System IRAS 16547{\textendash}4247}",
      journal = {\apj},
         year = 2024,
        month = oct,
       volume = {974},
       number = {2},
          eid = {257},
        pages = {257},
          doi = {10.3847/1538-4357/ad701d},
archivePrefix = {arXiv},
       eprint = {2408.10199},
 primaryClass = {astro-ph.SR},
       adsurl = {https://ui.adsabs.harvard.edu/abs/2024ApJ...974..257Z}
}

@ARTICLE{Zhang2019,
       author = {{Zhang}, Yichen and {Tan}, Jonathan C. and {Sakai}, Nami and {Tanaka}, Kei E.~I. and {De Buizer}, James M. and {Liu}, Mengyao and {Beltr{\'a}n}, Maria T. and {Kratter}, Kaitlin and {Mardones}, Diego and {Garay}, Guido},
        title = "{An Ordered Envelope-Disk Transition in the Massive Protostellar Source G339.88-1.26}",
      journal = {\apj},
         year = 2019,
        month = mar,
       volume = {873},
       number = {1},
          eid = {73},
        pages = {73},
          doi = {10.3847/1538-4357/ab0553},
archivePrefix = {arXiv},
       eprint = {1811.04381},
 primaryClass = {astro-ph.GA},
       adsurl = {https://ui.adsabs.harvard.edu/abs/2019ApJ...873...73Z}
}

@ARTICLE{Zhang2022,
       author = {{Zhang}, Yichen and {Tanaka}, Kei E.~I. and {Tan}, Jonathan C. and {Yang}, Yao-Lun and {Greco}, Eva and {Beltran}, Maria T. and {Sakai}, Nami and {De Buizer}, James M. and {Rosero}, Viviana and {Fedriani}, Rub{\'e}n and {Garay}, Guido},
        title = "{Massive Protostars in a Protocluster - A Multi-scale ALMA View of G35.20-0.74N.}",
      journal = {\apj},
         year = 2022,
        month = sep,
       volume = {936},
       number = {1},
          eid = {68},
        pages = {68},
          doi = {10.3847/1538-4357/ac847f},
archivePrefix = {arXiv},
       eprint = {2207.11320},
 primaryClass = {astro-ph.GA},
       adsurl = {https://ui.adsabs.harvard.edu/abs/2022ApJ...936...68Z}
}

@ARTICLE{Zhang2024,
       author = {{Zhang}, S. and {Cyganowski}, C.~J. and {Henshaw}, J.~D. and {Brogan}, C.~L. and {Hunter}, T.~R. and {Friesen}, R.~K. and {Bonnell}, I.~A. and {Viti}, S.},
        title = "{Filamentary mass accretion towards the high-mass protobinary system G11.92-0.61 MM2}",
      journal = {\mnras},
         year = 2024,
        month = sep,
       volume = {533},
       number = {1},
        pages = {1075-1094},
          doi = {10.1093/mnras/stae1844},
archivePrefix = {arXiv},
       eprint = {2407.19552},
 primaryClass = {astro-ph.GA},
       adsurl = {https://ui.adsabs.harvard.edu/abs/2024MNRAS.533.1075Z}
}

@ARTICLE{Zinnecker2007,
       author = {{Zinnecker}, Hans and {Yorke}, Harold W.},
        title = "{Toward Understanding Massive Star Formation}",
      journal = {\araa},
         year = 2007,
        month = sep,
       volume = {45},
       number = {1},
        pages = {481-563},
          doi = {10.1146/annurev.astro.44.051905.092549},
archivePrefix = {arXiv},
       eprint = {0707.1279},
 primaryClass = {astro-ph},
       adsurl = {https://ui.adsabs.harvard.edu/abs/2007ARA&A..45..481Z}
}

@software{2020zndo...4302846O,
       author = {{Olguin}, Fernando and {Sanhueza}, Patricio},
        title = "{GoContinuum: continuum finding tool}",
         year = 2020,
        month = dec,
          eid = {10.5281/zenodo.4302846},
          doi = {10.5281/zenodo.4302846},
      version = {v2.0.0},
    publisher = {Zenodo},
       adsurl = {https://ui.adsabs.harvard.edu/abs/2020zndo...4302846O}
}

@MISC{2018zndo...1216881C,
       author = {{Contreras}, Yanett},
        title = "{Automatic Line Clean}",
         year = 2018,
        month = apr,
          eid = {10.5281/zenodo.1216881},
          doi = {10.5281/zenodo.1216881},
      version = {1.0},
    publisher = {Zenodo},
  journal      = {Zenodo},
       adsurl = {https://ui.adsabs.harvard.edu/abs/2018zndo...1216881C}
}

@ARTICLE{2020SciPy-NMeth,
  author  = {Virtanen, Pauli and Gommers, Ralf and Oliphant, Travis E. and
            Haberland, Matt and Reddy, Tyler and Cournapeau, David and
            Burovski, Evgeni and Peterson, Pearu and Weckesser, Warren and
            Bright, Jonathan and {van der Walt}, St{\'e}fan J. and
            Brett, Matthew and Wilson, Joshua and Millman, K. Jarrod and
            Mayorov, Nikolay and Nelson, Andrew R. J. and Jones, Eric and
            Kern, Robert and Larson, Eric and Carey, C J and
            Polat, {\.I}lhan and Feng, Yu and Moore, Eric W. and
            {VanderPlas}, Jake and Laxalde, Denis and Perktold, Josef and
            Cimrman, Robert and Henriksen, Ian and Quintero, E. A. and
            Harris, Charles R. and Archibald, Anne M. and
            Ribeiro, Ant{\^o}nio H. and Pedregosa, Fabian and
            {van Mulbregt}, Paul and {SciPy 1.0 Contributors}},
  title   = {{{SciPy} 1.0: Fundamental Algorithms for Scientific
            Computing in Python}},
  journal = {Nature Methods},
  year    = {2020},
  volume  = {17},
  pages   = {261--272},
  adsurl  = {https://rdcu.be/b08Wh},
  doi     = {10.1038/s41592-019-0686-2},
}

@software{crameri_2023_8409685,
  author       = {Crameri, Fabio},
  title        = {Scientific colour maps},
  month        = oct,
  year         = 2023,
  publisher    = {Zenodo},
  journal      = {Zenodo},
  version      = {8.0.1},
  doi          = {10.5281/zenodo.8409685},
  url          = {https://doi.org/10.5281/zenodo.8409685},
}

@ARTICLE{2020NatCo..11.5444C,
       author = {{Crameri}, Fabio and {Shephard}, Grace E. and {Heron}, Philip J.},
        title = "{The misuse of colour in science communication}",
      journal = {Nature Communications},
         year = 2020,
        month = oct,
       volume = {11},
          eid = {5444},
        pages = {5444},
          doi = {10.1038/s41467-020-19160-7},
       adsurl = {https://ui.adsabs.harvard.edu/abs/2020NatCo..11.5444C}
}

@ARTICLE{2022PASP..134k4501C,
       author = {{CASA Team} and {Bean}, Ben and {Bhatnagar}, Sanjay and {Castro}, Sandra and {Donovan Meyer}, Jennifer and {Emonts}, Bjorn and {Garcia}, Enrique and {Garwood}, Robert and {Golap}, Kumar and {Villalba}, Justo Gonzalez and {Harris}, Pamela and {Hayashi}, Yohei and {Hoskins}, Josh and {Hsieh}, Mingyu and {Jagannathan}, Preshanth and {Kawasaki}, Wataru and {Keimpema}, Aard and {Kettenis}, Mark and {Lopez}, Jorge and {Marvil}, Joshua and {Masters}, Joseph and {McNichols}, Andrew and {Mehringer}, David and {Miel}, Renaud and {Moellenbrock}, George and {Montesino}, Federico and {Nakazato}, Takeshi and {Ott}, Juergen and {Petry}, Dirk and {Pokorny}, Martin and {Raba}, Ryan and {Rau}, Urvashi and {Schiebel}, Darrell and {Schweighart}, Neal and {Sekhar}, Srikrishna and {Shimada}, Kazuhiko and {Small}, Des and {Steeb}, Jan-Willem and {Sugimoto}, Kanako and {Suoranta}, Ville and {Tsutsumi}, Takahiro and {van Bemmel}, Ilse M. and {Verkouter}, Marjolein and {Wells}, Akeem and {Xiong}, Wei and {Szomoru}, Arpad and {Griffith}, Morgan and {Glendenning}, Brian and {Kern}, Jeff},
        title = "{CASA, the Common Astronomy Software Applications for Radio Astronomy}",
      journal = {\pasp},
         year = 2022,
        month = nov,
       volume = {134},
       number = {1041},
          eid = {114501},
        pages = {114501},
          doi = {10.1088/1538-3873/ac9642},
       adsurl = {https://ui.adsabs.harvard.edu/abs/2022PASP..134k4501C}
}

@article{astropy:2013,
Adsurl = {http://adsabs.harvard.edu/abs/2013A%26A...558A..33A},
Archiveprefix = {arXiv},
Author = {{Astropy Collaboration} and {Robitaille}, T.~P. and {Tollerud}, E.~J. and {Greenfield}, P. and {Droettboom}, M. and {Bray}, E. and {Aldcroft}, T. and {Davis}, M. and {Ginsburg}, A. and {Price-Whelan}, A.~M. and {Kerzendorf}, W.~E. and {Conley}, A. and {Crighton}, N. and {Barbary}, K. and {Muna}, D. and {Ferguson}, H. and {Grollier}, F. and {Parikh}, M.~M. and {Nair}, P.~H. and {Unther}, H.~M. and {Deil}, C. and {Woillez}, J. and {Conseil}, S. and {Kramer}, R. and {Turner}, J.~E.~H. and {Singer}, L. and {Fox}, R. and {Weaver}, B.~A. and {Zabalza}, V. and {Edwards}, Z.~I. and {Azalee Bostroem}, K. and {Burke}, D.~J. and {Casey}, A.~R. and {Crawford}, S.~M. and {Dencheva}, N. and {Ely}, J. and {Jenness}, T. and {Labrie}, K. and {Lim}, P.~L. and {Pierfederici}, F. and {Pontzen}, A. and {Ptak}, A. and {Refsdal}, B. and {Servillat}, M. and {Streicher}, O.},
Doi = {10.1051/0004-6361/201322068},
Eid = {A33},
Eprint = {1307.6212},
Journal = {\aap},
Month = oct,
Pages = {A33},
Primaryclass = {astro-ph.IM},
Title = {{Astropy: A community Python package for astronomy}},
Volume = 558,
Year = 2013}

@ARTICLE{astropy:2018,
       author = {{Astropy Collaboration} and {Price-Whelan}, A.~M. and
         {Sip{\H{o}}cz}, B.~M. and {G{\"u}nther}, H.~M. and {Lim}, P.~L. and
         {Crawford}, S.~M. and {Conseil}, S. and {Shupe}, D.~L. and
         {Craig}, M.~W. and {Dencheva}, N. and {Ginsburg}, A. and {Vand
        erPlas}, J.~T. and {Bradley}, L.~D. and {P{\'e}rez-Su{\'a}rez}, D. and
         {de Val-Borro}, M. and {Aldcroft}, T.~L. and {Cruz}, K.~L. and
         {Robitaille}, T.~P. and {Tollerud}, E.~J. and {Ardelean}, C. and
         {Babej}, T. and {Bach}, Y.~P. and {Bachetti}, M. and {Bakanov}, A.~V. and
         {Bamford}, S.~P. and {Barentsen}, G. and {Barmby}, P. and
         {Baumbach}, A. and {Berry}, K.~L. and {Biscani}, F. and {Boquien}, M. and
         {Bostroem}, K.~A. and {Bouma}, L.~G. and {Brammer}, G.~B. and
         {Bray}, E.~M. and {Breytenbach}, H. and {Buddelmeijer}, H. and
         {Burke}, D.~J. and {Calderone}, G. and {Cano Rodr{\'\i}guez}, J.~L. and
         {Cara}, M. and {Cardoso}, J.~V.~M. and {Cheedella}, S. and {Copin}, Y. and
         {Corrales}, L. and {Crichton}, D. and {D'Avella}, D. and {Deil}, C. and
         {Depagne}, {\'E}. and {Dietrich}, J.~P. and {Donath}, A. and
         {Droettboom}, M. and {Earl}, N. and {Erben}, T. and {Fabbro}, S. and
         {Ferreira}, L.~A. and {Finethy}, T. and {Fox}, R.~T. and
         {Garrison}, L.~H. and {Gibbons}, S.~L.~J. and {Goldstein}, D.~A. and
         {Gommers}, R. and {Greco}, J.~P. and {Greenfield}, P. and
         {Groener}, A.~M. and {Grollier}, F. and {Hagen}, A. and {Hirst}, P. and
         {Homeier}, D. and {Horton}, A.~J. and {Hosseinzadeh}, G. and {Hu}, L. and
         {Hunkeler}, J.~S. and {Ivezi{\'c}}, {\v{Z}}. and {Jain}, A. and
         {Jenness}, T. and {Kanarek}, G. and {Kendrew}, S. and {Kern}, N.~S. and
         {Kerzendorf}, W.~E. and {Khvalko}, A. and {King}, J. and {Kirkby}, D. and
         {Kulkarni}, A.~M. and {Kumar}, A. and {Lee}, A. and {Lenz}, D. and
         {Littlefair}, S.~P. and {Ma}, Z. and {Macleod}, D.~M. and
         {Mastropietro}, M. and {McCully}, C. and {Montagnac}, S. and
         {Morris}, B.~M. and {Mueller}, M. and {Mumford}, S.~J. and {Muna}, D. and
         {Murphy}, N.~A. and {Nelson}, S. and {Nguyen}, G.~H. and
         {Ninan}, J.~P. and {N{\"o}the}, M. and {Ogaz}, S. and {Oh}, S. and
         {Parejko}, J.~K. and {Parley}, N. and {Pascual}, S. and {Patil}, R. and
         {Patil}, A.~A. and {Plunkett}, A.~L. and {Prochaska}, J.~X. and
         {Rastogi}, T. and {Reddy Janga}, V. and {Sabater}, J. and
         {Sakurikar}, P. and {Seifert}, M. and {Sherbert}, L.~E. and
         {Sherwood-Taylor}, H. and {Shih}, A.~Y. and {Sick}, J. and
         {Silbiger}, M.~T. and {Singanamalla}, S. and {Singer}, L.~P. and
         {Sladen}, P.~H. and {Sooley}, K.~A. and {Sornarajah}, S. and
         {Streicher}, O. and {Teuben}, P. and {Thomas}, S.~W. and
         {Tremblay}, G.~R. and {Turner}, J.~E.~H. and {Terr{\'o}n}, V. and
         {van Kerkwijk}, M.~H. and {de la Vega}, A. and {Watkins}, L.~L. and
         {Weaver}, B.~A. and {Whitmore}, J.~B. and {Woillez}, J. and
         {Zabalza}, V. and {Astropy Contributors}},
        title = "{The Astropy Project: Building an Open-science Project and Status of the v2.0 Core Package}",
      journal = {\aj},
         year = 2018,
        month = sep,
       volume = {156},
       number = {3},
          eid = {123},
        pages = {123},
          doi = {10.3847/1538-3881/aabc4f},
archivePrefix = {arXiv},
       eprint = {1801.02634},
 primaryClass = {astro-ph.IM},
       adsurl = {https://ui.adsabs.harvard.edu/abs/2018AJ....156..123A}
}

@ARTICLE{astropy:2022,
       author = {{Astropy Collaboration} and {Price-Whelan}, Adrian M. and {Lim}, Pey Lian and {Earl}, Nicholas and {Starkman}, Nathaniel and {Bradley}, Larry and {Shupe}, David L. and {Patil}, Aarya A. and {Corrales}, Lia and {Brasseur}, C.~E. and {N{"o}the}, Maximilian and {Donath}, Axel and {Tollerud}, Erik and {Morris}, Brett M. and {Ginsburg}, Adam and {Vaher}, Eero and {Weaver}, Benjamin A. and {Tocknell}, James and {Jamieson}, William and {van Kerkwijk}, Marten H. and {Robitaille}, Thomas P. and {Merry}, Bruce and {Bachetti}, Matteo and {G{"u}nther}, H. Moritz and {Aldcroft}, Thomas L. and {Alvarado-Montes}, Jaime A. and {Archibald}, Anne M. and {B{'o}di}, Attila and {Bapat}, Shreyas and {Barentsen}, Geert and {Baz{'a}n}, Juanjo and {Biswas}, Manish and {Boquien}, M{'e}d{'e}ric and {Burke}, D.~J. and {Cara}, Daria and {Cara}, Mihai and {Conroy}, Kyle E. and {Conseil}, Simon and {Craig}, Matthew W. and {Cross}, Robert M. and {Cruz}, Kelle L. and {D'Eugenio}, Francesco and {Dencheva}, Nadia and {Devillepoix}, Hadrien A.~R. and {Dietrich}, J{"o}rg P. and {Eigenbrot}, Arthur Davis and {Erben}, Thomas and {Ferreira}, Leonardo and {Foreman-Mackey}, Daniel and {Fox}, Ryan and {Freij}, Nabil and {Garg}, Suyog and {Geda}, Robel and {Glattly}, Lauren and {Gondhalekar}, Yash and {Gordon}, Karl D. and {Grant}, David and {Greenfield}, Perry and {Groener}, Austen M. and {Guest}, Steve and {Gurovich}, Sebastian and {Handberg}, Rasmus and {Hart}, Akeem and {Hatfield-Dodds}, Zac and {Homeier}, Derek and {Hosseinzadeh}, Griffin and {Jenness}, Tim and {Jones}, Craig K. and {Joseph}, Prajwel and {Kalmbach}, J. Bryce and {Karamehmetoglu}, Emir and {Ka{l}uszy{'n}ski}, Miko{l}aj and {Kelley}, Michael S.~P. and {Kern}, Nicholas and {Kerzendorf}, Wolfgang E. and {Koch}, Eric W. and {Kulumani}, Shankar and {Lee}, Antony and {Ly}, Chun and {Ma}, Zhiyuan and {MacBride}, Conor and {Maljaars}, Jakob M. and {Muna}, Demitri and {Murphy}, N.~A. and {Norman}, Henrik and {O'Steen}, Richard and {Oman}, Kyle A. and {Pacifici}, Camilla and {Pascual}, Sergio and {Pascual-Granado}, J. and {Patil}, Rohit R. and {Perren}, Gabriel I. and {Pickering}, Timothy E. and {Rastogi}, Tanuj and {Roulston}, Benjamin R. and {Ryan}, Daniel F. and {Rykoff}, Eli S. and {Sabater}, Jose and {Sakurikar}, Parikshit and {Salgado}, Jes{'u}s and {Sanghi}, Aniket and {Saunders}, Nicholas and {Savchenko}, Volodymyr and {Schwardt}, Ludwig and {Seifert-Eckert}, Michael and {Shih}, Albert Y. and {Jain}, Anany Shrey and {Shukla}, Gyanendra and {Sick}, Jonathan and {Simpson}, Chris and {Singanamalla}, Sudheesh and {Singer}, Leo P. and {Singhal}, Jaladh and {Sinha}, Manodeep and {Sip{H{o}}cz}, Brigitta M. and {Spitler}, Lee R. and {Stansby}, David and {Streicher}, Ole and {Šumak}, Jani and {Swinbank}, John D. and {Taranu}, Dan S. and {Tewary}, Nikita and {Tremblay}, Grant R. and {Val-Borro}, Miguel de and {Van Kooten}, Samuel J. and {Vasovi{'c}}, Zlatan and {Verma}, Shresth and {de Miranda Cardoso}, Jos{'e} Vin{'i}cius and {Williams}, Peter K.~G. and {Wilson}, Tom J. and {Winkel}, Benjamin and {Wood-Vasey}, W.~M. and {Xue}, Rui and {Yoachim}, Peter and {Zhang}, Chen and {Zonca}, Andrea and {Astropy Project Contributors}},
        title = "{The Astropy Project: Sustaining and Growing a Community-oriented Open-source Project and the Latest Major Release (v5.0) of the Core Package}",
      journal = {\apj},
         year = 2022,
        month = aug,
       volume = {935},
       number = {2},
          eid = {167},
        pages = {167},
          doi = {10.3847/1538-4357/ac7c74},
archivePrefix = {arXiv},
       eprint = {2206.14220},
 primaryClass = {astro-ph.IM},
       adsurl = {https://ui.adsabs.harvard.edu/abs/2022ApJ...935..167A}
}

@software{olguin_2025_17197133,
  author       = {Olguin, Fernando},
  title        = {Improved and updated YCLEAN},
  month        = sep,
  year         = 2025,
  publisher    = {Zenodo},
  version      = {v2.3.0},
  doi          = {10.5281/zenodo.17197133},
  url          = {https://doi.org/10.5281/zenodo.17197133},
  swhid        = {swh:1:dir:e3c8520f31d87031b313a1fe46028dac55aac32c
                   ;origin=https://doi.org/10.5281/zenodo.17197132;vi
                   sit=swh:1:snp:bc826abd60d092669d1692d2c3dd759c702b
                   6b8a;anchor=swh:1:rel:3abd3e5700df8ab6850db1a26cf3
                   74a56a5405ff;path=folguinch-yclean-bb728c9
                  },
}

@software{2015ascl.soft02007M,
       author = {{Mohan}, Niruj and {Rafferty}, David},
        title = "{PyBDSF: Python Blob Detection and Source Finder}",
 howpublished = {Astrophysics Source Code Library, record ascl:1502.007},
         year = 2015,
        month = feb,
          eid = {ascl:1502.007},
archivePrefix = {ascl},
       eprint = {1502.007},
       adsurl = {https://ui.adsabs.harvard.edu/abs/2015ascl.soft02007M}
}
\bibliographystyle{aasjournalv7}

\end{document}